\documentclass[a4paper,fleqn,usenatbib,useAMS]{mnras}

\usepackage{graphicx}	
\usepackage{multicol}        
\usepackage{pdflscape}	
\usepackage{afterpage}
\usepackage{mathastext}
\usepackage{comment}
\usepackage{epstopdf}

\usepackage{tikz}

\bibpunct{(}{)}{;}{a}{,}{;}
\def\checkmark{\tikz\fill[scale=0.4](0,.35) -- (.25,0) -- (1,.7) -- (.25,.15) -- cycle;}
\newcommand{\citeb}[1]{\citeauthor{#1}, \citeyear{#1}}

\usepackage[T1]{fontenc}
\usepackage{ae,aecompl}

\usepackage{txfonts}

\title[Dust build-up and decline as galaxies evolve]{\textit{Herschel}-ATLAS:  Revealing dust build-up and decline across gas, dust and stellar mass selected samples:\\ I. Scaling relations}
\author[P. De Vis et al.]
{P.~De Vis$^{1,2}$\thanks{E-mail:
pieter.devis@pg.canterbury.ac.nz}, L.~Dunne$^{3,4}$, S.~Maddox$^{3,4}$, H.L.~Gomez$^{3}$, C.J.R.~Clark$^{3}$, A.E.~Bauer$^{5}$,
\newauthor
S.~Viaene$^{2}$, S.P.~Schofield$^{3}$, M.~Baes$^{2}$, A.J.~Baker$^{6}$, N.~Bourne$^{3}$,  S.P.~Driver$^{7}$, S.~Dye$^{8}$,
\newauthor
S.A.~Eales$^{2}$, C.~Furlanetto$^{8,9}$, R.J.~Ivison$^{4,10}$, A.S.G.~Robotham$^{7}$,
K.~Rowlands$^{11}$,
\newauthor
D.J.B.~Smith$^{12}$, M.W.L.~Smith$^{2}$, E.~Valiante$^{2}$, A.H.~Wright$^{7}$ \\
\\
$^{1}$Department of Physics \& Astronomy, University of Canterbury, Private Bag 4800, Christchurch, New Zealand\\
$^{2}$Sterrenkundig Observatorium, Universiteit Gent, Krijgslaan 281, B-9000 Gent, Belgium \\
$^{3}$School of Physics \& Astronomy, Cardiff University, Queen's Buildings, The Parade, Cardiff, CF24 3AA, UK \\
$^{4}$Institute for Astronomy, The University of Edinburgh, Royal Observatory, Blackford Hill, Edinburgh, EH9 3HJ, UK \\
$^{5}$Australian Astronomical Observatory, PO Box 915, North Ryde, NSW 1670, Australia \\
$^{6}$Department of Physics and Astronomy, Rutgers, The State University of New Jersey, \\
$\,$ 136 Frelinghuysen Road, Piscataway, NJ 08854-8019, U.S.A.\\
$^{7}$International Centre for Radio Astronomy Research, The University of Western Australia, Crawley, Perth, 6009, Australia\\
$^{8}$School of Physics \& Astronomy, University of Nottingham, University Park, Nottingham, NG7 2RD, UK \\
$^{9}$CAPES Foundation, Ministry of Education of Brazil, Braslia/DF, 70040-020, Brazil \\
$^{10}$European Southern Observatory, Karl Schwarzschild Strasse 2, Garching, D85748, Germany \\
$^{11}$School of Physics \& Astronomy, University of St Andrews, North Haugh, St Andrews, KY16 9SS, UK \\
$^{12}$Centre for Astrophysics, Science \& Technology Research Institute, University of Hertfordshire, Hatfield, Herts, AL10 9AB, UK}
\begin{document}

\pagerange{\pageref{firstpage}--\pageref{lastpage}}
\pubyear{2016}

\maketitle

\label{firstpage}
\begin{abstract}
   We present a study of the dust, stars and atomic gas (H{\sc i})
  in an H{\sc i}-selected sample of local galaxies ($z<0.035$) in the
  \textit{Herschel} Astrophysical Terahertz Large Area Survey (H-ATLAS)
  fields. This H{\sc i}-selected sample reveals a population of very
    high gas fraction ($>80$~per~cent), low stellar mass sources that
    appear to be in the earliest stages of their evolution.
    We compare this sample with dust and stellar mass selected samples to study the dust and gas scaling relations over a wide range of gas fraction (proxy for evolutionary state of a galaxy). The most robust scaling relations for gas and dust are those linked to NUV-\textit{r} (SSFR) and gas fraction, these do not depend on sample selection or environment. At the highest gas fractions, our additional sample shows the dust content is well below expectations from extrapolating scaling relations for more evolved sources, and dust is not a good tracer of the gas content. The specific dust mass for local galaxies peaks at a gas fraction of $\sim$75\,per\,cent. The atomic gas depletion time is also longer for high gas fraction galaxies, opposite to the trend found for molecular gas depletion timescale. We link this trend to the changing efficiency of conversion of H{\sc i} to $\rm{H_2}$ as galaxies increase in stellar mass surface density as they evolve. Finally, we show that galaxies start out barely obscured and increase in obscuration as they evolve, yet there is no clear and simple link between obscuration and global galaxy properties.
\end{abstract}


\section{Introduction}
About 30 to 50~per~cent of the optical/UV radiative energy produced by stars and AGN in galaxies is absorbed by dust and thermally re-emitted in the Far-InfraRed (FIR)
and submillimetre (submm) parts of the spectrum \citep{Fixsen1996,Driver2016,Viaene2016}. It is therefore
difficult to develop a thorough understanding of galaxy evolution
without also understanding the InterStellar Medium (ISM). Dust
properties have been investigated for several decades using space missions such
as {\em IRAS} \citep{Neugebauer1984}, {\em ISO} \citep{Kessler1996} and {\em Spitzer}
\citep{Werner2004} and ground based submillimetre
instruments such as SCUBA \citep{Holland1999}, MAMBO \citep{Kreysa1998}
and LABOCA \citep{Kreysa2003}. However, with the advent of the
\textit{Herschel Space Observatory}\footnote{\textit{Herschel} is an ESA space
observatory with science instruments
provided by European-led Principal Investigator consortia and
with important participation from NASA.} \citep{Pilbratt2010} we have
entered a new era for interstellar dust studies. \textit{Herschel} has
superior angular resolution and sensitivity compared to previous FIR
facilities and operates right across the peak of the dust SED ($70-500 \mu m$). This makes it
sensitive to the diffuse cold ($T<25$ K) dust component that dominates
the dust mass in galaxies \citep{Devereux1990B,Dunne2001,Draine2007,Clark2015}, as well as warmer ($T>30$ K) dust radiating at
shorter wavelengths which often dominates the dust luminosity.
The consensus is that the warm dust component is heated by star-forming regions
\citep{Devereux1990,Kennicutt1998,Calzetti2007,Boquien2010,Verley2010,Bendo2012},
and the cold dust component (which makes up the bulk of the dust mass) can
be heated by both star-forming regions and older stellar populations
\citep{Bendo2015}.

\textit{Herschel} is uniquely suited to studying the role played by
dust in the evolutionary history of galaxies. The first logical step
is to quantify how the dust content of galaxies
varies with galactic properties such as stellar mass, colour, gas
content, star formation rate (SFR), and other parameters. The
resulting scaling relations provide vital information about the
interplay of dust, gas and the star formation cycle, leading to
important insights into the physical processes regulating galaxy
evolution \citep[e.g.][]{Dunne2011} and providing strong constraints
on chemical evolution models \citep[e.g.][]{Rowlands2014b,Zhukovska2014}.  Before
{\em Herschel}, the main scaling relations studied were
global relations between dust, gas and stellar masses
\citep[e.g.][]{Devereux1990,Sanders1991,Dunne2000,Driver2007} and the evolution of the
dust-to-gas ratio with stellar mass and metallicity \citep[e.g.][]
{Issa1990,Lisenfeld1998,James2002,Draine2007}.  These studies showed
a strong correlation between dust and gas mass, and
found an increase of the dust-to-gas mass ratio as a function of
stellar mass and metallicity, though there is often disagreement in
the exact slope of the relationships. More recently,
\citet{daCunha2010} used {\em IRAS} data to show that the
dust-to-stellar mass ratio strongly correlates with specific star
formation rate (SSFR), as predicted by chemical evolution models. This
result has since been supported by further {\em Herschel} studies
\citep{SmithD2012,Sandstrom2013,Rowlands2014}.

Since then, {\em Herschel} has expanded on these studies to include
the cold dust component and explored a much wider range of galaxy
types and luminosities, in far greater numbers, than was possible previously.
The {\em Herschel} Reference
  survey (HRS, \citeb{Boselli2010}) is a quasi stellar mass selected
  sample of 323 local galaxies. Various HRS studies have derived
  scaling relations between the gas, dust and star formation
  properties as well as trends with FIR/submm and UV colours, stellar
  mass, morphology and environment. Next to providing benchmark scaling
  relations, these works found cluster galaxies are characterized by a significantly
  lower atomic, molecular, and dust mass content than similar stellar
  mass galaxies in the field.
  \citep{Cortese2011,Cortese2012,Cortese2012B,Cortese2014,Boselli2012,Boselli2013,Boselli2014B,
    Boselli2015}.  The {\em Herschel}-ATLAS (H-ATLAS,
  \citeb{Eales2010}) is a blind, large-area submm survey which
  provides an unbiased and unrivalled view of the nearby dusty
  Universe. Dust scaling relations in H-ATLAS have been studied by
  \citet{Bourne2012} through stacking $\sim 80000$ optically selected
  galaxies and also by \citep{SmithD2011} who used fits to the UV-FIR photometry of
  1402 $250 \mu m$-selected sources. More recently H-ATLAS has produced a local volume limited sample, and \citeauthor{Clark2015} (2015,
  hereafter C15) used it to study the dust properties of the first dust mass selected sample of galaxies in the local Universe.

C15 show that stellar mass selected samples are biased towards
galaxies that have converted a lot of their gas into stars,
i.e. towards more evolved galaxies, and thus under-represent immature
high gas fraction sources. Dust selection produces a more uniform
range of gas fractions but preferentially samples galaxies near the
peak of their dust content. In this work we compare a
  local, H{\sc i}-selected sample from the H-ATLAS equatorial fields
  to these stellar and dust mass selected samples. We will highlight
  scaling relations concerning dust properties as these have not been
  studied before for H{\sc i}-selected samples. Since H{\sc i}
  selection will bias us towards galaxies with high gas fractions, we
  can populate the scaling relations for these hitherto missing
  immature galaxies and, for the first time, study their dust
  properties. By comparing the three samples selected by stellar, dust
  and atomic gas content, we span a large range of gas fractions and
  can study the relationship of dust, gas and stars across as wide a
  range of evolutionary status as possible.

This paper is organized as follows. In Section 2 we describe the
observations, the sample selection, the extended source photometry
pipeline and the {\sc magphys} SED fitting code that was used to
obtain the galactic properties. In Section 3 we discuss the different
surveys used in this work. In Section 4 we compare the dust, gas and
stellar content of the H{\sc i}, dust and stellar mass selected samples. In
Section 5 we study the evolution of the star formation efficiency and in Section 6 we investigate the dust heating. Finally in Section 7 we study the obscuration for the different samples.
We adopt the cosmology of Planck
Collaboration et al. (2013), specifically H$_0 = 67.30 \rm \,km
s^{-1}\, Mpc^{-1}$, $\Omega_m = 0.315$, and $\Omega_{\Lambda} =
0.685$.

\section{The data}
In order to obtain a sample of galaxies with sufficient
multi-wavelength information to determine the stellar, dust and atomic
gas (H{\sc i}) content, it is necessary to select an area of sky which
has been surveyed in the optical, in the submillimeter and at 21 cm. The
ideal fields with the necessary multi-wavelength data are the three
equatorial fields ( $\sim $160 deg$^2$) of the \textit{Herschel}-ATLAS
(H-ATLAS; \citeb{Eales2010}) which have excellent multi-wavelength
auxiliary data and overlap with the Galaxy And Mass Assembly
spectroscopic survey (GAMA; \citeb{Driver2009}). The H{\sc i} Parkes
All-Sky Survey (HIPASS; \citeb{Barnes2001}), supplemented by the
Arecibo Legacy Fast ALFA Survey (ALFALFA; \citeb{Giovanelli2005}) is
used to determine the atomic gas properties.

\subsection{Observations}
The H{\sc i} Parkes All Sky Survey (HIPASS;
\citeb{Barnes2001}; \citeb{Meyer2004}) provides 21cm coverage over the
equatorial H-ATLAS/GAMA fields. The Parkes beamsize is 15.5 arcmin, the velocity
resolution is 18 km s$^{-1}$ and the rms noise is 13 mJy
beam$^{-1}$ in a channel of this width. The HIPASS catalogue (HICAT, \citeb{Meyer2004};
\citeb{Zwaan2004}; \citeb{Wong2006}) is used to identify our sources
and extract the basic H{\sc i}-parameters.

The HIPASS data are supplemented by observations from the Arecibo
Legacy Fast ALFA Survey (ALFALFA, \citeb{Giovanelli2005};
\citeb{Haynes2011}; Haynes, \textit{priv comm.}). With a beamsize of
$\sim 3.5$ arcmin and rms noise of $\sim 2$ mJy beam$^{-1}$ (for 11 km s$^{-1}$ channels), ALFALFA outperforms
HIPASS in both sensitivity and resolution. It does not however, have
full coverage over the three equatorial H-ATLAS/GAMA fields in this
study. For this reason we use HIPASS data supplemented with ALFALFA
where available.

The uniqueness and strength of this H{\sc i}-selected sample is that
it makes use of the H-ATLAS - the largest extra-galactic submm survey
covering $\sim$ 600 deg$^2$ in 5 bands from 100-500 $\mu m$. The
H-ATLAS observations were carried out in parallel mode using the
Photodetector Array Camera and Spectrometer (PACS,
\citeb{Poglitsch2010}) and the Spectral and Photometric Imaging
REceiver (SPIRE, \citeb{Griffin2010}) instruments on board the
\textit{Herschel Space Observatory}. This work makes use of the
H-ATLAS Phase 1 public data release, hereafter `DR1' \citep{Valiante2016,Bourne2016}.
(More details on the H-ATLAS data reduction can be found in \citet{Valiante2016}.)
To determine counterparts to our H{\sc i}-selected sources, we use the DR1 catalogue of $4\sigma$
detections at 250\,$\mu m$ \citep{Valiante2016} produced using the MAD-X algorithm (Maddox et al., \textit{in
prep.}). Optical counterparts to H-ATLAS sources were found by direct
comparison with the SDSS DR7 \citep{Abazajian2009} and DR9 \citep{Ahn2012} by means of
matching H-ATLAS sources to SDSS objects within a 10 arcsecond radius
using a likelihood ratio technique, where only SDSS
sources with a reliability $R>0.8$ are considered
to be likely matches to the H-ATLAS sources \citep{SmithD2011,Bourne2016}.

For ultraviolet, optical and near-infrared data, we use images
compiled by the Galaxy And Mass Assembly spectroscopic survey (GAMA;
\citeb{Driver2011}; \citeb{Hopkins2013}; \citeb{Liske2015}). GAMA provides spectroscopic redshifts, along with
supplementary reductions of ultraviolet (UV) {\em GALEX}
\citep{Morrissey2007,Seibert2012}, optical SDSS DR6 \citep{Adelman-McCarthy2008},
Near-InfraRed (NIR) VISTA VIKING \citep{Sutherland2012} and
Mid-InfraRed (MIR) WISE \citep{Wright2010,Cluver2014} data. Details
of these reprocessed maps can be found in \citet{Driver2016}.

Unfortunately we do not have CO data for our H{\sc i}-selected sample so we cannot measure
  the molecular gas mass present in these galaxies. However, we have
  estimated molecular gas masses for our sources based on scaling relations from
  \citet{Saintonge2011} and \citet{Bothwell2014}. We found that
  the estimated molecular gas component is small compared to the H{\sc i}
  masses for H{\sc i}-selected sources. Using these scaling relations to derive total (H{\sc i}+$H_2$) gas masses instead of H{\sc i} masses does not change the overall conclusions presented in this work.
  
\subsection{Sample selection}
Our sample consists of the 32 sources in the HIPASS catalogue (HICAT)
that overlap with the H-ATLAS/GAMA footprints. These H{\sc i}
sources are cross-matched to the H-ATLAS DR1 catalogue
and to SDSS DR9 \citep{Ahn2012}. The matching is done by identifying all
optical and submm sources that lie within the $15.5^{\prime}$ Parkes
beam and have spectroscopic
redshifts within the redshift-range of the HIPASS profile. To be
accepted, optical matches need to have a reliable redshift from
GAMA or SDSS and the H-ATLAS matches need to have a reliable SDSS
counterpart (R $>$ 0.8, \citeb{SmithD2011}). The H-ATLAS matches are
combined with their corresponding optical matches when possible.

We identified two additional sources by
checking the literature for bright HI sources that are located in the
H-ATLAS fields, but are not found by HIPASS.
Both UGC0700 \citep{Sulentic1983} and NGC5746 \citep{Popping2011} were detected with Arecibo ($3.5^{\prime}$ beam) and both are bright enough to be detected by HIPASS, yet
lie just outside the beam of the closest HIPASS source
(separation of about $20^{\prime}$). The close proximity to another bright
HIPASS source likely caused these sources to be missed. However they are still
bright enough to meet our selection\footnote{Without the close proximity to
another HIPASS source, both these sources would have been included in the
HIPASS sample.} and both these sources are added to our sample.

Multiple optical matches are found for 30 out of the 32 HIPASS sources due to two
different issues. The first is that the SExtractor \citep{Bertin1996} source detection
used by SDSS and GAMA was not optimised for the very extended local
sources in our sample (semi-major axis up to 3 arcmin). Most of the
sources in our sample also have clumpy optical distributions, which
together with their large angular sizes, leads to SExtractor
`shredding' galaxies into several components; 71~per~cent of the sources
in our sample are affected by
this shredding. For these sources we determine the correct central
position manually and reject the spurious listings. Fluxes were measured as described in Section \ref{photometrysection}.

After correcting the shredding issue, there are still a number of
HIPASS sources that have multiple distinct galaxy matches. These
galaxies are `confused' in the large H{\sc i} beam and there
is no sure way of determining how
much of the H{\sc i} signal corresponds to each of the sources without
obtaining higher angular resolution observations. The
galaxies for which the HIPASS signal is confused are labelled `a' in
Tables~\ref{table1} and \ref{tableHI}. The projected physical
distance between these confused galaxies is relatively small (
$\sim 100$ kpc) and they form groups (consisting of up to 5 sources).
In total, there are 49 matches to the 32 HIPASS sources in the sample.

In order to better determine the H{\sc i} properties for the
confused sources, we have supplemented the HIPASS data with ALFALFA
data where available ({$3.5^{\prime}$} beam size). ALFALFA only covers the more
northern sources in the sample (dec $>-0.05^\circ$) and we find an unconfused H{\sc i} source
for 23 out of the 49 optical matches. Because of its higher sensitivity, we use the ALFALFA
H{\sc i} measurements for all sources that lie in its footprint. The
ALFALFA data resolves 3 of the 9 confused HIPASS sources into 7 separate
H{\sc i} sources, each with its own optical counterpart. We are then left
with 6 confused HIPASS sources, containing 14 optical matches between
them. For these we have searched the literature for the highest
resolution 21 cm observations available, leading to the deconfusion of
5 HIPASS sources into 8 separate H{\sc i} sources with optical
counterparts (see Table \ref{tableHI} for relevant references).
This leaves us with 1 HIPASS source for which there
are two optical matches. For this source, we
estimate which of the two candidate counterparts contributes the vast
majority of the H{\sc i} signal. The dominant source was chosen by
comparing the stellar masses, the NUV-\textit{r} colours and the offsets
in optical positions and velocities of the optical counterparts from HIPASS.
Based on these, we are confident that one counterpart
(labelled `b' in Table \ref{table1} and Table \ref{tableHI}) has
nearly all the H{\sc i} mass, and the other is a small satellite galaxy
that can be discarded together with all the other optical matches without
H{\sc i} detections. In Table 1, we present the key characteristics
of the sources in our H{\sc i}-selected sample, such as their common
names, positions, redshifts, distances and sizes. Distances were
calculated using spectroscopic redshifts using $H_0=67.30 \, \rm{km
\,s^{-1}\, Mpc^{-1}}$, with velocities corrected by GAMA \citep{Baldry2012} to
account for bulk deviations from Hubble flow\footnote{C15 used
a redshift-independent distance for NGC5584 from measurements
of Cepheid variables \citep{Riess2011}. However there is a lot of scatter in redshift-independent distance estimates for NGC5584, and we have opted to use the same method as for the other sources in our sample.} \citep{Tonry2000}. We use the standard prescription to determine H{\sc i} masses: $$ M_{HI}=2.356 \times 10^5 \ S_{int} \ D^2 $$ where $M_{HI}$ is the H{\sc i} mass in solar units, $S_{int}$ is the integrated 21 cm line flux in $Jy \ km \ s^{-1} $ and D is the Hubble flow corrected distance in Mpc. The H{\sc i}-derived properties are listed in Table 2.

The H{\sc i} fluxes and masses in Table \ref{tableHI} have not been corrected for self-absorption
(which occurs if there is a high optical depth in the line of sight
for H{\sc i} clouds). \citet{Bourne2013} calculated
this correction for a sample of 20 galaxies over the H-ATLAS fields.
The average correction factor for the overlapping sources with our
H{\sc i}-selected sample is 1.09.
Neither of the comparative samples used in this work (see Section 3) have been
corrected for self-absorption. For this reason, and because of the uncertainty
associated with the correction, we do not account for self-absorption in this work,
but note that our gas masses, particularly for edge on galaxies, could therefore be
underestimated.

Adding in higher resolution H{\sc i} data for known HIPASS detections
could affect our H{\sc i} selection. Although we have found ALFALFA
counterparts to each of the HIPASS sources in the common region, we cannot be
confident that these individual counterparts would have made the
HIPASS detection limit by itself. This is possibly an issue for the 3 sources
in our sample with $S_{int}<1.7 \, Jy$ (labelled `c' in Table \ref{table2}) and we therefore
ignore these when we discuss selection effects later. These sources do not change
any of our conclusions, which is why we include them in our plots and do not discuss further.
Finally we arrive at a sample of 40 unconfused H{\sc i}-selected sources, 22 of
which overlap with the C15 dust-selected sample (HAPLESS ID given in
Table \ref{table1}). Note that this is more than the original number
of HIPASS sources due to the additional ALFALFA and literature data
for the confused sources. These 40 sources will form our sample of
`H{\sc i}-selected Galaxies in H-ATLAS', hereafter referred to
as the H{\sc i}GH sample.

\subsection{Extended-source photometry}
\label{photometrysection}
To study the extended galaxies in a consistent way across 21
bands ranging from FUV to 500$\mu$m, we consider the same
physical area for each wavelength. We perform our own aperture-matched
photometry across the entire UV-to-submm wavelength range, with
exceptions for the IRAS 60 $\mu m$ measurements and for the PACS 100
$\mu m$ and 160 $\mu m$ aperture fitting. The exceptions are described in Appendix \ref{photomexep}. 

The first stage of the process consists of determining the appropriate
aperture for each source. As described in C15, the optimal shape and size of the aperture
are automatically determined in each band from FUV to 22$\mu$m and the
largest aperture (after correcting the aperture size
for the PSF) is selected as the definitive photometric
aperture (typically FUV or NUV).
The semi-major axes of the apertures used are
listed in Table \ref{table1}.

Next we removed bright foreground stars as decribed in C15,
by using a curve of growth to measure the size
of each star and then replacing stellar pixels with pixels randomly
drawn from an annulus around the mask.
In addition to this, a similar technique was used to remove background galaxies and
remaining bright stars, yet using manually determined masking apertures. This was necessary for our H{\sc i}-selected sample as many of the most H{\sc i}-rich sources are low surface brightness and
more susceptible to the effects of contaminating sources in the apertures.

After contaminant removal, the aperture matched photometry was
performed in each band and the uncertainties determined. The photometry
from the FUV to K$_s$-band was corrected for Galactic extinction in the
same way as GAMA, using the method described in \citet{Adelman-McCarthy2008}.

\subsubsection{Uncertainties}
\label{uncertsection}
The aperture noise for the UV-IR bands was estimated as in C15, using random
apertures placed on non-target regions of the map and using a
clipping procedure to mimic the effects of the star subtraction
performed in the main target aperture and sky annulus. The aperture noise for \textit{Herschel}
was also estimated by placing random apertures, yet without additional clipping.
Instead all extended sources in the H-ATLAS DR1 catalogue were masked using the H-ATLAS
aperture and only randomly placed apertures not containing masked pixels were accepted.

Compared to C15, we include an additional source of uncertainty
related to the accuracy of the star subtraction
process. For each source and each band, we calculate the uncertainty
as the average relative difference between the original photometry and photometry
performed with a contaminant removal with stellar/galaxy radii that differ by
$\pm 10 $~per~cent. This uncertainty is small for most sources but can dominate the
total uncertainty for the few ($\sim 10$~per~cent of sample) sources with strong stellar contamination.


The above errors are added in quadrature to the aperture noise and the resulting photometry
is given in Table \ref{photo} in Appendix \ref{ApendixA}.
Before fitting SEDs, we apply an additional term of uncertainty to account for the calibration uncertainty,
model uncertainties in our SED fitting (see next Section) and contributions from
spectral lines. For this additional term we use either $10$~per~cent or the calibration error, whichever is larger.
%

\subsection{SED fitting}
\label{SEDfitS}
To interpret the resulting panchromatic SEDs of the galaxies in our
H{\sc i}GH sample in terms of their physical properties, we use the {\sc
magphys} code of \citet{daCunha2008}\footnote{The {\sc
magphys} package containing the models of \citet{daCunha2008} is
publicly available at: www.iap.fr/{\sc magphys}}.
{\sc magphys} uses libraries of $\sim 50000$ optical and $\sim
50000$ infrared models to describe the stellar and dust SED respectively.
These models are combined in such a way that the energy balance is maintained
in both the diffuse ISM and the birth clouds. For each combination, the model SED is compared
to the observed galaxy SED and a goodness-of-fit $\chi^2$ calculated.
Probability density functions (PDF) can then be made for any
of the model physical parameters by weighting the value of that
parameter by the probability $e^{\frac{-\chi^2}{2}}$.
Our most reliable estimate for each parameter is the
median value of its PDF and the corresponding uncertainties are the
16th and 84th percentiles of the PDF. For more details on the {\sc magphys} models,
we refer to \citet{daCunha2008}.

We made some adaptations to {\sc magphys} in order to tailor to our sample.  These include: \begin{itemize}
\item The cold dust temperature range needed to be extended to 10 - 30 K (instead of the standard 15 - 25 K) in order to fit some of the dusty sources in our sample.
\item Some H{\sc i}GH sources have bands with low SNR, and in some cases the measured fluxes in the FIR are negative, though with errors which are consistent with a zero or positive flux at the 1 sigma level. The standard {\sc magphys} version does not deal with negative fluxes, yet we have incorporated them in our $\chi^2$ calculation, as they still give statistical constraints.
\item Additionally we added a routine that allows to include IRAS 60 $\mu m$ upper limits (necessary for a third of our sample). For these upper limits, we only add a contribution to $\chi^2$ if the model fluxes are higher than the upper limit flux.
\item We have generated a PDF for the dust attenuation in the FUV ($A_{FUV}$) by comparing the attenuated and
unattenuated model FUV fluxes for each model.
\end{itemize} Before fitting the SEDs, we correct the SPIRE fluxes in Table \ref{photo} for $K_{beam}$ and apply the additional uncertainty term to all bands (see Section \ref{uncertsection}). {\sc magphys} intrinsically applies colour corrections to all fluxes, so the $K_{colP}$ colour corrections from the SPIRE handbook do not need to be applied.
The {\sc magphys} results for the H{\sc i}GH sample
are presented in Table \ref{table2} and multiwavelength images and SEDs
for H{\sc i}GH are given in Figures \ref{images} and \ref{SED}
respectively.

\begin{table*}
    		\caption{Properties of the 40 H{\sc i}GH sources derived with {\sc magphys} SED fitting. The columns are (from left to right): Index, galaxy name, stellar mass, dust luminosity, dust mass, $M_{d}/M_*$, temperature of the cold dust component, star formation rate (SFR) averaged over the last $10^8$ years, specific star formation rate (SSFR) averaged over the last $10^8$ years, $f_\mu$ (the fraction of the total dust luminosity contributed by the diffuse ISM), and the FUV attenuation by dust. Uncertainties are indicated by the 84th-16th percentile range from each of the individual PDF. }
    		\setlength{\tabcolsep}{5.5pt}
    		\renewcommand{\arraystretch}{1.25}
    		\begin{tabular}{llrrrrrrrrr}
    		\hline\hline
\# & Name & \multicolumn{1}{c}{log $ M_* $} & \multicolumn{1}{c}{log $ L_d $ } & \multicolumn{1}{c}{ log $ M_d $ } & \multicolumn{1}{c}{ log $ M_d/M_* $ } & \multicolumn{1}{c}{ \centering $ T_c $ } & \multicolumn{1}{c}{ log $ SFR $ } & \multicolumn{1}{c}{ log $ SSFR $ } & \multicolumn{1}{c}{$A_{FUV}$} & \multicolumn{1}{c}{$f_\mu$} \\
 &  & \multicolumn{1}{c}{ $(M_\odot)$ } & \multicolumn{1}{c}{ $(L_\odot)$} & \multicolumn{1}{c}{$(M_\odot)$}&  & \multicolumn{1}{c}{(K)} & \multicolumn{1}{c}{ $(M_\odot \, yr^{-1}) $} & \multicolumn{1}{c}{ $(yr^{-1})$} &  \multicolumn{1}{c}{ (mag)} &\\ \hline
1 &SDSSJ08... &$9.84_{-0.13}^{+0.13}$&$9.80_{-0.15}^{+0.12}$&$7.21_{-0.41}^{+0.40}$&$-2.63_{-0.43}^{+0.42}$&$16.1_{-3.3}^{+4.2}$&$-0.03_{-0.13}^{+0.10}$&$-9.88_{-0.18}^{+0.16}$&$0.81_{-0.23}^{+0.20}$&$0.29_{-0.14}^{+0.19}$ \\ 
2$^a$ &UGC04673 &$9.12_{-0.17}^{+0.20}$&$9.15_{-0.14}^{+0.10}$&$7.40_{-0.27}^{+0.23}$&$-1.74_{-0.29}^{+0.28}$&$12.5_{-1.4}^{+2.1}$&$-0.24_{-0.05}^{+0.04}$&$-9.36_{-0.21}^{+0.18}$&$0.34_{-0.07}^{+0.07}$&$0.49_{-0.21}^{+0.37}$ \\ 
3 &UGC04684 &$9.35_{-0.15}^{+0.14}$&$9.40_{-0.10}^{+0.07}$&$6.70_{-0.14}^{+0.16}$&$-2.64_{-0.21}^{+0.22}$&$17.5_{-2.0}^{+1.9}$&$-0.36_{-0.12}^{+0.10}$&$-9.72_{-0.19}^{+0.18}$&$0.61_{-0.13}^{+0.17}$&$0.36_{-0.13}^{+0.19}$ \\ 
4 &UGC04996 &$9.36_{-0.10}^{+0.12}$&$9.61_{-0.06}^{+0.05}$&$7.18_{-0.19}^{+0.21}$&$-2.18_{-0.24}^{+0.25}$&$14.8_{-2.1}^{+2.5}$&$-0.17_{-0.06}^{+0.06}$&$-9.53_{-0.13}^{+0.12}$&$0.69_{-0.10}^{+0.07}$&$0.23_{-0.14}^{+0.16}$ \\ 
5 &UGC06578 &$8.02_{-0.06}^{+0.10}$&$8.68_{-0.07}^{+0.04}$&$5.72_{-0.29}^{+0.32}$&$-2.31_{-0.32}^{+0.30}$&$16.1_{-2.5}^{+2.9}$&$-1.02_{-0.04}^{+0.03}$&$-9.04_{-0.11}^{+0.06}$&$0.51_{-0.07}^{+0.07}$&$0.16_{-0.05}^{+0.12}$ \\ 
6$^a$ &UGC06780 &$9.00_{-0.12}^{+0.18}$&$8.85_{-0.06}^{+0.04}$&$6.97_{-0.23}^{+0.20}$&$-2.06_{-0.24}^{+0.20}$&$12.8_{-1.2}^{+1.7}$&$-0.36_{-0.11}^{+0.11}$&$-9.36_{-0.21}^{+0.16}$&$0.24_{-0.00}^{+0.05}$&$0.41_{-0.10}^{+0.10}$ \\ 
7$^a$ &UM456 &$8.28_{-0.15}^{+0.15}$&$8.64_{-0.12}^{+0.10}$&$4.96_{-0.45}^{+0.59}$&$-3.33_{-0.46}^{+0.60}$&$22.4_{-5.8}^{+5.0}$&$-0.76_{-0.04}^{+0.04}$&$-9.04_{-0.16}^{+0.15}$&$0.34_{-0.07}^{+0.07}$&$0.18_{-0.14}^{+0.06}$ \\ 
8 &UM456A &$7.88_{-0.13}^{+0.11}$&$8.30_{-0.33}^{+0.21}$&$4.89_{-0.55}^{+0.65}$&$-2.98_{-0.56}^{+0.65}$&$20.7_{-5.8}^{+6.0}$&$-1.32_{-0.09}^{+0.12}$&$-9.19_{-0.14}^{+0.17}$&$0.49_{-0.25}^{+0.23}$&$0.19_{-0.12}^{+0.13}$ \\ 
9$^a$ &UGC06903 &$9.89_{-0.15}^{+0.09}$&$9.48_{-0.03}^{+0.04}$&$7.17_{-0.09}^{+0.10}$&$-2.68_{-0.14}^{+0.17}$&$16.4_{-1.1}^{+1.0}$&$-0.24_{-0.04}^{+0.04}$&$-10.13_{-0.10}^{+0.16}$&$0.56_{-0.03}^{+0.03}$&$0.58_{-0.11}^{+0.23}$ \\ 
10 &UGC06970 &$9.39_{-0.15}^{+0.12}$&$8.89_{-0.18}^{+0.14}$&$6.52_{-0.51}^{+0.35}$&$-2.86_{-0.54}^{+0.39}$&$14.6_{-2.5}^{+3.6}$&$-0.86_{-0.12}^{+0.10}$&$-10.26_{-0.16}^{+0.18}$&$0.56_{-0.20}^{+0.25}$&$0.32_{-0.16}^{+0.14}$ \\ 
11 &NGC4030b &$8.85_{-0.14}^{+0.16}$&$8.63_{-0.20}^{+0.13}$&$5.64_{-0.44}^{+0.53}$&$-3.22_{-0.47}^{+0.55}$&$19.9_{-5.6}^{+5.6}$&$-0.98_{-0.32}^{+0.09}$&$-9.83_{-0.35}^{+0.16}$&$0.39_{-0.15}^{+0.13}$&$0.32_{-0.17}^{+0.41}$ \\ 
12 &NGC4030 &$10.88_{-0.09}^{+0.12}$&$10.88_{-0.02}^{+0.03}$&$7.96_{-0.08}^{+0.04}$&$-2.93_{-0.14}^{+0.10}$&$20.9_{-0.5}^{+0.8}$&$0.78_{-0.05}^{+0.04}$&$-10.10_{-0.13}^{+0.10}$&$1.96_{-0.10}^{+0.10}$&$0.55_{-0.05}^{+0.04}$ \\ 
13$^a$ &UGC07053 &$8.19_{-0.10}^{+0.18}$&$7.98_{-0.34}^{+0.33}$&$4.80_{-0.54}^{+0.56}$&$-3.41_{-0.57}^{+0.58}$&$23.2_{-6.3}^{+4.5}$&$-1.03_{-0.07}^{+0.06}$&$-9.22_{-0.19}^{+0.12}$&$0.19_{-0.10}^{+0.15}$&$0.71_{-0.54}^{+0.23}$ \\ 
14$^a$ &UGC07332 &$7.70_{-0.13}^{+0.14}$&$7.78_{-0.28}^{+0.18}$&$4.31_{-0.40}^{+0.48}$&$-3.40_{-0.42}^{+0.50}$&$24.1_{-6.5}^{+4.2}$&$-1.39_{-0.04}^{+0.04}$&$-9.09_{-0.15}^{+0.13}$&$0.19_{-0.07}^{+0.07}$&$0.27_{-0.13}^{+0.66}$ \\ 
15 &NGC4202 &$10.30_{-0.10}^{+0.11}$&$10.29_{-0.03}^{+0.03}$&$7.46_{-0.06}^{+0.07}$&$-2.81_{-0.14}^{+0.11}$&$20.3_{-0.8}^{+0.8}$&$0.05_{-0.17}^{+0.06}$&$-10.25_{-0.20}^{+0.11}$&$1.74_{-0.07}^{+0.10}$&$0.67_{-0.05}^{+0.11}$ \\ 
16 &FGC1412 &$6.94_{-0.10}^{+0.13}$&$7.33_{-0.41}^{+0.20}$&$3.84_{-0.53}^{+0.78}$&$-3.11_{-0.55}^{+0.78}$&$21.3_{-7.3}^{+6.0}$&$-2.43_{-0.17}^{+0.14}$&$-9.37_{-0.21}^{+0.17}$&$0.69_{-0.38}^{+0.25}$&$0.18_{-0.10}^{+0.21}$ \\ 
17$^a$ &CGCG014-010 &$7.29_{-0.12}^{+0.13}$&$6.88_{-0.23}^{+0.34}$&$3.48_{-0.40}^{+0.45}$&$-3.82_{-0.39}^{+0.47}$&$25.0_{-5.8}^{+3.5}$&$-2.14_{-0.04}^{+0.04}$&$-9.42_{-0.14}^{+0.12}$&$0.14_{-0.05}^{+0.13}$&$0.57_{-0.48}^{+0.30}$ \\ 
18 &UGC07394 &$8.93_{-0.12}^{+0.14}$&$8.70_{-0.13}^{+0.17}$&$6.87_{-0.23}^{+0.21}$&$-2.07_{-0.27}^{+0.24}$&$12.2_{-1.3}^{+1.7}$&$-1.22_{-0.34}^{+0.17}$&$-10.15_{-0.36}^{+0.21}$&$0.69_{-0.17}^{+0.28}$&$0.39_{-0.19}^{+0.23}$ \\ 
19$^a$ &UGC07531 &$8.60_{-0.08}^{+0.15}$&$8.98_{-0.18}^{+0.09}$&$6.49_{-0.39}^{+0.39}$&$-2.13_{-0.42}^{+0.41}$&$13.7_{-2.6}^{+3.5}$&$-0.38_{-0.04}^{+0.04}$&$-8.97_{-0.16}^{+0.09}$&$0.26_{-0.07}^{+0.07}$&$0.16_{-0.05}^{+0.16}$ \\ 
20 &UM501 &$7.90_{-0.10}^{+0.10}$&$8.65_{-0.09}^{+0.24}$&$5.10_{-0.54}^{+0.71}$&$-2.80_{-0.53}^{+0.72}$&$19.4_{-5.6}^{+6.9}$&$-1.06_{-0.04}^{+0.12}$&$-8.95_{-0.11}^{+0.15}$&$0.54_{-0.07}^{+0.28}$&$0.10_{-0.04}^{+0.09}$ \\ 
21$^a$ &NGC5496 &$9.46_{-0.05}^{+0.14}$&$9.51_{-0.04}^{+0.04}$&$7.12_{-0.11}^{+0.14}$&$-2.35_{-0.14}^{+0.13}$&$16.9_{-2.0}^{+1.3}$&$-0.23_{-0.04}^{+0.04}$&$-9.69_{-0.14}^{+0.06}$&$0.61_{-0.10}^{+0.00}$&$0.66_{-0.37}^{+0.00}$ \\ 
22$^a$ &NGC5584 &$9.97_{-0.16}^{+0.09}$&$10.04_{-0.02}^{+0.04}$&$7.51_{-0.10}^{+0.07}$&$-2.45_{-0.15}^{+0.14}$&$17.0_{-1.2}^{+1.5}$&$0.26_{-0.04}^{+0.04}$&$-9.71_{-0.10}^{+0.16}$&$0.79_{-0.03}^{+0.13}$&$0.39_{-0.13}^{+0.40}$ \\ 
23 &UGC09215 &$9.31_{-0.04}^{+0.14}$&$9.57_{-0.05}^{+0.02}$&$6.95_{-0.10}^{+0.09}$&$-2.38_{-0.16}^{+0.12}$&$17.4_{-1.5}^{+0.9}$&$-0.24_{-0.06}^{+0.02}$&$-9.55_{-0.15}^{+0.04}$&$0.66_{-0.05}^{+0.07}$&$0.37_{-0.06}^{+0.08}$ \\ 
24 &2MASXJ14... &$9.60_{-0.07}^{+0.13}$&$9.98_{-0.08}^{+0.07}$&$7.26_{-0.15}^{+0.18}$&$-2.36_{-0.19}^{+0.21}$&$17.7_{-2.1}^{+1.7}$&$-0.03_{-0.07}^{+0.07}$&$-9.63_{-0.14}^{+0.11}$&$1.54_{-0.15}^{+0.17}$&$0.36_{-0.06}^{+0.07}$ \\ 
25 &IC1011 &$10.16_{-0.07}^{+0.13}$&$10.59_{-0.05}^{+0.04}$&$7.41_{-0.09}^{+0.08}$&$-2.77_{-0.14}^{+0.13}$&$21.9_{-1.2}^{+1.3}$&$0.60_{-0.04}^{+0.06}$&$-9.56_{-0.14}^{+0.09}$&$1.66_{-0.15}^{+0.13}$&$0.40_{-0.07}^{+0.10}$ \\ 
26 &IC1010 &$10.82_{-0.25}^{+0.08}$&$10.29_{-0.02}^{+0.04}$&$7.93_{-0.13}^{+0.12}$&$-2.85_{-0.20}^{+0.21}$&$16.1_{-0.9}^{+1.4}$&$0.44_{-0.06}^{+0.04}$&$-10.38_{-0.10}^{+0.25}$&$0.69_{-0.07}^{+0.10}$&$0.52_{-0.15}^{+0.05}$ \\ 
27$^a$ &UGC09299 &$8.61_{-0.04}^{+0.19}$&$8.82_{-0.01}^{+0.03}$&$6.39_{-0.14}^{+0.15}$&$-2.24_{-0.20}^{+0.18}$&$17.3_{-1.4}^{+1.4}$&$-0.55_{-0.04}^{+0.04}$&$-9.16_{-0.20}^{+0.05}$&$0.31_{-0.00}^{+0.03}$&$0.62_{-0.04}^{+0.14}$ \\ 
28$^a$ &SDSSJ14... &$7.77_{-0.18}^{+0.19}$&$7.66_{-0.41}^{+0.35}$&$4.52_{-0.63}^{+0.73}$&$-3.24_{-0.68}^{+0.76}$&$21.3_{-6.9}^{+5.9}$&$-1.69_{-0.04}^{+0.04}$&$-9.46_{-0.19}^{+0.19}$&$0.26_{-0.15}^{+0.28}$&$0.31_{-0.19}^{+0.54}$ \\ 
29 &NGC5690 &$10.38_{-0.09}^{+0.11}$&$10.48_{-0.02}^{+0.03}$&$7.61_{-0.05}^{+0.05}$&$-2.78_{-0.12}^{+0.12}$&$20.5_{-0.8}^{+0.6}$&$0.31_{-0.04}^{+0.05}$&$-10.07_{-0.12}^{+0.10}$&$2.59_{-0.10}^{+0.10}$&$0.59_{-0.03}^{+0.05}$ \\ 
30 &NGC5691 &$10.01_{-0.17}^{+0.10}$&$10.15_{-0.04}^{+0.03}$&$6.85_{-0.04}^{+0.07}$&$-3.15_{-0.14}^{+0.17}$&$24.1_{-1.2}^{+0.4}$&$-0.06_{-0.05}^{+0.06}$&$-10.07_{-0.11}^{+0.18}$&$1.76_{-0.10}^{+0.13}$&$0.60_{-0.04}^{+0.03}$ \\ 
31$^a$ &UGC09432 &$8.19_{-0.14}^{+0.02}$&$7.76_{-0.94}^{+0.17}$&$4.40_{-0.51}^{+0.48}$&$-3.76_{-0.53}^{+0.51}$&$24.4_{-6.0}^{+3.9}$&$-1.06_{-0.06}^{+0.05}$&$-9.24_{-0.06}^{+0.15}$&$0.09_{-0.07}^{+0.05}$&$0.58_{-0.50}^{+0.20}$ \\ 
32$^a$ &NGC5705 &$9.33_{-0.12}^{+0.08}$&$9.34_{-0.03}^{+0.01}$&$7.35_{-0.13}^{+0.12}$&$-1.98_{-0.15}^{+0.17}$&$14.3_{-1.3}^{+1.1}$&$-0.24_{-0.04}^{+0.04}$&$-9.58_{-0.09}^{+0.12}$&$0.41_{-0.03}^{+0.03}$&$0.54_{-0.16}^{+0.23}$ \\ 
33 &NGC5725 &$9.13_{-0.13}^{+0.08}$&$9.14_{-0.09}^{+0.07}$&$6.45_{-0.19}^{+0.19}$&$-2.68_{-0.20}^{+0.21}$&$17.3_{-2.2}^{+2.6}$&$-0.65_{-0.07}^{+0.07}$&$-9.78_{-0.10}^{+0.14}$&$0.71_{-0.15}^{+0.10}$&$0.32_{-0.11}^{+0.14}$ \\ 
34 &NGC5713 &$10.56_{-0.11}^{+0.14}$&$10.94_{-0.03}^{+0.03}$&$7.54_{-0.05}^{+0.05}$&$-3.02_{-0.14}^{+0.12}$&$24.8_{-0.9}^{+0.6}$&$0.72_{-0.05}^{+0.06}$&$-9.84_{-0.15}^{+0.12}$&$2.71_{-0.13}^{+0.10}$&$0.57_{-0.03}^{+0.04}$ \\ 
35 &NGC5719 &$10.79_{-0.08}^{+0.09}$&$10.45_{-0.04}^{+0.03}$&$7.43_{-0.06}^{+0.07}$&$-3.36_{-0.11}^{+0.12}$&$22.0_{-1.0}^{+0.8}$&$-0.17_{-0.06}^{+0.04}$&$-10.96_{-0.11}^{+0.09}$&$3.06_{-0.17}^{+0.10}$&$0.78_{-0.02}^{+0.02}$ \\ 
36 &UGC09482 &$8.72_{-0.14}^{+0.10}$&$8.55_{-0.15}^{+0.12}$&$6.09_{-0.27}^{+0.31}$&$-2.62_{-0.29}^{+0.32}$&$15.5_{-2.6}^{+2.6}$&$-1.30_{-0.18}^{+0.14}$&$-10.02_{-0.21}^{+0.19}$&$0.56_{-0.13}^{+0.17}$&$0.29_{-0.12}^{+0.31}$ \\ 
37$^a$ &UGC09470 &$8.90_{-0.13}^{+0.07}$&$8.86_{-0.03}^{+0.03}$&$6.22_{-0.19}^{+0.17}$&$-2.65_{-0.17}^{+0.16}$&$18.2_{-1.8}^{+2.3}$&$-0.68_{-0.04}^{+0.04}$&$-9.58_{-0.09}^{+0.13}$&$0.39_{-0.03}^{+0.07}$&$0.50_{-0.24}^{+0.16}$ \\ 
38 &NGC5740 &$10.28_{-0.07}^{+0.11}$&$10.03_{-0.04}^{+0.04}$&$7.16_{-0.07}^{+0.07}$&$-3.13_{-0.10}^{+0.12}$&$19.9_{-0.9}^{+0.8}$&$-0.05_{-0.04}^{+0.04}$&$-10.33_{-0.12}^{+0.08}$&$1.54_{-0.13}^{+0.10}$&$0.50_{-0.05}^{+0.11}$ \\ 
39$^a$ &UGC07000 &$9.11_{-0.16}^{+0.08}$&$9.15_{-0.04}^{+0.07}$&$6.43_{-0.11}^{+0.12}$&$-2.67_{-0.14}^{+0.17}$&$18.6_{-1.5}^{+1.8}$&$-0.45_{-0.04}^{+0.04}$&$-9.56_{-0.09}^{+0.16}$&$0.49_{-0.13}^{+0.07}$&$0.48_{-0.25}^{+0.20}$ \\ 
40 &NGC5746 &$11.31_{-0.10}^{+0.07}$&$10.34_{-0.01}^{+0.02}$&$8.00_{-0.07}^{+0.07}$&$-3.30_{-0.10}^{+0.10}$&$17.1_{-0.5}^{+0.4}$&$-0.41_{-0.70}^{+0.36}$&$-11.72_{-0.71}^{+0.37}$&$1.46_{-0.38}^{+0.10}$&$0.87_{-0.12}^{+0.08}$ \\  \hline
\multicolumn{2}{l}{Mean}&$9.20$&$9.22$&$6.40$&$-2.80$&$18.9$&$-0.51$&$-9.71$&$0.86$&$0.43$ \\
\multicolumn{2}{l}{$M_*<10^9$}&$8.17$&$8.27$&$5.21$&$-2.96$&$19.8$&$-1.18$&$-9.35$&$0.38$&$0.38$ \\
\multicolumn{2}{l}{$M_*>10^9$}&$9.89$&$9.85$&$7.19$&$-2.69$&$18.3$&$-0.07$&$-9.95$&$1.18$&$0.50$ \\
\hline
\end{tabular}
\begin{flushleft}
$^a$ For these sources, we use SFR and SSFR estimates using the same method for SFR as C15 since the {\sc magphys} SFR and SSFR PDFs show two peaks. The two peaks occur because the model SFR (averaged over the last $10^8$ years) will be quite different if it includes a burst (that ended nearly $10^8$ years ago), compared to if the burst ended just before $10^8$ years ago, even though there are only very small differences to the SEDs. Schofield et al., (\textit{in prep.}) will explore this issue in more detail. We note that the C15 SFR estimates would be biased when the SSFR is small and the dust luminosity has a large contribution from heating by old stars \citep{Boquien2016}. This is not the case for these galaxies.
\end{flushleft}
\label{table2}
\end{table*}

\section{Surveys used in this work}
Here we introduce the dust-selected and stellar mass selected samples of local galaxies to compare with our H{\sc i}-selected sample.

\subsection{Dust-selected sample}
\label{HAPLESSsec}
The best comparative dust-selected sample is the \textit{Herschel}-ATLAS
Phase-1 Limited-Extent Spatial Survey (HAPLESS) described in the
companion paper to this work (C15). HAPLESS is a volume limited sample
consisting of 42 H-ATLAS galaxies detected at 250$\mu m$ in the
equatorial H-ATLAS fields with $0.0035 < \mathit{z} < 0.01$. Throughout the rest of this
work we will refer to HAPLESS as a dust-selected sample to indicate this
250$\mu m$ flux selection.

HAPLESS has 22 sources in common with H{\sc i}GH and the
photometry was performed using the same pipeline. For consistency, we have redetermined the \textit{Herschel} photometry for HAPLESS using the H-ATLAS DR1 maps that have since become available.
Additionally we redetermined the galaxy properties for HAPLESS using {\sc magphys} instead
of the combination of different techniques at different wavelengths used by C15. The {\sc magphys} cold dust temperatures are, on average, 3 K warmer and the dust masses smaller by 0.25 dex than the results in C15, and the offset is largest for sources with cold ($T_c < 15$ K) dust temperature in C15. The differences originate in part from differences in the SED fitting technique and in part from changes to the \textit{Herschel} photometry due to using the H-ATLAS DR1 data release instead of Phase 1 version 3. In contrast to modified blackbody fits in C15, {\sc magphys} limits the warm dust to $30<T_w<60$ K, and at least half of the dust luminosity in the diffuse ISM must originate from the cold dust component. Therefore {\sc magphys} assigns low probabilities to poorly constrained cold dust components that make up a tiny fraction of the total luminosity but peak at the longest wavelengths, therefore making up the bulk of the dust mass.  Additionally, {\sc magphys} uses the median Tc from the PDF whereas C15 used the best fit to the data; when comparing C15 with the best-fit {\sc magphys} result, we find overall a better agreement between the two estimates. C15 compiled literature atomic gas
masses, including HIPASS \citep{Meyer2004} and ALFALFA (Haynes, \textit{priv. comm.)}.

\subsection{Stellar mass selected sample}
For a stellar mass selected sample we follow C15 and use the {\em Herschel} Reference
Survey (HRS, \citeb{Boselli2010}) which targeted 323 local
galaxies. The HRS is a volume-limited sample (between 15 and 25 Mpc)
and uses K$_S$-band selection because this band suffers least from
extinction and is known to be a good proxy for stellar mass;
throughout the rest of this work, we will refer to the HRS as a
stellar mass selected sample. The HRS contains both
Late Type Galaxies (LTG) and 75 Early Type Galaxies (ETG) and includes many galaxies in cluster environments. 
We do not include ETGs when determining best fit relations and correlations. Instead we highlight
them as a separate sub-sample in the plots.

Again, for consistency, we derived properties for HRS sources using
{\sc magphys}. Our results are slightly different to the {\sc magphys}
HRS results in \citet{Viaene2016} because they did not apply any
corrections for galactic extinction and $K_{beam}$, and used smaller
uncertainties. For the H{\sc i} masses of the HRS galaxies, we used
the unconfused results from \citet{Boselli2014}. For HRS, CO derived $H_2$ masses are presented in \citet{Boselli2014} and can be relatively large compared to their H{\sc i}. However, they are still small compared to the total baryon mass and using total (H{\sc i}+$H_2$) gas masses instead of H{\sc i} masses only gives small differences to the overall scaling relations for HRS in this work.

\section{Dust, gas and stars}
\begin{figure*}
  \includegraphics[width=\textwidth]{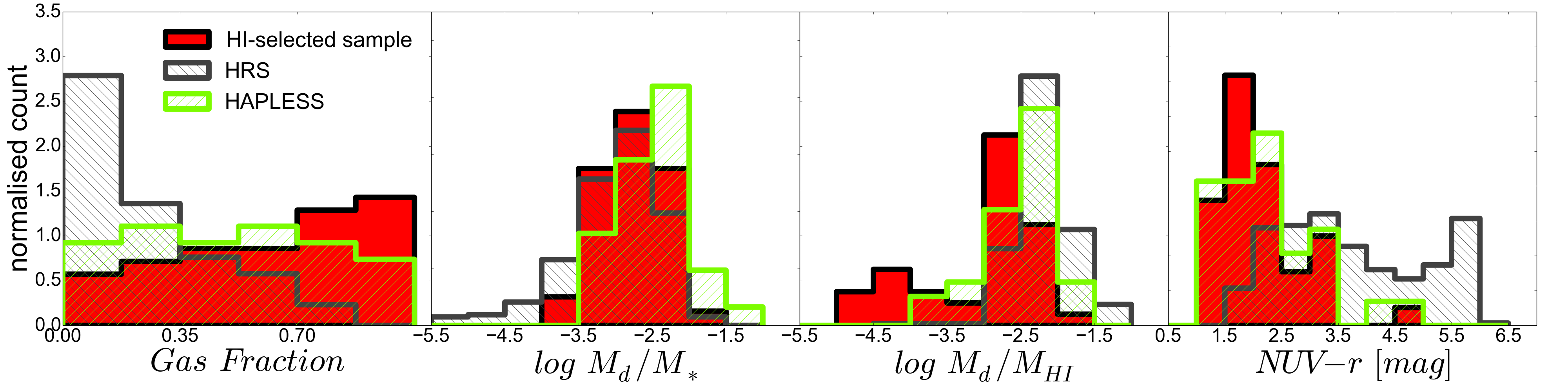}
  \caption{\textit{From left to right:} Histogram showing the distribution of gas fraction ($\frac{M_{HI}}{M_*+M_{HI}}$), specific dust mass, dust-to-HI ratio and NUV-\textit{r} colour (proxy for specific star formation rate) for the H{\sc i}GH sample, HAPLESS and HRS.}
  \label{histograms2}
\end{figure*}

We first investigate the distribution of the relative masses of stars,
dust and atomic gas. The distribution of the gas fractions
($\frac{M_{HI}}{M_*+M_{HI}}$) in the left panel of Figure
\ref{histograms2} shows that the H{\sc i}GH sample is more gas rich
than the HRS, while the HAPLESS gas fractions are relatively uniformly
distributed. In this paper, we will define the evolutionary status of
a galaxy using its gas fraction as a measure of how much of the
available gas reservoir has been converted into stars\footnote{The
  most important caveat to this method is that we do not take into
  account interactions like inflows, outflows and merging.}. HRS then
consists mainly of evolved sources, HAPLESS consists of galaxies at a
range of stages of evolution and the H{\sc i}GH \textit{sample
  consists mainly of relatively unevolved sources}. Our H{\sc i}
selection therefore gives us vital insights into the `youthful'
sources which were previously under-represented in samples like the HRS.

Figure \ref{histograms2} also shows that the specific dust mass
($M_d/M_*$) is highest for HAPLESS, followed by H{\sc i}GH and then
HRS. When we look at the distribution of dust-to-H{\sc i} ratio we now
find that H{\sc i}GH has the lowest $M_d/M_{HI}$, followed by HAPLESS
and then HRS. Finally we show the NUV-\textit{r} colour distribution
in the right panel of Figure \ref{histograms2}. This colour is closely
related to the specific star formation rate and we find that both the HAPLESS and H{\sc i}GH
samples are much bluer and thus more actively forming stars than the
HRS. The large tail of red sources in HRS is because it
contains a larger fraction of more evolved, passive sources.

\begin{figure*}
  \includegraphics[width=\textwidth]{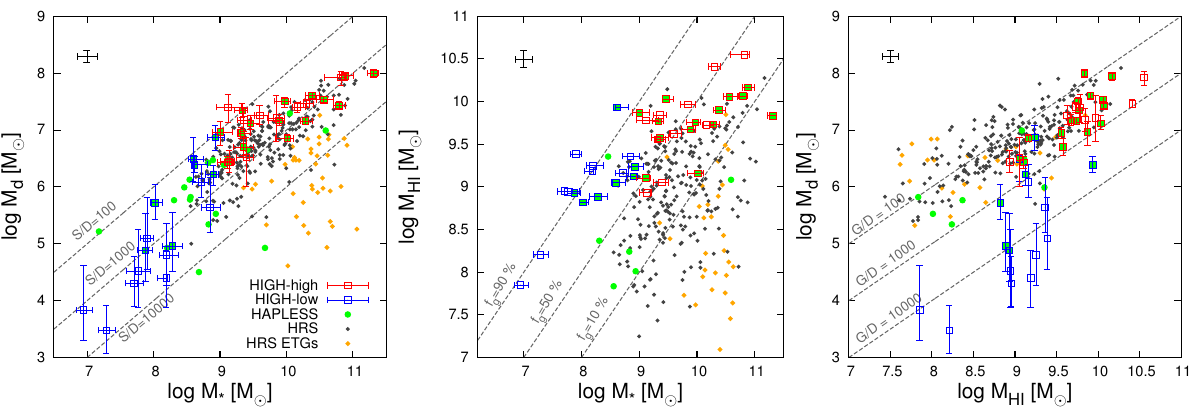}
  \caption{Scaling relations showing how the stellar, dust, and gas
    masses depend on each other. A representative error bar for HRS is
    shown in the upper-left corner. Note the selection effects towards
    higher dust and gas masses for the HI-selected H{\sc i}GH-low (blue squares),
    H{\sc i}GH-high (red squares) samples and the dust-selected
    HAPLESS sample (green circles) compared to the stellar mass selected
    HRS sample (grey dots). The common sources between HAPLESS and our
    H{\sc i}-selected sub-samples are shown as green filled squares
    with red/blue borders. Lines of constant $M_*/M_d$ (S/D), gas
    fraction ($f_g$) and $M_{HI}/M_d$ (G/D) are shown in grey.}
  \label{bigisbig}
\end{figure*}

The H{\sc i}GH sample consists mainly of very
  blue, low surface brightness gas rich sources which have irregular
  or flocculent morphologies and are actively forming stars.
  The blue sources in our sample divide into dust rich and dust poor
subsamples at $M_{\ast}\sim 10^9\,\rm{M_{\odot}}$ (see also Figure \ref{bigisbig}). The blue dust rich
sources were already discussed in C15 and constitute around half of all
dust mass selected galaxies in the local volume. In the rest of this
work we will highlight the new population of blue, dust poor sources
as a separate sub-sample, using a stellar mass cut of $M_*<10^9 \,
M_{\odot}$. We will use this criterion to split the H{\sc i}GH sample
into H{\sc i}GH-low (for $M_*<10^9 \, M_{\odot}$) and H{\sc i}GH-high
(for $M_*>10^9 \, M_{\odot}$) throughout the rest of this work.

In Figure \ref{bigisbig} we show the stellar, dust and H{\sc i} mass
scaling relations. In the left panel we find a strong correlation
between dust and stellar mass for both H{\sc i}GH (Spearman rank correlation
coefficient $r= 0.93$) and HAPLESS ($r= 0.81$). For HRS there is a
strong correlation ($r= 0.88$) for the Late Type Galaxies (LTGs),
yet the correlation weakens significantly ($r=0.30$) if the ETGs are included. Comparing the H{\sc i} and stellar masses
(centre panel of Figure \ref{bigisbig}), we find the strongest correlation
for H{\sc i}GH ($r=0.77$), a weaker one for HAPLESS ($r=0.67$) and the weakest for the HRS LTGs ($r=0.63$). When the ETGs are included, there is no significant correlation for HRS. The HRS and the H{\sc i}GH sample segregate in this plot because they
intrinsically consist of galaxies in different stages of evolution
(stellar mass selection favours lower gas fractions and vice versa). In the right panel of Figure \ref{bigisbig} we find a strong
correlation between the H{\sc i} and dust mass for HAPLESS, the HRS
and the H{\sc i}GH-high sample (Spearman rank coefficient of $r=0.74$
for the combined samples). However H{\sc i}GH-low lies significantly
below this relation and we will investigate the reasons for this in
the following sections. Interestingly, The HRS ETGs now follow the same trend as the
LTGs. For a given H{\sc i} mass, HAPLESS and HIGH have lower dust
masses on average than the HRS.

\subsection{Gas richness and specific star formation rate}
\label{HI-section}

\begin{figure}
  \includegraphics[width=\columnwidth]{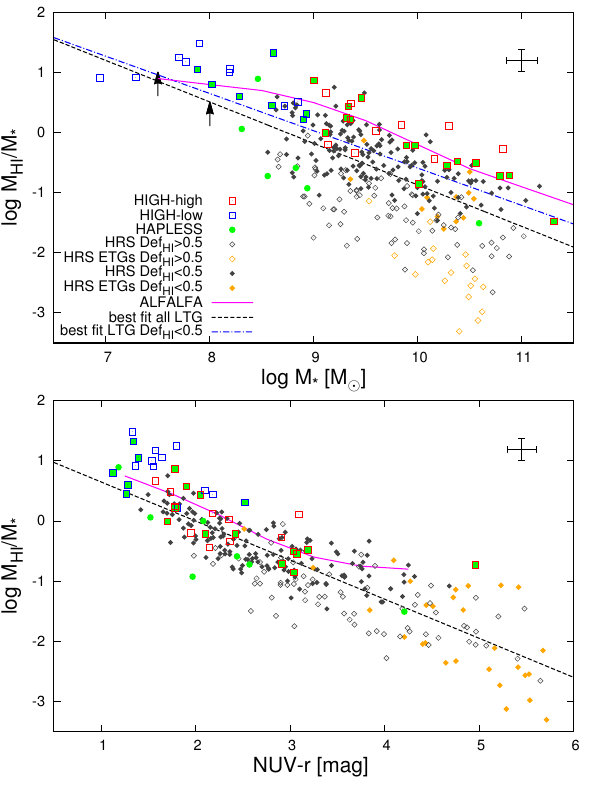}
  \caption{Trends with $M_{HI}/M_*$ and SSFR. Symbols are as in Figure \ref{bigisbig}, with open
    symbols for H{\sc i} deficient ($Def_{HI}\ge 0.5$) LTGs in HRS and filled symbols for
    H{\sc i} normal HRS LTGs. The best fit power law relationship
      for the combined samples (excluding ETGs) is shown as a black dashed
      line, and the best relation also excluding H{\sc i} deficient HRS galaxies as a blue dash-dot line.
      \textit{Top:} $M_{HI}/M_*$ against stellar mass. The ALFALFA relation \citep{Huang2012} is shown in
      magenta. The HIPASS detection limits at $M_*=10^{7.5}\, M_{\odot}$ and
      $M_*=10^{8}\, M_{\odot}$ are shown as black arrows.
      \textit{Bottom:} $M_{HI}/M_*$ against NUV-\textit{r}
      colour (proxy for SSFR). This strong correlation is applicable
      to all samples and thus is a very useful scaling relation.}
  \label{HIscaling}
\end{figure}

Figure \ref{HIscaling} show how gas richness ($M_{HI}/M_*$) scales with stellar mass and NUV-\textit{r}, which is known to be a good proxy for SSFR \citep[e.g.][]{Schiminovich2007}. These
relations have been extensively studied for HRS \citep{Cortese2011}, ALFALFA \citep{Huang2012}, H$\alpha$3 \citep{Gavazzi2013}, GASS \citep{Catinella2013} and in earlier work \citep{Gavazzi1996,Boselli2001}. As seen in Figure \ref{HIscaling},
the H{\sc i}GH sample follows the same relations as determined for other H{\sc i} selected samples, such as ALFALFA.

\begin{table*}
\caption{The Spearman rank correlation coefficients (r) and lines of best fit ($y = ax + b$, where a is the slope and b is the intercept) for the significant correlations in the form of a powerlaw. The best fitting relations were determined using a BCES linear regression method \citep{Akritas1996} using the H{\sc i}GH, HAPLESS and HRS samples combined. For HRS, only late type galaxies are included (both H{\sc i} deficient and H{\sc i} normal). The first two columns specify the x and y parameters, the last columns specify whether H{\sc i}GH-low is offset and whether the derived relation is dependent on the selection used. We caution the use of relations which are strongly dependent on the selection criterea. H{\sc i}GH-low is not included in the combined sample if it is offset (lower dust mass) compared to the other samples.} \label{correlations}
    		\begin{tabular}{ccrrrrccc}
    		\hline\hline
    		y & x & \hspace{3mm} & $r$\hspace{2mm} & Slope\hspace{4mm} & Intercept\hspace{2mm} & \hspace{3mm} & \shortstack{H{\sc i}GH-low \\ offset} & \shortstack{strong selection \\ dependence} \\ \hline

log $M_{HI}/M_*$ & log $M_*$ & & -0.59 & $ -0.69 \pm 0.05$ & $6.02 \pm 0.47$ & &  & \checkmark   \\
log $M_{HI}/M_*$ & NUV-\textit{r}& & -0.84 & $ -0.65 \pm 0.03$ & $1.30 \pm 0.09$ & &  &   \\
log $M_d/M_*$ & log $M_*$& & -0.44 & $ -0.26 \pm 0.03$ & $-0.44 \pm 0.34$ & &\checkmark & \checkmark \\
log $M_d/M_*$ & log SSFR& & 0.72 &  $ 0.51 \pm 0.03$ & $2.30 \pm 0.30$ & &\checkmark &  \\
log $M_d/M_*$ & NUV-\textit{r}& & -0.77 & $ -0.33 \pm 0.02$ & $-1.92 \pm 0.06$ & &\checkmark &  \\
log $M_d/M_*$ & log $M_{HI}/M_*$& & 0.87 &  $ 0.47 \pm 0.02$ & $-2.59 \pm 0.02$ & &\checkmark &  \\
log $M_d/M_{HI}$ & log $M_*$& & 0.47 &  $ 0.32 \pm 0.04$ & $-5.33 \pm 0.37$ & &\checkmark & \checkmark \\
log $M_d/M_{HI}$  & NUV-\textit{r}& & 0.66  & $ 0.28 \pm 0.02$ & $-3.06 \pm 0.07$ & &\checkmark &  \\
log $M_d/M_{HI}$  & log $M_{HI}/M_*$& & -0.86  & $ -0.52 \pm 0.02$ & $-2.57 \pm 0.02$ & &\checkmark &  \\
log $SFR/M_d$ & log $M_{HI}/M_*$  & & 0.37 & $ 0.25 \pm 0.03$ & $-7.19 \pm 0.03$ & &\checkmark & \\
log $SFR/M_{HI}$ & log $M_{HI}/M_*$& & -0.53  & $ -0.29 \pm 0.03$ & $-9.80 \pm 0.02$ & &  &    \\
log $SFR/M_{HI}$ & log $\Sigma_*$& & 0.58  & $ 0.50 \pm 0.06$ & $-10.28 \pm 0.07 $ & &  &  \\
  \hline
   	\end{tabular}
   	  \linebreak
\end{table*}

Some of the selection effects for the different
  samples are evident in the top panel. The H{\sc i} selection of
  H{\sc i}GH (and ALFALFA) selects higher $M_{HI}/M_{*}$ at fixed
  $M_*$ compared to the stellar mass selection of HRS.
    This is due both to the H{\sc i} selection favouring gas rich galaxies
    (and {\em vice-versa\/} for stellar mass selection), and also in
    part to a fraction ($\sim 25$~per~cent) of the HRS sources being in the Virgo
    cluster. 
    In Figure \ref{HIscaling} we have used open
    symbols for H{\sc i} deficient LTGs in HRS. Following \citet{Cortese2011}, we
    consider galaxies to be H{\sc i} deficient if $Def_{HI}\ge 0.5$ (this
    corresponds to galaxies with 70 per cent less hydrogen than isolated systems
    with the same diameter and morphological type). Next to
    our best fit relation for all samples combined (excluding ETGs), we have also
    plotted the best fit line excluding H{\sc i} deficient galaxies to
    illustrate the effect of including H{\sc i} deficient galaxies in
    scaling relations.

We have also added the HIPASS detection limits
at $M_*=10^{7.5} \, M_{\odot}$ and $M_*=10^{8}\, M_{\odot}$ as black arrows to Figure
\ref{HIscaling} (top), assuming a distance equal to the
average distance for H{\sc i}GH-low (29.2 Mpc). The lack of sources below
the dashed line at low $M_*$ is due to this limit.
However, the upper bound of the
trend in Figure \ref{HIscaling} does not suffer these selection
effects. The large range of gas fractions found
at a given stellar mass indicates that, although the star formation
history has a well known dependence on halo mass (more massive galaxies are more evolved; e.g. \citeb{Cowie1996}),
local factors such as environment and gas supply play an important role \citep[e.g.][]{DeLucia2006} and thus cause
the scatter in the top panel of Figure \ref{HIscaling}. Here we clearly see that the
$M_{HI}/M_{*}$ vs $M_*$ scaling relation fit depends on the sample used. The relation for HRS (or any stellar mass-selected sample) is offset to that derived for an H{\sc i}-selected sample. It is also sensitive to the environment, with samples from high density regions lying below the scaling relations.

Figure \ref{HIscaling} (bottom) shows the gas fraction is more
strongly correlated with NUV-\textit{r} colour than with
  $M_*$ ($r= -0.84$ and $r= -0.59$ respectively for all samples
  combined). The different samples collated here
  ({\em including the H{\sc i} deficient sources\/}) now
lie on the same best fit relation (contrary to the top panel). The
range in gas fraction at fixed NUV-\textit{r} is thus significantly smaller
than at fixed $M_*$. This can be understood by realising that the parameters
  on both axes scale with the cold gas content, and shows that $M_{HI}/M_{*}$
is a strong driver of the specific star formation rate (either directly, or
indirectly through scaling relations with the molecular gas, which is directly involved in
star formation; e.g. \citeb{Bigiel2011}; \citeb{Schruba2011}; \citeb{Saintonge2011}; \citeb{Bothwell2014}).
This is in line with the scatter in the main sequence of star forming galaxies (SSFR vs $M_*$) being driven by the gas supply \citep{Cortese2011,Catinella2010,Huang2012,Santini2014}. We have checked this is the case for our samples by colour-coding data by H{\sc i} mass in the relation between SSFR and $M_*$.


\subsection{Specific Dust scaling relations}
\label{dustscalingS}
\begin{figure*}
  \includegraphics[width=\textwidth]{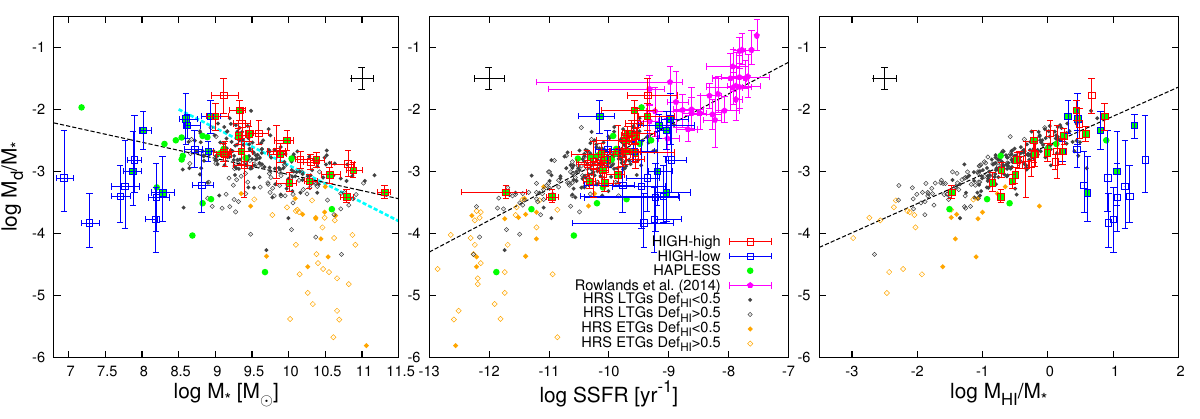}
  \caption{$M_d/M_*$ scaling relations for for the two H{\sc i}-selected sub-samples, HRS and HAPLESS. \textit{From left to right:} Scaling relations with stellar mass, SSFR and $M_{HI}/M_*$. Correlations are found for each of the scaling relations for the high stellar mass H{\sc i}-selected sample, HAPLESS and the HRS and the line of best fit for these `evolved' samples combined is shown as a black dashed line. The relation from \citet{Bourne2012} for blue optically selected galaxies is given as a cyan dashed line. The ETGs are not included in our best fit relations. The symbols are as in Figure \ref{HIscaling}.}
  \label{Dustscaling}
\end{figure*}

\citet{Cortese2012} and \citet{Bourne2012} have studied
specific dust ($M_d/M_*$) scaling relations for HRS and for H-ATLAS stacks on optically selected
sources respectively. They found a strong anti-correlation between $M_{d}/M_*$
with NUV-\textit{r} colour and a weaker anti-correlation with stellar mass,
similar to the scaling relations for $M_{HI}/M_{*}$ in the previous section
\citep[see also:][]{daCunha2010,SmithD2012,Rowlands2014}.
\citet{Viaene2014} also note a similar trend for regions inside M31,
indicating that the driving processes for these scaling relations
(most likely star formation) are local processes.
\citet{Cortese2012} also found a strong correlation of $M_{d}/M_*$  with gas fraction.
Figure \ref{Dustscaling} shows the specific dust scaling relations for the
different samples. We find the scaling relations for HRS, HAPLESS and H{\sc i}GH-high are
consistent with those in \citet{Cortese2012} and \citet{Bourne2012}.

For H{\sc i}GH-low however, we find that the sources lie
significantly below the trends for the other samples in each of the scaling relations
in Figure \ref{Dustscaling}. \textit{The benchmark dust scaling relations identified by
\citet{Cortese2012} and \citet{Bourne2012}\footnote{Note that the H-ATLAS stacks
only extend down to $M_*=10^{8.5} \, M_\odot	$ so the drop in $M_{d}/M_*$ for
our low stellar mass sources does not contradict the statistically significant
trend for the stacks.} based on optically selected samples, do not hold for gas rich, low
$M_*$ (unevolved) sources.} We note that a larger sample is necessary to determine
the appropriate relationship for these low $M_*$ sources.


For the HRS, HAPLESS and H{\sc i}GH-high samples, we find that $M_{d}/M_*$ correlates most strongly with $M_{HI}/M_{*}$ ($r=0.87$), followed by SSFR ($r=0.72$) and then stellar mass ($r=-0.44$). For the scaling relations with
stellar mass in Figure \ref{Dustscaling} (left), we find an offset towards higher
$M_{d}/M_*$ for the H{\sc i}GH-high and HAPLESS samples compared to HRS (similar to Figure
\ref{HIscaling}). This offset is absent in the scaling relations with SSFR and
$M_{HI}/M_{*}$, which appear to be the more fundamental parameters driving the
specific dust mass.

In the centre panel of Figure \ref{Dustscaling} we have plotted
$M_{d}/M_*$ against SSFR, and have added the sample of
high-redshift SubMillimetre Galaxies (SMGs) from
\citet{Rowlands2014}, which were also fitted with {\sc magphys}.
These galaxies are forming stars at a remarkably high rate and lie on a relation that extends
the trend for H{\sc i}GH-high, HAPLESS and the HRS (the best fit
relation was not fitted to the SMGs). The correlation of $M_{d}/M_*$
with SSFR holds over 5 orders of magnitude. This is consistent with
the general idea that dust likely traces the molecular ISM where
star-formation occurs \citep{Dunne2000,Cortese2012,SmithD2012,Rowlands2014}.
Despite the large differences in galaxy properties among the H{\sc
  i}GH-high, HAPLESS and HRS samples, there is no evidence that they
are forming stars in a fundamentally different way, they just have
more or less star formation occurring as a result of their varying gas
fractions. Figure \ref{Dustscaling} (centre) also shows that, for all but the most
immature sources in H{\sc i}GH-low, dust mass is a reasonable
indicator of SFR across a very wide range of $M_*$ and galaxy type.

Since $M_{HI}/M_*$ is a proxy for how far a galaxy is through
its evolution, the correlation seen in the right panel of Figure
\ref{Dustscaling} implies that $M_{d}/M_*$ depends on the evolutionary state.
As galaxies evolve, they
move from high to low $M_{HI}/M_{*}$ and (for H{\sc i}GH-high,
the HRS and HAPLESS) $M_{d}/M_*$ decreases. The unevolved sources
in H{\sc i}GH-low clearly lie below the relation for the other samples
and imply a rising $M_{d}/M_*$ at the earliest stages of evolution
($M_{HI}/M_{*}>10^{0.5}$). \textit{At high gas fractions, dust is not a
good tracer of the the H{\sc i} content}. These
 galaxies must be increasing their dust content at a faster
fractional rate than their stellar content. The combined samples have allowed us
to find a peak in the specific dust mass ($M_{d}/M_*$) in the
local Universe occuring at a gas fraction of $\sim 75$~per~cent and a stellar mass of
$\rm{M_* =10^{8.5}}$. H{\sc i}GH-low is the first sample of galaxies to be identified as
preceding this peak $M_{d}/M_*$ in an evolutionary sequence.

\subsection{Dust enrichment relations}
\label{dust-to-HI-section}

\begin{figure*}
  \includegraphics[width=\textwidth]{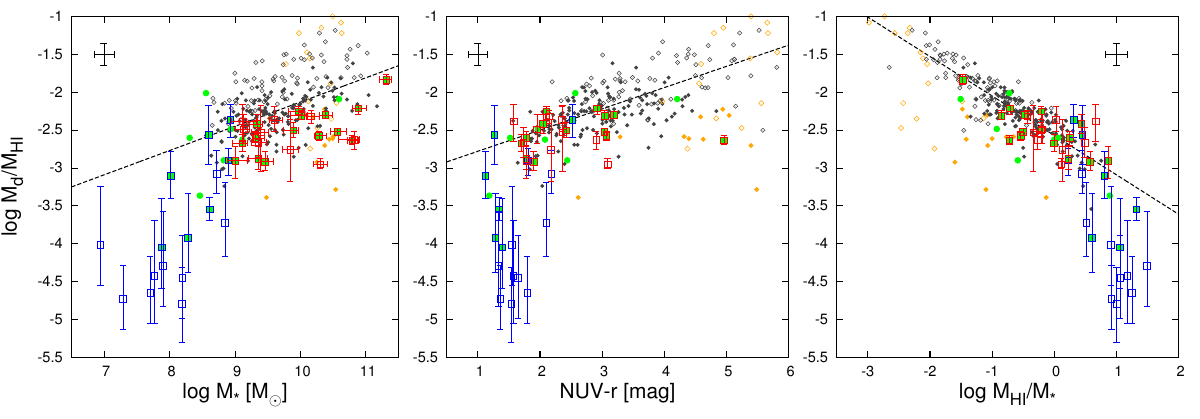}
  \caption{Dust enrichment relations for HRS, HAPLESS and H{\sc i}GH. \textit{From left to
      right:} Scaling relations with stellar mass, NUV-\textit{r}
    colour (proxy for SSFR) and $M_{HI}/M_*$. The line of best fit for the combined
    `evolved' samples (H{\sc i}GH-high, HAPLESS and HRS) is shown as a black dashed line. The H{\sc i}GH-low
    sample lies significantly below the trend for the other samples
    and has a steeper slope. The symbols are as in Figure \ref{HIscaling}.}
  \label{HIDustper2}
\end{figure*}

We next look at the dust content of the ISM as a function of
stellar mass, NUV-\textit{r} colour and $M_{HI}/M_{*}$ (Figure
\ref{HIDustper2}), where we find different scaling relations
for H{\sc i}GH-low. For H{\sc i}GH-high, HAPLESS and
the HRS there is a weak but significant correlation between
$M_{d}/M_{HI}$ and $M_*$ ($r=0.47$). For H{\sc i}GH-low, however, we find a
steeper slope (Table \ref{correlations}) and a significantly smaller
$M_d/M_{HI}$ than expected from extrapolating the relation for the other
samples. We find tighter scaling relations with NUV-\textit{r} colour
($r=0.66$) and gas richness ($r=-0.86$) for H{\sc i}GH-high, HAPLESS and the
HRS and again an offset towards lower dust enrichment for
H{\sc i}GH-low. Note again
the offset between the H{\sc i}GH-high/HAPLESS samples and the HRS
for the stellar mass scaling relations (cf. Figures
\ref{HIscaling} \& \ref{Dustscaling}). Once again H{\sc i} deficient galaxies are offset when $M_{d}/M_{HI}$ is plotted against $M_*$, yet this offset disappears for the more fundamental relations of $M_{d}/M_{HI}$ with NUV-\textit{r} colour and $M_{HI}/M_{*}$. The offset between the samples is once again caused by
differences in gas fractions at fixed stellar mass. Our interpretation of these dust enrichment relations is as follows:

Gas is continuously converted into stars and dust is produced at the
endpoints of stellar evolution (supernovae and AGB stars). Yet at the same time dust is destroyed by shocks and also via astration as the ISM at the
ambient dust-to-gas ratio forms the next generation of stars. For
H{\sc i}GH-high, HAPLESS and the HRS we have found positive
correlations of the dust-to-gas ratio with stellar mass and
NUV-\textit{r} colour, together with a negative correlation with the
gas richness, showing that $M_{d}/M_{HI}$ increases monotonically
as galaxies move through their evolution. This implies that the
dust destruction budget is not dominated by
dust destruction through shocks or sputtering. If it was,
we would observe a decrease in the dust-to-gas ratio as galaxies evolve.
Some of the ETGs in HRS may be an exception to this. These ETGs
are bright X-ray sources and some have AGN in their centres. The hot gas
in these sources erodes and breaks up the dust grains (sputtering),
significantly reducing the dust mass and resulting in the outliers towards
low $M_{d}/M_{HI}$ for HRS in Figure \ref{HIDustper2}.

\begin{figure}
  \includegraphics[width=\columnwidth]{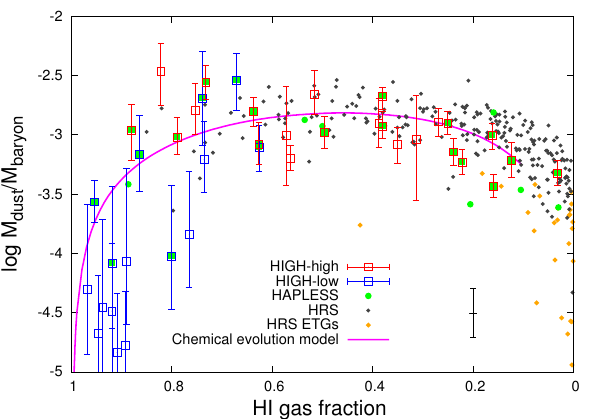}
  \caption{$M_{d}/M_{bary}$ against gas fraction (without molecular hydrogen) reveals the evolution of dust. As galaxies evolve, the dust content first increases (high gas fraction), then reaches its peak for a gas fraction of $\sim 0.5$ and afterwards decreases as dust is consumed together with the gas (low gas fraction tail). A chemical evolution model (C15) is also shown.}
  \label{chemev}
\end{figure}

In Figure \ref{chemev}, we follow C15 in plotting
  $M_{d}/M_{bary}$ mass ratio vs gas fraction $f_g$ (their Figure 21) where
  we define the baryon mass as $M_{bary} = M_{HI}+M_*$ and
  $f_g=M_{HI}/(M_{HI}+M_*)$. Note that we
  do not have CO data for H{\sc i}GH and HAPLESS so we cannot measure
  the molecular gas mass present in these galaxies. We also follow C15 in comparing the
  observations with a simple, closed box chemical evolution
  track\footnote{Further details on the model are presented in
    Rowlands et al. (\citeyear{Rowlands2014b}; see also
    \citeb{Morgan2003})} showing the expected change in
  $M_{d}/M_{bary}$ with gas fraction for a Milky Way type star
  formation history \citep{Yin2009}. The track (solid line; same as C15)
  shows the evolution of a galaxy as it evolves from gas rich to gas
  poor, with gas consumed by star formation.

  Combining H{\sc i}GH with HAPLESS and HRS allows us to sample
  a wider range of $f_g$. As in C15, we see $M_{d}/M_{bary}$ first rises
  steeply, then levels off and then drops again as galaxies evolve
  from high to low gas fractions. This supports the idea of the dust
  content being built up as galaxies move through the early stages of
  their evolution (gas fraction > 0.8). The dust content then plateaus
  as dust destruction through astration balances the dust
  production. Note that while the position of a galaxy in
  Figure \ref{chemev} does not depend on its total mass, since both
  axes are ratios, when sampling at the current epoch we find that the
  highest $M_*$ galaxies are at the right of the plot and the lowest
  $M_*$ are at the left, because massive galaxies go through their
  evolution faster. Including the H{\sc i}GH sample provides crucial
  information at the highest gas fractions compared to the initial
  study in C15.

Of course, galaxies are far more complex than our simple model, with
inflows and outflows and dust destruction expected to be important
factors. A more detailed study of the build up of dust at high gas
fractions will be presented in paper II (De Vis et al., \textit{in prep.}), and
trends with metallicity will also be studied (high gas fraction sources have
significantly lower metallicities than low gas fraction sources).
However, even with the simplistic approach in C15, \textit{the model is still able to match
  the observed overall shape of the build-up and destruction of dust
  as a galaxy evolves.}

\section{The evolution of star formation efficiency}
\label{SFE}

\begin{figure}
  \includegraphics[width=\columnwidth]{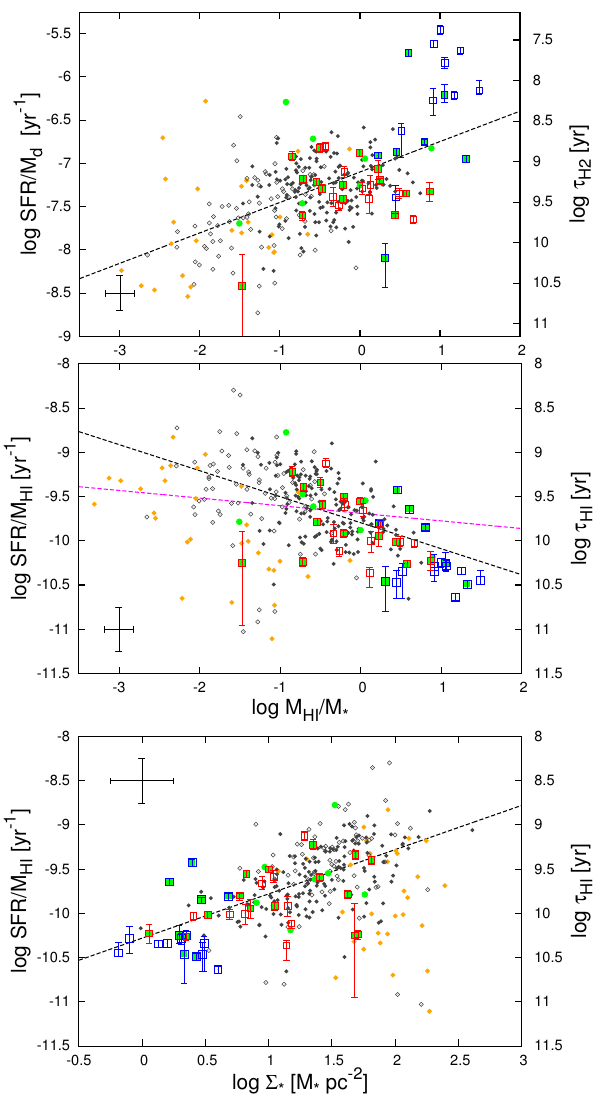}
  \caption{\textit{Top:} $SFR/M_{d}$ against $M_{HI}/M_*$ showing a
   slow decline of $SFR/M_{d}$ as galaxies evolve.
   	H{\sc i}GH-low is significantly offset towards
    higher $SFR/M_{d}$. The right axis shows the molecular gas depletion timescale using $M_d$ as a proxy. \textit{Centre:} $SFR/M_{HI}$ against $M_{HI}/M_*$. There is a clear
    evolution towards higher $SFR/M_{HI}$ for more
    evolved sources (best-fit line including all samples is shown in
    dashed black). The correlation resulting from a typical H{\sc i} error of 0.1 dex has been determined using MC simulations (dashed magenta).
    \textit{Bottom:} Star formation rate per H{\sc i} mass ($SFR/M_{HI}$) against
    	stellar mass surface density $\Sigma_{\ast}$.}
  \label{SFRdust}
\end{figure}

In Figure \ref{SFRdust}, we look at two measures of
  the star formation efficiency of the galaxies as a function of their
  gas fraction (or evolutionary status). In the top panel we consider
  $SFR/M_{d}$, while in the bottom two panels we show $SFR/M_{HI}$. Our
  interpretation of these quantities is that $SFR/M_{d}$ is a proxy
  for the molecular star formation efficiency to the extent that dust
  is a good tracer of molecular gas in galaxies
  \citep{Dunne2000,Planck2011XIX,Corbelli2012,Rowlands2014,Scoville2014,Santini2014}.
  As molecular gas is directly involved in star
  formation (e.g. \citet{Bigiel2008, Schruba2011}), SFR/$M_{\rm{H2}}$
  (aka SFR/$M_d$) is an indicator of the efficiency with which gas is
  converted to stars inside the radius at which molecular clouds are
  present in the galaxy.\footnote{For a true measure of the efficiency
    of converting dense gas into stars within star forming regions, it
    is necessary to choose a high density molecular tracer (e.g. HCN)
    \citeb{Gao2004}; \citeb{Papadopoulos2012}.}

On the other hand, atomic hydrogen does not directly form stars, it
must first make a transition to molecular form. $SFR/M_{HI}$ is
therefore not a true star formation efficiency but rather an
indication how effectively the H{\sc i} is able to turn into molecular
form and subsequently form stars. With this distinction in mind, we now turn to the trends shown in
Figure \ref{SFRdust}. Taken at face value, and assuming a canonical value for
$M_d/M_{H2}$ of 0.007 (\citeb{Draine2007}; see also \citeb{Corbelli2012}),
the top panel of Fig
\ref{SFRdust} shows that the star formation efficiency in galaxies
declines as they evolve, with the relationship in Table \ref{correlations}
indicating a rise in the molecular gas depletion timescale
($\tau_{H2}$) from 1.7 Gyr to 4 Gyr over a range in gas fraction from
80-10~per~cent. The H{\sc i}GH-low sample lies well above this relationship
indicating either a much shorter molecular gas depletion time (average of 140
Myr), or a much lower $M_d/M_{H2}$ ratio (by a factor $\sim 10$). Resolving this issue
would require resolved CO + H{\sc i} maps for these sources.

Studies of the other main molecular gas tracer (CO) in
  local galaxies selected over a range of stellar mass from $8.5<
  \rm{Log \,M_{\ast}} < 11.5$, find a similar result; that the star formation efficiency
  increases (or the gas depletion time decreases) as the stellar mass
  decreases and as SSFR increases
  \citep{Saintonge2011,Bothwell2014,Boselli2014}.\footnote{Earlier
    studies of the molecular gas depletion times in local spiral
    galaxies found a constant $\tau_{H2}$ of $\sim 2$ Gyr
    \citep[e.g.][]{Bigiel2008,Leroy2008}, however these studies probed
    a much smaller range of intrinsic stellar mass or gas fraction and
    so are not thought to contradict the later findings of these
    larger studies.} These studies find a range of $\tau_{H2}$ from
  100 Myr -- 5 Gyr over the same range of stellar mass as sampled
  here although our study contains three
  samples selected in very different ways (dust, gas and stellar
  content)\footnote{One of the samples (HRS) is the same as that used by
    \citet{Boselli2014} though we are using dust as a tracer of
    $\rm{H_2}$ rather than CO.}.

  In the centre panel of Figure \ref{SFRdust} we find that there is
  considerable evolution in $SFR/M_{HI}$ ($r=-0.53$), such that more
  evolved galaxies have higher star formation per H{\sc i} mass
  (shorter H{\sc i} depletion times, $\rm{\tau_{HI}}$, assuming
  constant star formation rate and no re-supply of gas).  We must be
  cautious in interpreting Figure \ref{SFRdust} (centre) as the
  quantity $\rm{M_{HI}}$ is present in both the x and y axes. Monte
  Carlo simulations were used to confirm that this relation cannot be
  due to biases introduced by the errors in $M_{HI}$. For each source in
  the sample, we generated an artificial H{\sc i} mass so that its
  $SFR/M_{HI}$ is equal to the average $SFR/M_{HI}$ in the whole
  sample (the null hypothesis is that there is no evolution in
  $SFR/M_{HI}$) and then added Gaussian scatter with a standard
  deviation of 0.1 dex (typical $\rm{M_{HI}}$ error). This process was
  repeated 100 times and the resulting average trend is shown by the
  magenta line in Figure \ref{SFRdust} (centre). The error on
  $\rm{M_{HI}}$ does introduce an artificial correlation, however, the
  observed slope in Figure \ref{SFRdust} (centre) is significantly
  steeper and we believe this is a real effect.

The galaxies with the highest gas fractions, which were previously
found to be the most actively star forming galaxies in terms of their
stellar mass (SSFR) and their dust mass ($SFR/M_{d}$), are now least
active with respect to their H{\sc i} mass (they have the lowest
$SFR/M_{HI}$). The H{\sc i} depletion timescales range from 0.2 -- 63
Gyr, with the most gas rich ($\rm{M_{HI}>M_{\ast}}$) sources capable
of sustaining their current star formation rates for longer than the
Hubble time. Previous studies find a comparable range in the value of
$\rm{\tau_{HI}}$ but no trend with any of the parameters which
correlate with $\tau_{H2}$ e.g. stellar mass, SSFR
\citep{Saintonge2011,Bothwell2014,Boselli2014}. Similarly, we do not
find a correlation of $\rm{\tau_{HI}}$ with either stellar mass or
SSFR.

  There is, however, a known relationship between $\tau_{HI}$ and
  stellar mass surface density ($\Sigma_{\ast}$) {\em within}
  galaxies. The THINGS survey \citep{Walter2008, Leroy2008} found
  $SFR/M_{HI}$ to be a strong and almost linear function of stellar
  mass surface density in the outer regions of spirals and in dwarfs,
  where the ISM is dominated by H{\sc i}. Within the inner regions of
  spiral galaxies, the higher stellar mass surface density produces a
  higher hydrostatic pressure in the ISM
  \citep{Elmegreen1989,Wong2002} favouring the conversion of H{\sc i}
  to $\rm{H_2}$ and results in a molecular dominated region where the
  star formation efficiency ($\tau_{H2}$) is constant. We find a
  correlation ($r=0.58$) between the global $\tau_{HI}$ and stellar
  mass surface density in the bottom panel of Figure
  \ref{SFRdust}. This is the first time that such a relationship has
  been reported for global values {\em between\/} galaxies.

  We can use Figure \ref{SFRdust} (bottom) to interpret the top two
  panels as being the result of an increasing efficiency of conversion of
  $\rm{H_I\rightarrow H_2}$ as galaxies become more dominated by their
  stellar mass. As galaxies build up their stellar mass and
  increase in $\Sigma_{\ast}$ they create the conditions for
  $\rm{H_2}$ formation across a wider area; and their H{\sc i}
  reservoirs are depleted due to conversion to $\rm{H_2}$ and
  thence to stars. As galaxies become very dominated by stars and have
  large bulges, they can be $\rm{H_2}$ dominated over large areas and
  their H{\sc i} reservoirs will be relegated to the outskirts of the
  galaxy. In very evolved galaxies (e.g. ETGs) the presence of gas and
  star formation will be more strongly correlated with recent
  interactions or environmental effects \citep{Davis2011,Kaviraj2012,Kaviraj2013,Davies2015}.
  This may explain the very large scatter in $\tau_{HI}$ for the lowest gas fraction galaxies.  

This general picture is not strongly dependent on an assumption of a
constant dust/$\rm{H_2}$ ratio, that ratio would need to vary by
several orders of magnitude to invalidate this
interpretation. Detailed observations of resolved CO, H{\sc
  i} and dust would be required to further elaborate on this.

\section{Dust heating in the diffuse ISM}
\label{dustheatingS}
Dust in the diffuse ISM is heated by the interstellar radiation
field (ISRF), which has contributions from both old and young
stellar populations. Dust in birth clouds experiences
more intense and harsh radiation fields in the photo-dissociation
regions (PDRs). To account for this {\sc magphys} has a parameter,
$f_\mu$, which represents the fraction of the total dust luminosity
arising in the diffuse ISM. The majority of the dust mass resides in
a cold ($10<T<30$ K) diffuse dust component, whereas in most
actively star forming galaxies a large fraction of the dust luminosity is
due to a warm ($30<T<60$ K) dust component arising in birth clouds
(futher details of the {\sc magphys} components are found in Section
\ref{SEDfitS}).

In a typical galaxy in the local universe much of the stellar mass is
in low mass stars, yet the small fraction of massive, short-lived
stars radiate much more strongly at UV wavelengths. This UV radiation
is highly susceptible to absorption by dust and the high energy UV
photons can cause much of the dust heating (e.g. \citeb{Draine2007})
even though they only make up a very small fraction of the photons in
the ISRF. In birth clouds the UV photons from young stars
dominate the dust heating, but even in the diffuse ISM the PAHs, small grains
(stochastically heated and emitting at MIR) and warm dust components are still
mostly heated by UV photons that leak from the birth clouds and
form part of the diffuse ISRF
\citep{Devereux1990,Kennicutt1998,Calzetti2005,Calzetti2007,Boquien2010,Bendo2012,Kirkpatrick2014}.
Many literature works have studied the sources of dust heating for the bulk of the dust mass in the diffuse ISM, and found that both the young stars in star forming regions and the diffuse evolved populations heat the diffuse dust \citep[][C15]{Bendo2010,Boselli2010B,Boselli2012,Foyle2013,Ciesla2014,Cortese2014,Kirkpatrick2014,Draine2014,Bendo2015}. In this section we will study which parameters drive change in $f_\mu$ and cold dust temperature and investigate which source of dust heating dominates in a particular galaxy for a wide range of gas fraction.

\begin{figure}
\center
  \includegraphics[width=\columnwidth]{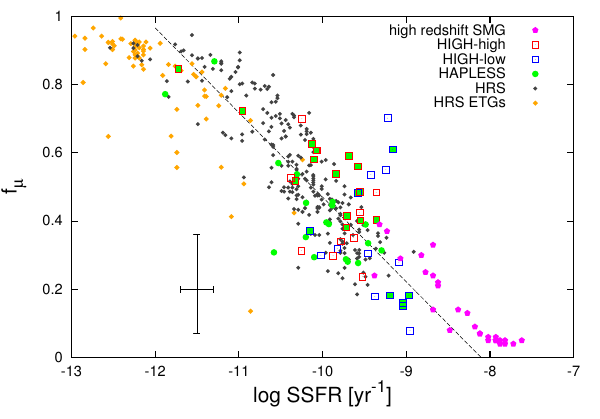}
  \caption{Influence of SSFR on the fraction of the total luminosity
    that originates in the diffuse ISM ($f_\mu$). Less actively star
    forming galaxies have a larger fraction of their dust luminosity
    originating in the diffuse ISM. The line of best fit for combined
    HAPLESS, H{\sc i}GH and SMG \citep{Rowlands2014} samples is shown as a
    dashed black line.}
  \label{f_mu}
\end{figure}

Figure \ref{f_mu} shows an anti-correlation of $f_\mu$ with SSFR
for all the samples. In order to probe to the highest SSFR, we have
included the high redshift SMGs from
\citet{Rowlands2014}. As expected, for most galaxies the fraction of the total dust luminosity originating in
the birth-clouds ($1-f_\mu$) is proportional to the star forming
activity of the galaxy.
This would be the case if a reasonable
  fraction of the energy in the birth clouds was being absorbed
  locally and re-radiated by dust (i.e. at least moderate $A_{FUV}$). Outliers
  can occur if the UV attenuation in the birth clouds is very low,
  allowing most of the UV energy to escape and heat the dust in the
  diffuse dust component. We indeed find that the outliers towards high $f_\mu$ in Figure \ref{f_mu}
  are all amongst the least attenuated sources in the sample ($A_{FUV}<0.35$; see next section).
   On the other hand, outliers can also occur if a considerable fraction of the dust is heated by AGN activity or the hot X-ray halo that is often present in ETGs, as these sources of heating are not included in the {\sc magphys} prescription. All the outliers towards low $f_\mu$ are ETGs and the strongest outliers are known bright X-ray sources.  

\begin{figure}
\center
  \includegraphics[width=\columnwidth]{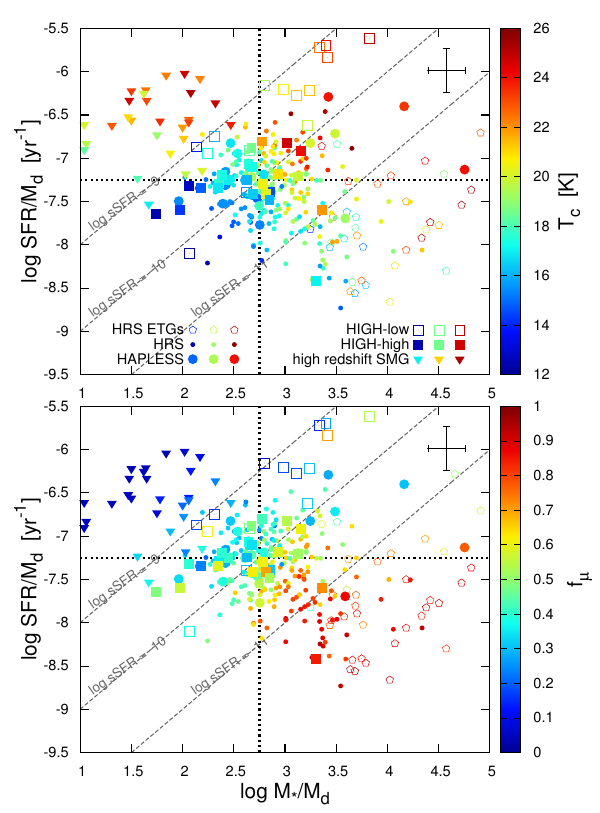}
  \caption{$M_*/M_d$ against $SFR/M_d$, colour-coded by cold dust
    temperature $T_{c}$ (\textit{top}) and the fractional contribution of
    diffuse dust to the total dust luminosity, $f_\mu$ (\textit{bottom}) in
    order to study the sources of dust heating. These plots have been
    divided in 4 `heating' quadrants to highlight the differences. Lines of constant SSFR are shown in
    dashed gray.}
  \label{dustheating}
\end{figure}

Both the young stars (traced by the SFR) and the old
  stars (traced by $M_*$) play a role in heating the diffuse
  dust. Figure \ref{dustheating} provides us with a graphical way to
  understand the contributions from the old and young stellar
  populations to the total and cold dust heating. We have plotted
  $M_*/M_d$ against $SFR/M_d$ and colour-coded the data by cold dust
  temperature and $f_\mu$ respectively (again including
  the SMGs from \citeb{Rowlands2014}).

Even discounting the SMGs\footnote{Including the SMGs leads to an even
  weaker correlation.}, which generally lie off the main sequence of
star formation, we find only a very weak correlation between SFR and
$\rm{M_{\ast}}$ {\em after normalising by dust mass}. There is a clear
trend towards higher temperatures as one goes to higher $M_*/M_d$ or
$SFR/M_d$. For a fixed $M_*/M_d$, we find the spread in temperatures
largely follows the differences in $SFR/M_d$ and the same when fixing
$SFR/M_d$ and varying $M_*/M_d$. This explains why ETGs have
warm $T_c$, as despite having low SFR their $M_*/M_d$ are the highest,
and so their old stellar radiation fields are intense enough to heat
the diffuse dust to warmer temperatures.  At $SFR/M_d >10^{-6.5}
yr^{-1}$, there no longer seems to be any dependence of $T_c$ on
$\rm{M_*/M_d}$, probably because the dust heating is completely
dominated by the young stellar population for these galaxies and the
old stellar population has little effect. For galaxies with $SFR/M_d <
10^{-6.5} yr^{-1}$, both the young and old stellar populations heat
the cold dust component, with some dominated by one and some by the
other.

Figure \ref{dustheating}(bottom) shows that the direction of increasing $f_\mu$ (also direction of increasing SSFR) is
nearly orthogonal to the direction of increasing $T_c$. This means
that the cold dust temperature is more or less independent of the
fraction of the total dust luminosity originating in the diffuse ISM
(i.e. $T_c$ is not affected by the SSFR). However the fraction of the dust luminosity originating from heating by old stars is inversely proportional to SSFR \citep{Boquien2016}.
Galaxies in the upper left quadrant of Figure \ref{dustheating} have most of the dust luminosity originating in birth clouds and the dust heating is dominated by photons from young stars. Galaxies with moderate SSFR have dust emission from both birth clouds and the diffuse ISM as well as significant contributions to the heating of the diffuse dust from both young and old stars. These galaxies can be dust rich and cold (lower left quadrant) or dust poor and warmer (upper right quadrant). Finally, quiescent galaxies (lower right quadrant) have most of their dust luminosity originating in the diffuse ISM and this is heated mostly by the old stars.

\begin{figure*}
  \includegraphics[width=\textwidth]{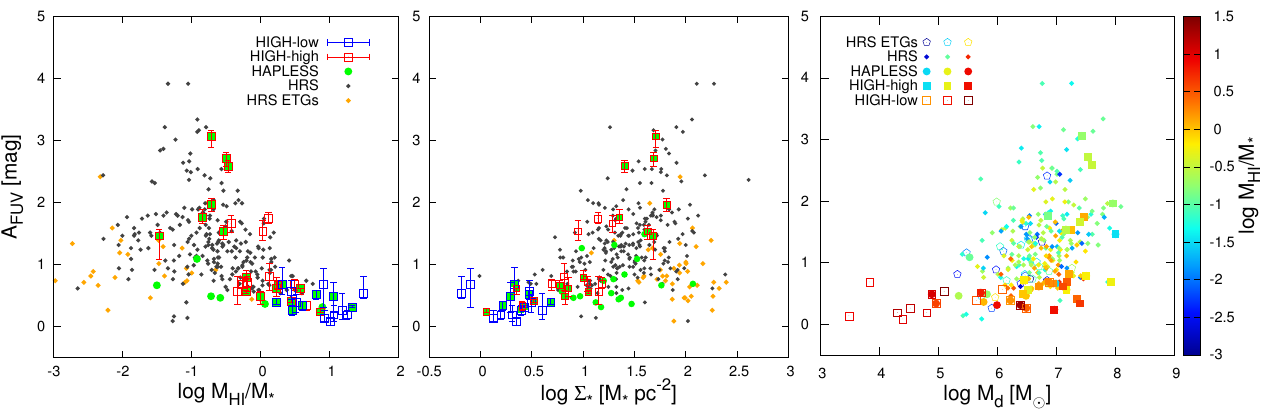}
  \caption{The variation of FUV attenuation $A_{FUV}$ with galaxy
    parameters. \textit{Left:} The obscuration increases as one
    moves from high to low gas fraction (i.e. from less to more
    evolved sources). \textit{Centre:} Relationship between $A_{FUV}$ and
    stellar mass surface density $\Sigma_{\ast}$.  \textit{Right:} For low dust
    masses ($M_d<10^5 M_{\odot}$) there is little to no
    obscuration. We find a positive correlation between the
    obscuration and the dust mass above this value for the
    HAPLESS and H{\sc i}GH samples. The outliers towards high obscuration
    at moderate dust masses are evolved sources in HRS ($log\, M_{HI}/M_{*} < -0.6$).
    For all three plots there is a large amount of scatter
    that correspond to a wide range of obscuration for a given $M_d$
    (likely due to different dust and stellar geometries).}
  \label{SFR_FUV}
\end{figure*}

\section{Obscuration}

To study how the UV obscuration depends on other galaxy properties we
have plotted the {\sc magphys} $A_{FUV}$ parameter against $M_{HI}/M_*$, $\Sigma_{\ast}$ and dust
mass in Figure \ref{SFR_FUV}. In the left panel we find an anti-correlation
($r=-0.54$) for $A_{FUV}$ with
$M_{HI}/M_{*}$. As galaxies move through their evolution, from
gas rich to gas poor (right to left on this plot) the obscuration initially
increases. This makes sense as dust is continuously produced and mixed
with the ISM. Note that all galaxies with $\rm{log \, M_{HI}/M_* > 0.5}$
approach $A_{FUV}=0$, which corresponds to the limit of no
obscuration. From Figure \ref{HIDustper2} this corresponds to $log \, M_d/M_{HI}<-3$.

The sources with the highest obscuration have the highest cold dust
temperatures and are on average slightly more inclined than the less
obscured galaxies at the same $M_{HI}/M_*$. In the latest stages of evolution, the obscuration decreases again as most of the dust mass is consumed due to astration (Figure \ref{chemev}) or removed.

The large scatter in $A_{FUV}$ at lower gas fractions is at least partly due to differences in the
intrinsic stellar and dust geometries and inclinations
of these galaxies. Attenuation strongly
depends on how much of the dust is mixed into the diffuse ISM as
opposed to being distributed in a more clumpy geometry, and on other
geometric differences like scale heights and scale lengths of the
stellar and dust disks \citep{Baes2001,Bianchi2008,Holwerda2012,Popescu2011}.
Investigating whether or not the star-dust geometry is the main factor that drives this scatter is a difficult task. One potential way to do that is by including realistic recipes for dust attenuation in hydrodynamical models of galaxy evolution, and comparing the attenuation properties of simulated mock galaxies to observed data. As both cosmological hydrodynamical simulations \citep[e.g. ][]{Vogelsberger2014,Schaye2015} and 3D dust radiative transfer techniques \citep{Steinacker2013} have reached a level of maturity, this combination has recently become possible (e.g., \citeb{Camps2016}; Trayford et al., \textit{in prep.}). Such an investigation is beyond the scope of this paper, but will be the subject of future work.

\citet{Grootes2013} found a relationship between the optical depth and stellar mass surface density $\Sigma_{\ast}$ of nearby spiral galaxies. We find a similar relationship ($r=0.53$) when we plot $A_{FUV}$ against $\Sigma_{\ast}$ in Figure \ref{SFR_FUV} (centre).
The increased stellar mass potential associated with higher $\Sigma_{\ast}$ creates instabilities in the cold ISM, which lead to the formation of a thin dust disk \citep{Dalcanton2004}. This changes the relative geometries of dust and stars which provides a possible explanation for the changes in obscuration.

The attenuation by dust is expected to depend on the total column of dust along a photon's
trajectory. We show how $A_{FUV}$ varies with the total dust mass\footnote{Plotting
$A_{FUV}$ against the dust surface density (which could be argued to be a
better tracer of the dust mass along a photon's path), does not change the
results in any way.}, colour coded by gas fraction, in the right panel of
Figure \ref{SFR_FUV}. We find a positive
correlation ($r=0.38$) but the relationship
is not a simple power law and there is a lot of scatter.
At $\rm{M_d<10^{5.5}}$ the obscuration tends to zero, while at higher
dust masses, there is a large
range in obscuration and again this is likely due to different
stellar and dust geometries or different extinction laws.
In summary we find no clear and simple link between UV obscuration and global galaxy properties.

\section{Conclusions}
We have studied the interplay of dust, gas and star formation for combined samples made up of local H{\sc i}-, dust- and stellar mass selected galaxies, using {\sc magphys} to determine physical properties from UV to submm photometry.
The combined samples cover a wide range of gas fractions (proxy for the evolutionary state of a galaxy). Our main results are: \begin{itemize}
\item We have identified a sub-sample of H{\sc i}-selected sources
  (H{\sc i}GH-low) with very high gas fractions ($f_g>80$~per~cent) and low
  stellar masses ($M_*<10^{9} M_{\odot}$). These probe the earliest
  stages of evolution, and have a much smaller dust content than expected
  from extrapolating published scaling relations for more
  evolved sources.
\item In the earliest stages of evolution ($f_g>80$~per~cent), dust is not a good tracer of the gas content.
The dust content relative to stellar mass first rises steeply with decreasing gas fraction, reaches a peak at
a gas fraction of $\sim$75\,per\,cent (which for local galaxies is equivalent to a stellar mass of $\sim 10^{8.5}\rm \, M_{\odot}$), and then decreases together with gas fraction.
\item The galaxies with the highest gas fractions are the most
  actively star forming galaxies relative to their stellar masses
  (SSFR) and relative to their $\rm{H_2}$ content (using dust as a
  proxy for $\rm{H_2}$).
\item We find a trend of decreasing H{\sc i} depletion time with
  decreasing gas fraction, such that the most gas rich galaxies have
  the longest $\tau_{HI}$. We interpret this, together with the opposing behaviour of
  $\tau_{H2}$, as being due to the increasing efficiency with which
  H{\sc i} can be converted to $\rm{H_2}$ as galaxies increase in
  stellar mass surface density with decreasing gas fraction.
\item We confirm literature results that both old and young stellar populations play an important role in heating the diffuse dust component, and both can be the dominant
  contributor in individual systems. The SSFR determines which dominates.
\item The FUV obscuration of galaxies shows no clear and simple link with global galaxy properties. Galaxies start out barely obscured and increase in obscuration as they evolve until the dust mass decreases significantly in the latest stages of evolution.
\end{itemize}

The derived scaling relations for the combined samples in this work span a wider range in gas fraction than previous relations in the literature, yet admittedly have somewhat complex selection biases. Since the sample size of the stellar mass selected sample (HRS) is 8 times larger than the H{\sc i}- and dust-selected samples, the scaling relations are therefore heavily weighted towards this sample. This especially affects the scaling relations with stellar mass, which show significant offsets between the differently selected samples. However using the combined sample including the high gas fraction sources, we show that the most robust scaling relations for gas and dust are those linked to NUV-\textit{r} (SSFR) and gas fraction. These are tight relations which do not depend on sample selection or environment and are thus not affected by the complex selection criteria of the combined sample.

\section*{Acknowledgments}
We thank the anonymous referee for useful suggestions.
The authors gratefully acknowledge Martha Haynes,
Riccardo Giovanelli, and the ALFALFA team for supplying
the latest ALFALFA survey data. The authors also thank
Ariadna Manilla Robles, Asantha Cooray, Paul van
der Werf and Michal Michalowski for helpful discussions.
LD, SJM and RJI acknowledge support from the ERC in the
form of the Advanced Investigator Program, COSMICISM.
HLG, LD and SJM acknowledge support from
the European Research Council in the form of Consolidator Grant
CosmicDust.
CJRC acknowledges support from the European Research Council
(ERC) FP7 project DustPedia (PI Jon Davies) and the STFC
Doctoral Training Grant scheme.
The H-ATLAS is a project with \textit{Herschel}, which is an
ESA space observatory with science instruments provided
by European-led Principal Investigator consortia and with
important participation from NASA. The H-ATLAS website
is http://www.h-atlas.org/. GAMA is a joint European-
Australasian project based around a spectroscopic campaign
using the Anglo-Australian Telescope. The GAMA
input catalogue is based on data taken from the Sloan
Digital Sky Survey and the UKIRT Infrared Deep Sky Survey.
Complementary imaging of the GAMA regions is being
obtained by a number of independent survey programs
including {\em GALEX} MIS, VST KIDS, VISTA VIKING, WISE,
\textit{Herschel}-ATLAS, GMRT, and ASKAP, providing UV to
radio coverage. GAMA is funded by the STFC (UK), the
ARC (Australia), the AAO, and the participating institut
The GAMA website is: http://www.gama-survey.org/.
This research has made use of Astropy\footnote{http://www.astropy.org/},
a community-developed core Python package for Astronomy \citep{Astropy2013}.
This research has made use of TOPCAT\footnote{http://www.star.bris.ac.uk/~mbt/topcat/}
\citep{Taylor2005}, which was initially developed under the UK Starlink project,
and has since been supported by PPARC, the VOTech project,
the AstroGrid project, the
AIDA project, the STFC, the GAVO project, the European
Space Agency, and the GENIUS project. This research has
made use of of APLpy\footnote{http://aplpy.github.io/},
an open-source astronomical image
plotting package for Python. This research has made use of
NumPy\footnote{http://www.numpy.org/} \citep{Walt2011},
SciPy\footnote{http://www.scipy.org/}, and MatPlotLib\footnote{http://matplotlib.org/}
\citep{Hunter2007A}. This research has made use of the
SIMBAD\footnote{http://simbad.u-strasbg.fr/simbad/} database
\citep{Wenger2000} and the VizieR\footnote{http://vizier.u-strasbg.fr/viz-bin/VizieR}
catalogue access tool \citep{Ochsenbein2000}, both
operated at CDS, Strasbourg, France. This research has made
use of SAOImage DS9\footnote{http://ds9.si.edu/site/Home.html},
developed by the Smithsonian
Astrophysical Observatory with support from the Chandra X-ray
Science Center (CXC), the High Energy Astrophysics
Science Archive Center (HEASARC), and the JWST Mission
office at the Space Telescope Science Institute (STSI).
This research has made use of the NASA/IPAC Extragalactic
Database (NED\footnote{http://ned.ipac.caltech.edu/}),
operated by the Jet Propulsion
Laboratory, California Institute of Technology, under contract
with the National Aeronautics and Space Administration.

\bibliographystyle{mnras}
\bibliography{Library}

\appendix
\section{Exceptions to the aperture photometry}
\label{photomexep}
\subsection{IRAS SCANPI photometry}
\label{scanpi}
We used the Scan Processing and Integration Tool
(SCANPI\footnote{Provided by the NASA/IPAC Infrared Science Archive:
  http://irsa.ipac.caltech.edu/applications/Scanpi/}), following the
procedure of \citet{Sanders2003} to measure 60 $\mu m$ fluxes and
uncertainties for our sources. For a third of the sample, no reliable
detection could be found at the location of the source. For these
sources the scans were inspected manually and an upper limit was
defined for the flux as 3 times the local rms.

\subsection{\textit{Herschel} PACS photometry}
\label{PACS}
We dealt with PACS photometry as described in C15,
using apertures based on the 250$\mu$m source size.
In contrast to C15, we use the H-ATLAS DR1 Nebulised\footnote{{\sc Nebuliser} is
an algorithm to remove background emission \citep{Irwin2010}} maps
for all our sources. The filtering applied to the maps could lead to a localised negative background for very
extended sources \citep{Valiante2016}. By limiting the PACS aperture
to the obvious extent of the dust emission we are minimising the
effects of these large scale background issues and increasing the
accuracy and reliability of the flux measurements. Where we do not
have a strong enough 250 $\mu m$ detection to reliably determine an
aperture, we use an aperture 0.8 times the largest aperture size from
the other bands. This factor was determined to be the average ratio of
$r_{ap(\rm 250)}/r_{ap(\rm max)}$ for sources with $SNR_{250} > 5$ within
the aperture. We have performed tests that there are no significant
systematic differences in the fluxes obtained when either using a larger
aperture or when the raw {\sc Scanamorphos} \citep{Roussel2013} maps are used instead.

\section{Properties of the H{\sc i}GH galaxies}
\label{ApendixA}
Basic properties for our H{\sc i}-selected sample, such as identifiers, positions and sizes are given in Table \ref{table1}. The H{\sc i} fluxes, H{\sc i} masses, references and other H{\sc i}-derived properties are given in Table \ref{tableHI}. The UV to FIR photometry for the H{\sc i}-selected H{\sc i}GH sample can be found in Table \ref{photo}. Multiwavelength imagery of the H{\sc i}GH galaxies is shown in Figure \ref{images} . The {\sc magphys} fits to the spectral energy distributions of the H{\sc i}GH sources are shown in Figure \ref{SED}.

	\begin{landscape}
	\begin{table}
    		\caption{Basic properties for our H{\sc i}-selected sample. Velocities and distances are corrected for bulk deviation from Hubble flow \citep{Baldry2012}. Semi major axes were calculated using the bespoke photometry pipeline (See Section \ref{photometrysection} and C15). HAPLESS ID (C15) given for overlapping sources.}
    		\setlength{\tabcolsep}{9pt}
    		\begin{tabular}{llllclllrrr}
    		\hline\hline
    		\# & Common name	& HIPASS ID & H-ATLAS IAU ID & HAPLESS & RA & DEC & z &velocity&Distance&Semi-maj \\
 & & & & & {(J200 deg)} & {(J200 deg)} & {(helio)} & (km s$^{-1}$) & (Mpc) & (")  \\   \hline
 1 & SDSSJ084258.35+003838.5 & HIPASSJ0842+00 & HATLASJ084258.4+003838  & &  130.74318 & 0.64408 & 0.03464 & 10696 & 158.9 & 36.4 \\
 2 & UGC04673 & HIPASSJ0855+02 & HATLASJ085552.3+023125 & &  133.967 & 2.52426 & 0.01277 & 4020 & 59.7 & 77.2 \\
 3 & UGC04684 & HIPASSJ0856+00 & HATLASJ085640.5+002229 & 39 &  134.17066 & 0.37591 & 0.00859 & 2730 & 40.6 & 62.4 \\
 4$^{a,b}$ & UGC04996 & HIPASSJ0923-00 & HATLASJ092315.6-004342 & &  140.81604 & -0.72945 & 0.01174 & 3852 & 57.3 & 72.6 \\
 5 & UGC06578 & HIPASSJ1136+00b & HATLASJ113636.7+004901 & 41 & 174.153 & 0.81678 & 0.00378 & 1374 & 20.4 & 56.8 \\
 6 & UGC06780 & HIPASSJ1148-02 & HATLASJ114850.4-020156 & 19 & 177.20993 & -2.03249 & 0.00578 & 2298 & 34.2 & 135.0 \\
 7$^a$ & UM456 & HIPASSJ1150-00 & HATLASJ115036.2-003406  & 17 & 177.65105 & -0.56613 & 0.00574 & 2270 & 33.7 & 40.3 \\
 8$^a$ & UM456A & HIPASSJ1150-00 & HATLASJ115033.8-003213 & 24 & 177.6415 & -0.53795 & 0.006 & 2391 & 35.5 & 32.1 \\
 9 & UGC06903 & HIPASSJ1155+01 & HATLASJ115536.9+011417 & 31 & 178.9025 & 1.23817 & 0.00635 & 2534 & 37.7 & 109.4 \\
 10 & UGC06970 & HIPASSJ1158-01 &   & & 179.69101 & -1.46169 & 0.005 & 2040 & 30.3 & 72.6 \\
 11 & NGC4030b & HIPASSJ1200-00 &  & & 180.19873 & -0.02333 & 0.0065 & 2581 & 38.4 & 73.1 \\
 12 & NGC4030 & HIPASSJ1200-01 & HATLASJ120023.7-010553 & 6 & 180.09841 & -1.10033 & 0.00477 & 1978 & 29.4 & 180.7 \\
 13 & UGC07053 & HIPASSJ1204-01 &   & & 181.0863 & -1.53071 & 0.00488 & 2028 & 30.1 & 85.3 \\
 14 & UGC07332 & HIPASSJ1217+00 &   & & 184.48653 & 0.43491 & 0.00318 & 936 & 13.9 & 109.8 \\
 15$^a$ & NGC4202 & HIPASSJ1218-01 & HATLASJ121808.4-010350 & & 184.53574 & -1.06413 & 0.019 & 6272 & 93.2 & 62.4 \\
 16$^a$ & FGC1412 & HIPASSJ1220+00 &    & & 184.85783 & 0.21197 & 0.00302 & 761 & 11.3 & 48.5 \\
 17$^a$ & CGCG014-010 & HIPASSJ1220+00 &   & & 185.08868 & 0.36769 & 0.00306 & 796 & 11.8 & 60.9 \\
 18 & UGC07394 & HIPASSJ1220+01 & HATLASJ122027.6+012812  & 11 & 185.11652 & 1.46789 & 0.00526 & 2197 & 32.6 & 81.2 \\
 19$^a$ & UGC07531 & HIPASSJ1226-01 & HATLASJ122611.1-011813 & 34 & 186.55054 & -1.30325 & 0.00675 & 2654 & 39.4 & 32.5 \\
 20$^{a}$ & UM501 & HIPASSJ1226-01 &   & & 186.59463 & -1.2534 & 0.00676 & 2658 & 39.5 & 32.5 \\
 21 & NGC5496 & HIPASSJ1411-01 & HATLASJ141137.7-010928 & 7 & 212.9082 & -1.15909 & 0.00488 & 1840 & 27.4 & 174.8 \\
 22 & NGC5584 & HIPASSJ1422-00 & HATLASJ142223.4-002313  & 14 & 215.59857 & -0.3869 & 0.00548 & 2033 & 30.2 & 154.3 \\
 23 & UGC09215 & HIPASSJ1423+01 & HATLASJ142327.2+014335  & 3 & 215.86342 & 1.7243 & 0.00457 & 1726 & 25.6 & 124.5 \\
 24$^{a}$ & SDSSJ142653.06+005746.2 & HIPASSJ1427+00 & HATLASJ142653.0+005745 &  & 216.72078 & 0.96285 & 0.02618 & 8099 & 120.4 & 62.4 \\
 25$^{a}$ & IC1011 & HIPASSJ1427+00 & HATLASJ142804.4+010023 & & 217.01885 & 1.00607 & 0.02564 & 7938 & 118.0 & 62.4 \\
 26$^a$ & IC1010 & HIPASSJ1427+00 & HATLASJ142720.5+010132  & & 216.83483 & 1.02589 & 0.02566 & 7954 & 118.2 & 93.5 \\
 27 & UGC09299 & HIPASSJ1429-00 & HATLASJ142934.8-000105  & 9 & 217.39393 & -0.01906 & 0.00516 & 1904 & 28.3 & 113.8 \\
 28 & SDSSJ143353.30+012905.6 & HIPASSJ1433+01 &    & & 218.47167 & 1.48543 & 0.00609 & 2220 & 33.0 & 60.9 \\
 29 & NGC5690 & HIPASSJ1437+02 & HATLASJ143740.9+021729  & 23 & 219.42 & 2.29162 & 0.00583 & 2160 & 32.1 & 154.3 \\
 30 & NGC5691 & HIPASSJ1437-00 & HATLASJ143753.3-002354  & 28 & 219.47216 & -0.39846 & 0.00626 & 2244 & 33.4 & 78.5 \\
 31 & UGC09432 & HIPASSJ1439+02 &   &  & 219.766 & 2.94708 & 0.00513 & 1920 & 28.5 & 69.0 \\
 32$^a$ & NGC5705 & HIPASSJ1439-00 & HATLASJ143949.5-004305 &  26 &  219.95623 & -0.71874 & 0.00589 & 2110 & 31.4 & 117.8 \\
 33 & NGC5725 & HIPASSJ1440+02 &   &  &  220.24298 & 2.18655 & 0.00543 & 1980 & 29.4 & 52.7 \\
 34$^a$ & NGC5713 & HIPASSJ1440-00 & HATLASJ144011.1-001725 &  29 & 220.04759 & -0.28933 & 0.00633 & 2260 & 33.6 & 124.5 \\
 35$^a$ & NGC5719 & HIPASSJ1440-00 & HATLASJ144056.2-001906 &  20 &  220.23393 & -0.31856 & 0.00575 & 2067 & 30.7 & 171.0 \\
 36$^a$ & UGC09482 & HIPASSJ1442+00 & HATLASJ144247.1+003942 &  &  220.69539 & 0.66151 & 0.00606 & 2179 & 32.4 & 65.0 \\
 37$^a$ & UGC09470 & HIPASSJ1442+00 & HATLASJ144148.7+004121 &  30 &  220.45274 & 0.68756 & 0.00637 & 2290 & 34.0 & 60.9 \\
 38$^a$ & NGC5740 & HIPASSJ1444+01 & HATLASJ144424.3+014046 &  10 &  221.10171 & 1.68019 & 0.0052 & 1890 & 28.1 & 139.4 \\
 39$^c$ & UGC0700 &  & HATLASJ120110.4-011750 &  8 &  180.295 & -1.29751 & 0.00501 & 2070 & 30.8 & 69.1 \\
 40$^c$ & NGC5746 &  & HATLASJ144455.9+015719 &  21 &  221.23292 & 1.955 & 0.00575 & 2070 & 30.8 & 276.6 \\
 \hline
    \end{tabular}

    $^a$ The HIPASS signal for this source is confused. Higher resolution H{\sc i} data from ALFALFA or the literature were supplemented to resolve confusion. Note that only unconfused counterparts are listed in this table.

    $^b$ Based on its colour, stellar mass and position, we identified this galaxy as the likely source of most of the HI flux in HIPASSJ0923-00.

     $^c$ UGC0700 and NGC5746 are both sources that are bright enough to make it into HIPASS and our sample, yet they were missed in HIPASS due to their close proximity to other, brighter HIPASS sources.
    \label{table1}
\end{table}
\end{landscape}

\begin{table*}
    		\caption{H{\sc i} Properties for all the sources in the H{\sc i}-selected H{\sc i}GH sample. The gas fraction does not include molecular gas (i.e. $f_g=M_{HI}/(M_{HI}+M_*$). The $ M_{HI}/M_* $ and $ M_{d}/M_{HI} $ ratios use {\sc magphys} derived stellar and dust masses (See Table \ref{table2}).}
    		\setlength{\tabcolsep}{6pt}
    		\renewcommand{\arraystretch}{1.25}
    		\begin{tabular}{llrrrrrr}
		\hline\hline
\# & common name & $S_{int}$ & log $ M_{HI} $ & gas fraction & log $ M_{HI}/M_* $ & log $ M_{d}/M_{HI} $ & H{\sc i} origin \\
 & & (Jy km/s) & ($M_{\odot}$) &  & & & \\
    		\hline
 1$^c$  &  SDSSJ084258.35+003838.5  &  1.56  & 9.97 & 0.58 & 0.14 & -2.81 &  ALFALFA \\
2  &  UGC04673  &  7.18  & 9.78 & 0.82 & 0.67 & -2.41 &  ALFALFA \\
3  &  UGC04684  &  9.72  & 9.58 & 0.63 & 0.23 & -2.91 &  ALFALFA \\
4$^{a,b}$  &  UGC04996  &  9.00  & 9.84 & 0.75 & 0.48 & -2.71 &  HIPASS \\
5  &  UGC06578  &  6.72  & 8.82 & 0.86 & 0.8 & -3.11 &  ALFALFA \\
6  &  UGC06780  &  26.90  & 9.87 & 0.88 & 0.87 & -2.94 &  HIPASS \\
7$^a$  &  UM456  &  2.86  & 8.89 & 0.8 & 0.6 & -3.93 &  Taylor et al. (1995) \\
8$^a$  &  UM456A  &  2.86  & 8.93 & 0.92 & 1.05 & -4.06 &  Taylor et al. (1995) \\
9  &  UGC06903  &  14.11  & 9.68 & 0.38 & -0.21 & -2.53 &  ALFALFA \\
10  &  UGC06970  &  5.20  & 9.05 & 0.31 & -0.34 & -2.56 &  HIPASS \\
11  &  NGC4030b  &  6.61  & 9.36 & 0.78 & 0.55 & -3.76 &  ALFALFA \\
12  &  NGC4030  &  72.00  & 10.17 & 0.16 & -0.71 & -2.27 &  HIPASS \\
13  &  UGC07053  &  8.30  & 9.25 & 0.92 & 1.06 & -4.46 &  HIPASS \\
14  &  UGC07332  &  19.61  & 8.95 & 0.95 & 1.25 & -4.65 &  ALFALFA \\
15$^a$  &  NGC4202  &  12.56  & 10.41 & 0.56 & 0.11 & -2.97 &  Richer et al. (1987) \\
16$^a$  &  FGC1412  &  2.35  & 7.85 & 0.89 & 0.92 & -4.02 &  ALFALFA \\
17$^a$  &  CGCG014-010  &  4.87  & 8.21 & 0.89 & 0.93 & -4.74 &  ALFALFA \\
18  &  UGC07394  &  6.86  & 9.24 & 0.67 & 0.31 & -2.4 &  ALFALFA \\
19$^a$  &  UGC07531  &  3.05  & 9.05 & 0.74 & 0.45 & -2.58 &  Taylor et al. (1995) \\
20$^a$  &  UM501  &  6.60  & 9.39 & 0.97 & 1.49 & -4.31 &  Taylor et al. (1995) \\
21  &  NGC5496  &  60.90  & 10.03 & 0.79 & 0.58 & -2.95 &  HIPASS \\
22  &  NGC5584  &  27.10  & 9.76 & 0.38 & 0.76 & -2.3 &  HIPASS \\
23  &  UGC09215  &  23.18  & 9.56 & 0.64 & 0.25 & -2.64 &  ALFALFA \\
24$^{a,c}$ &  SDSSJ142653.06+005746.2  &  1.23  & 9.62 & 0.52 & 0.03 & -2.4 &  ALFALFA \\
25$^{a,c}$  &  IC1011  &  1.62  & 9.73 & 0.27 & -0.43 & -2.34 &  ALFALFA \\
26$^a$  &  IC1010  &  10.80  & 10.55 & 0.36 & -0.26 & -2.65 &  ALFALFA \\
27  &  UGC09299  &  45.54  & 9.94 & 0.95 & 1.32 & -3.57 &  ALFALFA \\
28  &  SDSSJ143353.30+012905.6  &  3.42  & 8.94 & 0.94 & 1.18 & -4.43 &  ALFALFA \\
29  &  NGC5690  &  32.97  & 9.9 & 0.25 & -0.48 & -2.33 &  ALFALFA \\
30  &  NGC5691  &  5.50  & 9.16 & 0.12 & -0.85 & -2.33 &  HIPASS \\
31  &  UGC09432  &  8.03  & 9.19 & 0.91 & 1.0 & -4.8 &  ALFALFA \\
32$^a$  &  NGC5705  &  25.30  & 9.77 & 0.73 & 0.44 & -2.46 &  Fisher et al. (1981)\\
33  &  NGC5725  &  4.20  & 8.93 & 0.39 & -0.2 & -2.53 &  ALFALFA \\
34$^a$  &  NGC5713  &  42.79  & 10.06 & 0.24 & -0.5 & -2.55 &  Schneider et al. (1986)\\
35$^a$  &  NGC5719  &  52.45  & 10.07 & 0.16 & -0.72 & -2.67 &  Schneider et al. (1986)\\
36$^a$  &  UGC09482  &  5.86  & 9.16 & 0.74 & 0.45 & -3.1 &  ALFALFA \\
37$^a$  &  UGC09470  &  4.84  & 9.12 & 0.62 & 0.22 & -2.94 &  ALFALFA \\
38$^a$  &  NGC5740  &  29.23  & 9.74 & 0.22 & -0.54 & -2.61 &  ALFALFA \\
39$^d$  &  UGC0700  &  5.7  & 9.1 & 0.5 & -0.01 & -2.7 &   Sulentic et al. (1983) \\
40$^d$  &  NGC5746  &  30.7  & 9.84 & 0.03 & -1.47 & -1.87 &  Popping et al. (2011) \\
\hline
Mean &  & 16.25 & 9.46 & 0.61 &	0.28 	&	-3.05& \\
$M_*<10^9$ &  & 9.72 & 9.07 & 0.87 &	0.87 	&	-3.74& \\
$M_*>10^9$ &  & 21.33 & 9.76 & 0.44 &	-0.09 	&	-2.56& \\  \hline
    \end{tabular}
    \begin{flushleft}
    $^a$ The HIPASS signal for this source is confused. Higher resolution H{\sc i} data from ALFALFA or the literature were supplemented to resolve confusion. Note that only unconfused counterparts are listed in this table.

    $^b$ Based on its colour, stellar mass and position, we identified this galaxy as the likely source of most of the HI flux in HIPASSJ0923-00.

    $^c$ The individual H{\sc i}-flux is lower than HIPASS detection limit.

    $^d$ UGC0700 and NGC5746 are both sources that are bright enough to make it into HIPASS and our sample, yet they were missed in HIPASS due to their close proximity to other, brighter HIPASS sources.
    \end{flushleft}

    \label{tableHI}
\end{table*}

	\begin{landscape}
	\begin{table}
    		\caption{Photometry for the H{\sc i}-selected sample. UV-IR photometry has been corrected for Galactic extinction in line with \citet{Adelman-McCarthy2008}. The \textit{Herschel}-SPIRE fluxes were measured using maps reduced for extended sources, but have not been colour-corrected. The given uncertainties do not include calibration uncertainties. Before fitting an SED, we have added in quadrature the calibration error or $10$~per~cent, whichever is larger. The semi-major axis of the aperture is denoted by \textit{a}, the position angle by $\theta$ (east of north) and the axial ratio by \textit{a/b}. }
    		\setlength{\tabcolsep}{6pt}
    		\renewcommand{\arraystretch}{1.05}
    		\begin{tabular}{llrrrrrrrrrrrrrrr}
    		\hline\hline
    		\# & Common name	 & \multicolumn{3}{c}{Aperture dimensions} & \multicolumn{4}{c}{{\em GALEX} (mJy)} & \multicolumn{8}{c}{SDSS (mJy)} \\
    		 & & \textit{a}(arcsec) & $\theta$ (deg) & \textit{a/b} & \hspace{7mm} FUV & $\Delta$FUV & NUV & $\Delta$NUV & \hspace{10mm} \textit{u} & \textit{$\Delta$u} & \textit{g} & \textit{$\Delta$g} & \textit{r} & \textit{$\Delta$r} & \textit{u} & \textit{$\Delta$u} \\ \hline
1 & SDSSJ084258.35+003838.5 & 36.56 & 2.91 & 1.33 & 0.211 & 0.004 & 0.265 & 0.023 & 0.59 & 0.36 & 1.37 & 0.13 & 1.97 & 0.15 & 2.63 & 0.35 \\
2 & UGC04673 & 77.27 & -10.73 & 1.2 & 1.061 & 0.011 & 1.390 & 0.051 & 0.46 & 1.06 & 4.86 & 0.65 & 5.93 & 1.00 & 6.73 & 1.81 \\
3 & UGC04684 & 60.99 & 42.35 & 1.05 & 1.938 & 0.007 & 2.533 & 0.051 & 3.33 & 0.94 & 9.56 & 0.51 & 12.99 & 0.72 & 16.13 & 0.99 \\
4 & UGC04996 & 60.99 & 30.86 & 1.53 & 1.304 & 0.006 & 1.877 & 0.017 & 3.04 & 0.37 & 7.12 & 0.27 & 9.25 & 0.41 & 10.19 & 0.83 \\
5 & UGC06578 & 56.92 & 43.02 & 1.41 & 1.812 & 0.005 & 2.067 & 0.006 & 2.74 & 0.76 & 5.07 & 0.23 & 5.80 & 0.25 & 6.39 & 0.59 \\
6 & UGC06780 & 125.97 & 107.49 & 2.8 & - & - & 2.790 & 0.022 & 5.00 & 1.28 & 11.97 & 0.55 & 14.35 & 0.83 & 17.06 & 1.61 \\
7 & UM456 & 40.58 & 91.72 & 1.16 & 0.978 & 0.004 & 1.189 & 0.011 & 1.76 & 0.16 & 3.34 & 0.24 & 3.88 & 0.46 & 3.92 & 0.97 \\
8 & UM456A & 32.42 & -13.84 & 1.87 & 0.284 & 0.002 & 0.321 & 0.004 & 0.48 & 0.07 & 0.94 & 0.10 & 1.16 & 0.18 & 1.17 & 0.33 \\
9 & UGC06903 & 105.75 & 46.69 & 1.16 & 2.875 & 0.011 & 3.016 & 0.064 & 4.38 & 3.31 & 19.49 & 1.02 & 28.01 & 1.91 & 34.10 & 3.91 \\
10 & UGC06970 & 65.06 & -10.33 & 1.42 & 1.081 & 0.008 & 1.491 & 0.017 & 3.20 & 2.03 & 7.82 & 0.56 & 13.15 & 0.38 & 22.51 & 0.82 \\
11 & NGC4030b & 73.2 & 15.52 & 1.12 & 0.644 & 0.005 & 0.781 & 0.015 & 0.98 & 1.42 & 3.60 & 0.17 & 5.30 & 0.49 & 6.55 & 1.44 \\
12 & NGC4030 & 170.84 & 134.08 & 1.41 & 11.979 & 0.031 & 22.401 & 0.027 & 56.87 & 3.58 & 191.95 & 1.99 & 327.87 & 2.71 & 447.49 & 4.62 \\
13 & UGC07053 & 85.4 & 102.19 & 1.56 & 0.610 & 0.007 & 0.801 & 0.014 & 1.72 & 1.65 & 3.08 & 0.38 & 3.64 & 1.12 & 4.07 & 1.32 \\
14 & UGC07332 & 109.82 & 16.85 & 1.09 & 1.682 & 0.013 & 2.039 & 0.020 & 3.15 & 1.31 & 6.72 & 1.70 & 3.63 & 2.03 & 8.93 & 2.05 \\
15 & NGC4202 & 56.92 & 49.65 & 1.88 & 0.282 & 0.006 & 0.749 & 0.012 & 2.64 & 0.10 & 7.46 & 0.20 & 12.90 & 0.26 & 17.02 & 0.45 \\
16 & FGC1412 & 52.81 & 98.99 & 3.57 & 0.188 & 0.002 & 0.243 & 0.004 & 0.41 & 0.09 & 0.86 & 0.05 & 1.01 & 0.07 & 1.14 & 0.22 \\
17 & CGCG014-010 & 60.99 & -42.33 & 2.73 & 0.414 & 0.004 & 0.551 & 0.013 & 0.83 & 0.16 & 1.80 & 0.06 & 1.93 & 0.09 & 2.29 & 0.23 \\
18 & UGC07394 & 81.31 & 54.57 & 3.45 & 0.411 & 0.005 & 0.705 & 0.008 & 1.97 & 0.33 & 5.04 & 0.45 & 7.18 & 0.53 & 8.17 & 0.85 \\
19 & UGC07531 & 45.0 & 0.0 & 1.0 & 1.994 & 0.007 & 2.143 & 0.018 & 3.25 & 0.49 & 6.10 & 0.30 & 6.87 & 0.50 & 7.17 & 0.86 \\
20 & UM501 & 31.59 & 120.0 & 1.57 & 0.506 & 0.002 & 0.410 & 0.007 & 0.79 & 0.15 & 1.36 & 0.08 & 1.39 & 0.17 & 1.33 & 0.24 \\
21 & NGC5496 & 210.0 & 83.07 & 4.2 & 5.311 & 0.038 & 8.017 & 0.029 & 12.69 & 6.03 & 34.00 & 0.64 & 46.02 & 1.52 & 54.49 & 1.87 \\
22 & NGC5584 & 138.3 & 63.47 & 1.26 & 9.901 & 0.019 & 13.541 & 0.061 & 26.66 & 4.84 & 64.37 & 2.38 & 93.91 & 2.45 & 105.27 & 4.38 \\
23 & UGC09215 & 109.82 & 71.65 & 1.47 & 5.778 & 0.014 & 7.441 & 0.066 & 13.62 & 4.88 & 29.33 & 1.73 & 38.32 & 2.31 & 45.71 & 3.69 \\
24 & 2MASXJ14265308+0057462 & 24.25 & -3.35 & 1.38 & 0.175 & 0.003 & 0.260 & 0.010 & 0.60 & 0.09 & 1.65 & 0.05 & 2.27 & 0.08 & 2.47 & 0.12 \\
25 & IC1011 & 36.5 & -17.74 & 1.14 & 0.687 & 0.004 & 1.105 & 0.006 & 2.32 & 0.20 & 5.35 & 0.15 & 7.95 & 0.25 & 9.85 & 0.33 \\
26 & IC1010 & 97.61 & 91.4 & 1.19 & 1.165 & 0.011 & 1.550 & 0.075 & 3.29 & 1.80 & 12.54 & 1.47 & 22.46 & 1.60 & 28.49 & 2.13 \\
27 & UGC09299 & 113.89 & -18.47 & 1.53 & 2.492 & 0.029 & 3.114 & 0.051 & 3.89 & 1.06 & 10.65 & 0.64 & 10.68 & 1.19 & 12.89 & 2.36 \\
28 & SDSSJ143353.30+012905.6 & 60.99 & 98.24 & 2.96 & 0.148 & 0.004 & 0.193 & 0.013 & 0.21 & 0.80 & 0.67 & 0.30 & 0.82 & 0.31 & 0.96 & 0.76 \\
29 & NGC5690 & 122.02 & 60.0 & 2.32 & 1.994 & 0.014 & 3.811 & 0.143 & 13.17 & 2.79 & 42.80 & 4.03 & 71.51 & 5.20 & 98.39 & 3.55 \\
30 & NGC5691 & 60.99 & 63.13 & 1.16 & 2.079 & 0.007 & 3.285 & 0.033 & 14.31 & 1.14 & 36.52 & 0.49 & 53.98 & 0.56 & 65.48 & 1.04 \\
31 & UGC09432 & 69.13 & -44.31 & 1.2 & 0.890 & 0.006 & 1.042 & 0.019 & 1.86 & 1.85 & 3.82 & 0.75 & 4.28 & 0.95 & 4.23 & 1.12 \\
32 & NGC5705 & 117.94 & -20.27 & 1.38 & 4.186 & 0.016 & 5.362 & 0.036 & 5.61 & 3.34 & 23.41 & 1.24 & 35.32 & 1.84 & 42.05 & 2.11 \\
33 & NGC5725 & 52.85 & 117.69 & 1.14 & 1.573 & 0.008 & 2.222 & 0.014 & 4.25 & 0.63 & 9.59 & 0.29 & 13.38 & 0.53 & 15.67 & 0.88 \\
34 & NGC5713 & 89.49 & 92.25 & 1.17 & 4.667 & 0.015 & 9.649 & 0.039 & 33.50 & 2.05 & 95.45 & 0.95 & 157.65 & 1.42 & 201.46 & 1.92 \\
35 & NGC5719 & 185.0 & 29.36 & 4.11 & 0.508 & 0.020 & 0.992 & 0.041 & 10.16 & 1.75 & 41.36 & 0.82 & 95.67 & 2.2 & 146.43 & 3.80 \\
36 & UGC09482 & 65.06 & -22.6 & 3.2 & 0.409 & 0.004 & 0.635 & 0.005 & 1.47 & 0.39 & 3.36 & 0.09 & 4.72 & 0.15 & 5.60 & 0.29 \\
37 & UGC09470 & 60.99 & -42.57 & 1.47 & 1.243 & 0.009 & 1.578 & 0.011 & 3.00 & 0.55 & 6.12 & 0.36 & 8.31 & 0.41 & 9.37 & 0.92 \\
38 & NGC5740 & 126.08 & 63.69 & 2.16 & 3.154 & 0.014 & 4.986 & 0.019 & 11.58 & 4.01 & 47.29 & 0.31 & 84.12 & 1.16 & 112.22 & 0.97 \\
39 & UGC07000 & 69.1 & -29.7 & 1.39 & 2.617 & 0.007 & 3.286 & 0.011 & 5.94 & 1.18 & 11.99 & 0.62 & 15.68 & 0.87 & 19.12 & 1.05 \\
40 & NGC5746 & 276.61 & 78.56 & 4.49 & - & - & - & - & 32.63 & 5.29 & 137.90 & 2.34 & 323.15 & 2.77 & 496.51 & 7.12 \\
  \hline
  \label{photo}
   	\end{tabular}
\end{table}
\end{landscape}
\addtocounter{table}{-1}

\begin{landscape}
\begin{table}
    \caption{ - \textit{continued}}
    		\setlength{\tabcolsep}{5pt}
    		\renewcommand{\arraystretch}{1.1}
    		\begin{tabular}{lrrrrrrrrrrrrrrrrrr}
    		\hline\hline
    		\# & \multicolumn{10}{c}{VIKING (mJy)} & \multicolumn{8}{c}{WISE (mJy)}\\
    		\hspace{8mm} & \textit{Z} & \textit{$\Delta$Z} & \textit{Y} & \textit{$\Delta$Y} & \textit{J} & \textit{$\Delta$J} & \textit{H} & \textit{$\Delta$H} & \textit{K$_S$} & \textit{$\Delta$K$_S$} & \hspace{7mm} 3.4$\mu m$&$\Delta$3.4$\mu m$&4.6$\mu m$&$\Delta$4.6$\mu m$&12$\mu m$&$\Delta$12$\mu m$&22$\mu m$&$\Delta$22$\mu m$ \\ \hline
1 & 2.76 & 0.69 & 3.20 & 0.96 & 3.33 & 0.48 & 3.54 & 1.77 & 2.72 & 1.04 & 1.64 & 0.26 & 0.97 & 0.24 & 2.29 & 0.66 & 4.45 & 1.90 \\
2 & 6.87 & 1.66 & 6.51 & 2.58 & 6.29 & 2.70 & 2.58 & 8.06 & 4.80 & 1.83 & 4.75 & 0.98 & 3.03 & 0.84 & 4.47 & 1.88 & 13.97 & 3.92 \\
3 & 22.35 & 3.53 & 20.38 & 2.93 & 21.82 & 3.44 & 18.09 & 1.21 & 16.17 & 1.81 & - & - & - & - & - & - & - & - \\
4 & 11.39 & 0.54 & 12.70 & 0.56 & 13.30 & 0.59 & 11.72 & 2.86 & 11.52 & 0.66 & 4.95 & 0.53 & 3.40 & 0.38 & 9.32 & 0.79 & 15.93 & 3.68 \\
5 & 7.20 & 0.40 & 6.21 & 1.29 & 6.73 & 1.43 & 2.62 & 1.50 & 2.95 & 1.82 & 3.82 & 0.44 & 2.38 & 0.34 & 8.56 & 1.07 & 42.56 & 3.42 \\
6 & 15.95 & 1.02 & 13.51 & 1.64 & 19.49 & 1.81 & 18.41 & 4.37 & 11.54 & 3.36 & 9.73 & 0.63 & 5.16 & 0.83 & 5.57 & 1.02 & 21.24 & 4.19 \\
7 & 4.18 & 0.67 & 4.39 & 1.20 & 4.37 & 1.79 & 4.31 & 2.2 & 3.13 & 1.07 & 1.46 & 0.71 & 1.03 & 0.48 & 1.70 & 0.82 & 15.35 & 2.24 \\
8 & 1.23 & 0.16 & 1.23 & 0.33 & 1.37 & 0.48 & 1.32 & 0.68 & 1.29 & 0.71 & 0.58 & 0.35 & 0.20 & 0.26 & 0.51 & 0.41 & 5.41 & 1.65 \\
9 & 44.43 & 7.06 & 47.44 & 4.50 & 49.94 & 7.47 & 51.89 & 4.71 & 40.11 & 4.46 & 23.03 & 3.44 & 14.66 & 2.72 & 31.08 & 1.87 & 26.26 & 6.79 \\
10 & 23.01 & 0.38 & 24.74 & 0.48 & 25.27 & 0.71 & 24.81 & 1.62 & 24.31 & 1.09 & - & - & - & - & - & - & - & - \\
11 & 6.13 & 0.75 & 6.58 & 0.58 & 6.21 & 0.87 & 6.58 & 2.00 & 5.35 & 2.31 & 3.09 & 0.87 & 2.08 & 0.73 & 1.85 & 1.62 & -1.32 & 5.94 \\
12 & 543.82 & 2.68 & 667.79 & 4.32 & 759.13 & 4.85 & 900.11 & 5.07 & 759.58 & 6.35 & 458.13 & 2.41 & 285.00 & 1.40 & 1283.52 & 3.56 & 1936.21 & 9.48 \\
13 & 2.67 & 1.01 & 2.41 & 1.40 & 4.13 & 1.70 & 3.18 & 4.00 & 2.75 & 1.29 & 2.26 & 0.52 & 1.99 & 0.79 & -1.02 & 1.38 & 12.11 & 4.6 \\
14 & 5.87 & 0.76 & 6.16 & 1.64 & 10.54 & 3.83 & 6.88 & 4.18 & 1.13 & 3.70 & 3.93 & 1.15 & 3.43 & 2.76 & 10.64 & 11.69 & -75.52 & 26.70 \\
15 & 20.20 & 0.40 & 24.48 & 0.64 & 27.06 & 0.57 & 30.77 & 0.75 & 26.62 & 1.03 & 14.32 & 0.25 & 8.69 & 0.37 & 32.78 & 2.11 & 43.17 & 4.51 \\
16 & 1.37 & 0.11 & 1.47 & 0.20 & 1.41 & 0.26 & - & - & 1.13 & 0.81 & 0.47 & 0.16 & 0.16 & 0.23 & 0.18 & 1.49 & 10.27 & 4.44 \\
17 & 2.71 & 0.12 & 2.77 & 0.17 & 2.80 & 0.20 & - & - & 2.16 & 1.00 & 1.14 & 0.18 & 0.31 & 0.42 & -0.17 & 3.80 & -2.09 & 4.57 \\
18 & 10.07 & 0.30 & 11.07 & 0.55 & 11.31 & 0.95 & 17.30 & 4.25 & 9.45 & 2.26 & 4.87 & 0.41 & 2.95 & 0.95 & 4.72 & 3.16 & 10.63 & 4.9 \\
19 & 7.11 & 0.97 & 7.28 & 0.70 & 7.26 & 1.08 & 7.22 & 0.99 & 5.76 & 0.79 & 2.93 & 0.48 & 1.81 & 0.43 & 3.84 & 0.73 & 13.99 & 2.97 \\
20 & 1.34 & 0.15 & 1.36 & 0.34 & 1.28 & 0.54 & 1.22 & 0.40 & 0.98 & 0.41 & 0.71 & 0.14 & 0.36 & 0.16 & 0.24 & 0.41 & 10.64 & 1.76 \\
21 & - & - & - & - & 55.54 & 2.42 & 80.08 & 4.97 & 45.65 & 1.81 & 36.09 & 0.70 & 22.83 & 0.67 & 41.49 & 1.64 & 64.31 & 5.39 \\
22 & 121.54 & 2.49 & 130.23 & 2.83 & 129.62 & 2.82 & 131.19 & 8.34 & 117.33 & 3.82 & 72.53 & 2.27 & 45.16 & 1.34 & 159.81 & 3.65 & 330.71 & 5.27 \\
23 & 49.26 & 4.24 & 53.24 & 2.96 & 55.53 & 4.9 & 54.45 & 3.73 & 38.67 & 4.22 & 28.14 & 2.58 & 17.87 & 1.34 & 51.30 & 1.28 & 117.04 & 3.86 \\
24 & 3.34 & 0.16 & 3.81 & 0.21 & 4.11 & 0.40 & 4.16 & 0.57 & 3.99 & 0.34 & 1.97 & 0.15 & 1.28 & 0.12 & 5.86 & 0.23 & 13.65 & 0.84 \\
25 & 11.88 & 0.65 & 13.82 & 0.87 & 15.34 & 0.60 & 17.52 & 0.59 & 14.30 & 0.54 & 9.14 & 0.22 & 5.87 & 0.19 & 34.00 & 0.40 & 56.25 & 1.86 \\
26 & 23.99 & 3.54 & 39.29 & 9.35 & 40.06 & 5.66 & 52.65 & 4.94 & 37.98 & 4.64 & 24.35 & 3.72 & 13.72 & 2.79 & 24.67 & 1.67 & 27.05 & 3.52 \\
27 & 11.60 & 4.54 & 12.65 & 4.38 & 11.81 & 4.55 & 4.20 & 6.19 & 5.87 & 5.71 & 5.66 & 1.56 & 3.16 & 1.99 & 7.87 & 1.23 & 22.07 & 5.33 \\
28 & 1.09 & 0.39 & 1.15 & 0.43 & 1.18 & 0.71 & 0.95 & 1.18 & 1.62 & 2.38 & 0.38 & 0.30 & -0.20 & 0.28 & 0.23 & 0.56 & -0.54 & 1.98 \\
29 & 125.93 & 2.74 & 157.51 & 2.91 & 179.73 & 2.73 & 221.01 & 4.33 & 190.26 & 7.66 & 118.96 & 1.04 & 77.84 & 0.82 & 399.63 & 2.74 & 609.96 & 4.34 \\
30 & 76.51 & 0.78 & 86.41 & 1.16 & 92.99 & 0.81 & 102.01 & 1.32 & 80.4 & 2.68 & 47.70 & 0.67 & 30.55 & 0.49 & 130.42 & 0.76 & 304.88 & 3.10 \\
31 & 4.53 & 1.71 & 3.34 & 1.67 & 4.95 & 1.38 & -10.54 & 2.21 & 3.74 & 2.75 & 2.00 & 0.82 & 0.59 & 0.59 & 0.54 & 0.79 & 0.94 & 3.10 \\
32 & 42.30 & 0.79 & 42.46 & 1.29 & 41.18 & 1.23 & 32.60 & 1.73 & 29.97 & 4.17 & 22.93 & 0.65 & 14.21 & 0.75 & 26.47 & 1.38 & 33.35 & 4.65 \\
33 & 17.51 & 0.43 & 19.16 & 0.62 & 20.46 & 0.85 & 23.72 & 1.19 & 17.08 & 1.06 & 10.47 & 0.33 & 6.33 & 0.44 & 17.79 & 0.63 & 24.98 & 2.42 \\
34 & 241.54 & 1.43 & 288.94 & 1.75 & 325.88 & 2.22 & 370.08 & 1.85 & 312.09 & 11.10 & 190.49 & 0.90 & 129.90 & 0.77 & 914.84 & 1.71 & 2362.8 & 4.81 \\
35 & 193.50 & 5.43 & 256.49 & 6.77 & 319.21 & 3.52 & 399.96 & 3.57 & 345.82 & 12.40 & 191.03 & 3.69 & 116.08 & 2.18 & 373.55 & 2.2 & 679.42 & 5.66 \\
36 & 6.18 & 0.21 & 6.60 & 0.33 & 6.80 & 0.24 & 7.15 & 0.47 & 5.58 & 0.55 & 2.60 & 0.21 & 1.62 & 0.21 & 1.61 & 0.44 & 7.30 & 1.62 \\
37 & 10.16 & 0.44 & 10.31 & 0.52 & 10.31 & 0.52 & 9.98 & 1.86 & 8.71 & 1.14 & 5.05 & 0.40 & 2.69 & 0.47 & 2.84 & 0.46 & 12.77 & 1.77 \\
38 & 143.49 & 0.46 & 173.28 & 0.49 & 197.03 & 0.86 & 228.03 & 0.90 & 178.12 & 1.62 & 101.88 & 0.53 & 58.83 & 0.59 & 181.98 & 0.96 & 327.74 & 3.64 \\
39 & 19.76 & 1.76 & 21.22 & 2.04 & 22.54 & 1.80 & 25.29 & 2.38 & 19.25 & 2.60 & 11.33 & 0.49 & 7.15 & 0.62 & 15.52 & 1.05 & 27.18 & 3.64 \\
40 & 673.93 & 18.87 & 910.96 & 20.16 & 1124.89 & 17.99 & 1401.90 & 26.53 & 1219.78 & 19.23 & 622.88 & 16.93 & 347.73 & 10.10 & 391.11 & 4.71 & 429.59 & 5.20 \\
 \hline
    	\end{tabular}
\end{table}
\end{landscape}
\addtocounter{table}{-1}

\begin{landscape}
\begin{table}
    		\caption{ - \textit{continued}}
    		\setlength{\tabcolsep}{8pt}
    		\renewcommand{\arraystretch}{1.03}
    		\begin{tabular}{lrrrrrrrrrrrr}
    		\hline\hline
    		\# & \multicolumn{2}{c}{IRAS SCANPI (mJy)} & \multicolumn{4}{c}{\textit{Herschel}-PACS (mJy)} & \multicolumn{6}{c}{\textit{Herschel}-SPIRE (mJy)}  \\
    		\hspace{15mm} & 60 $\mu m$&$\Delta$60 $\mu m$ & \hspace{15mm}100 $\mu m$ &$ \Delta$100 $\mu m$ & 160 $\mu m$ & $\Delta$160 $\mu m$ & \hspace{10mm}250 $\mu m$ & $\Delta$250 $\mu m $ & 350 $\mu m$ & $\Delta$350 $\mu m $ & 500 $\mu m$ & $\Delta$500 $\mu m $ \\ \hline
1 & 140.0$ ^a$ & -38.7$ ^a$ & 114.0 & 120.4 & 146.0 & 105.9 & 103.6 & 28.5 & 63.8 & 26.4 & 21.1 & 19.9 \\
2 & 60.0 & 60.0 & 207.1 & 207.3 & 302.2 & 158.8 & 317.5 & 58.8 & 215.6 & 46.5 & 132.0 & 32.8 \\
3 & 350.0 & 129.6 & 851.1 & 200.1 & 1006.2 & 157.4 & 563.3 & 45.4 & 280.8 & 40.2 & 126.5 & 26.6 \\
4 & 340.0 & 111.8 & 876.8 & 195.7 & 775.5 & 145.7 & 542.0 & 37.2 & 313.7 & 33.1 & 131.7 & 27.6 \\
5 & 380.0 & 132.6 & 262.8 & 121.4 & 231.7 & 98.0 & 188.8 & 39.1 & 97.9 & 35.8 & 52.9 & 24.2 \\
6 & 160.0 & 85.4 & 228.8 & 186.8 & 370.8 & 144.2 & 393.7 & 51.6 & 267.2 & 43.6 & 133.2 & 33.4 \\
7 & 110.0 & 67.8 & 228.0 & 148.1 & 58.9 & 119.7 & 49.1 & 32.3 & 18.3 & 29.9 & 8.9 & 23.8 \\
8 & 350.0$ ^a$ & -99.8$ ^a$ & 4.3 & 87.1 & 133.7 & 84.2 & 33.3 & 22.7 & 13.7 & 21.4 & -4.8 & 20.0 \\
9 & 310.0 & 111.0 & 1120.0$ ^b$ & 98.0$ ^b$ & 2739.7 & 288.8 & 1349.2 & 63.7 & 808.6 & 49.3 & 327.3 & 38.7 \\
10 & 280.0 & 119.2 & 127.9 & 198.4 & 371.4 & 163.0 & 406.1 & 42.0 & 220.5 & 35.2 & 96.1 & 26.5 \\
11 & 190.0$ ^a$ & -57.5$ ^a$ & 227.6 & 264.6 & 499.0 & 213.1 & 76.5 & 52.8 & 78.1 & 45.4 & 18.8 & 33.4 \\
12 & 16550.0 & 3376.4 & 60459.7 & 479.0 & 70131.2 & 362.0 & 33414.7 & 93.8 & 13566.3 & 70.7 & 4556.4 & 51.4 \\
13 & 141.3$ ^a$ & -47.1$ ^a$ & 253.5 & 292.9 & 306.7 & 249.4 & 28.3 & 48.4 & 5.2 & 41.5 & -10.0 & 31.7 \\
14 & 140.0 & 77.5 & 159.5 & 491.8 & 122.7 & 361.9 & 3.3 & 69.7 & -4.2 & 61.6 & 78.7 & 46.2 \\
15 & 290.0 & 124.0 & 1455.2 & 220.0 & 1910.3 & 195.2 & 996.9 & 32.0 & 410.4 & 28.7 & 146.1 & 21.6 \\
16 & 156.9$ ^a$ & -52.3$ ^a$ & 81.8 & 117.6 & -30.6 & 98.6 & 17.8 & 25.7 & 17.7 & 25.7 & 35.1 & 20.1 \\
17 & 160.1$ ^a$ & -53.4$ ^a$ & 87.3 & 148.3 & -38.3 & 127.9 & -34.6 & 33.5 & -28.3 & 28.2 & -28.5 & 22.9 \\
18 & - & - & 260.7 & 143.7 & 385.3 & 125.8 & 260.5 & 32.9 & 206.1 & 29.1 & 126.0 & 22.6 \\
19 & 260.0 & 68.0 & 141.7 & 180.5 & 139.6 & 156.1 & 138.7 & 32.8 & 88.6 & 30.0 & 67.0 & 21.5 \\
20 & - & - & 63.9 & 93.9 & 7.3 & 84.8 & 42.4 & 20.4 & 21.6 & 21.2 & 14.7 & 16.4 \\
21 & 1000.0 & 268.8 & 2992.0 & 322.3 & 3283.1 & 269.0 & 2466.9 & 72.9 & 1373.4 & 63.4 & 636.2 & 51.2 \\
22 & 2345.0 & 519.9 & 7795.4 & 505.4 & 8443.8 & 423.0 & 5770.6 & 77.2 & 2940.0 & 62.8 & 1180.2 & 53.1 \\
23 & 1420.0 & 327.9 & 2768.1 & 341.9 & 3759.4 & 271.6 & 2049.2 & 66.7 & 1159.7 & 55.1 & 503.4 & 43.9 \\
24 & 210.0 & 81.6 & 328.2 & 107.7 & 374.4 & 87.9 & 228.2 & 21.5 & 121.7 & 19.4 & 55.9 & 16.4 \\
25 & 780.0 & 208.3 & 1501.2 & 138.1 & 1632.1 & 122.0 & 696.5 & 27.1 & 281.9 & 25.5 & 99.4 & 19.5 \\
26 & 310.0$ ^a$ & -50.6$ ^a$ & 1066.2 & 207.8 & 927.0 & 180.6 & 913.9 & 66.8 & 433.1 & 58.1 & 187.3 & 39.1 \\
27 & 230.0 & 86.6 & 540.4 & 159.1 & 779.9 & 139.5 & 487.4 & 55.1 & 263.6 & 49.0 & 132.5 & 39.7 \\
28 & 184.8$ ^a$ & -61.6$ ^a$ & 3.1 & 146.1 & 43.1 & 124.1 & 9.1 & 27.4 & 7.0 & 25.7 & 6.9 & 21.6 \\
29 & 6460.0 & 1336.3 & 18807.2 & 429.2 & 22254.8 & 375.0 & 11158.0 & 52.2 & 4901.4 & 48.3 & 1759.8 & 37.7 \\
30 & 3480.0 & 796.3 & 7906.1 & 310.9 & 6943.5 & 221.5 & 2772.8 & 38.1 & 1166.0 & 33.3 & 412.6 & 22.9 \\
31 & 159.4$ ^a$ & -53.1$ ^a$ & - & - & - & - & 9.5 & 40.5 & -38.2 & 34.0 & -26.3 & 25.7 \\
32 & 440.0 & 129.6 & 1325.0 & 378.5 & 2188.6 & 310.7 & 1787.3 & 65.9 & 1041.4 & 64.7 & 542.4 & 43.4 \\
33 & 430.0 & 126.0 & - & - & - & - & 617.5 & 36.4 & 299.7 & 30.1 & 108.8 & 24.7 \\
34 & 21290.0 & 4326.3 & 43106.1 & 376.5 & 38327.6 & 356.8 & 15351.7 & 55.7 & 5812.5 & 47.3 & 1881.3 & 32.8 \\
35 & 8535.0 & 1747.2 & 19072.1 & 356.9 & 19922.1 & 273.0 & 9776.1 & 63.8 & 4275.7 & 52.1 & 1500.2 & 37.4 \\
36 & 430.7$ ^a$ & -143.6$ ^a$ & 84.8 & 99.9 & 185.6 & 85.1 & 144.0 & 29.5 & 79.0 & 24.9 & 29.4 & 20.7 \\
37 & 170.0 & 68.8 & 489.1 & 124.3 & 545.1 & 109.4 & 243.1 & 35.2 & 156.7 & 29.0 & 82.3 & 23.0 \\
38 & 3170.0 & 711.0 & 7917.2 & 366.4 & 8956.7 & 305.6 & 4825.9 & 52.2 & 2190.6 & 47.7 & 811.2 & 34.6 \\
39 & 370.0 & 137.0 & 973.8 & 174.6 & 1276.2 & 138.3 & 624.7 & 43.9 & 317.9 & 35.8 & 112.4 & 28.0 \\
40 & 2600.0 & 570.0 & 12903.9 & 604.7 & 24687.9 & 452.6 & 17211.5 & 71.8 & 8278.6 & 57.7 & 3203.6 & 48.0 \\
 \hline
  \end{tabular}
 \\ $ ^a$ Sources with no reliable detection at the location of the source are shown with a negative $\Delta$60 $\mu m$ ($=-rms$). For these sources, the scans were inspected manually and an upper limit was defined as 3 times the local rms and listed as the flux. \\
  $ ^b$ The PACS 100$\mu m$ flux poorly matched the SED for UGC06903. We have instead used the IRAS 100$\mu m$ photometry from \citet{Lisenfeld2007} which has a significantly higher SNR and better matches the SED.
\end{table}
\end{landscape}

\begin{figure*}
  \includegraphics[width=0.49\textwidth]{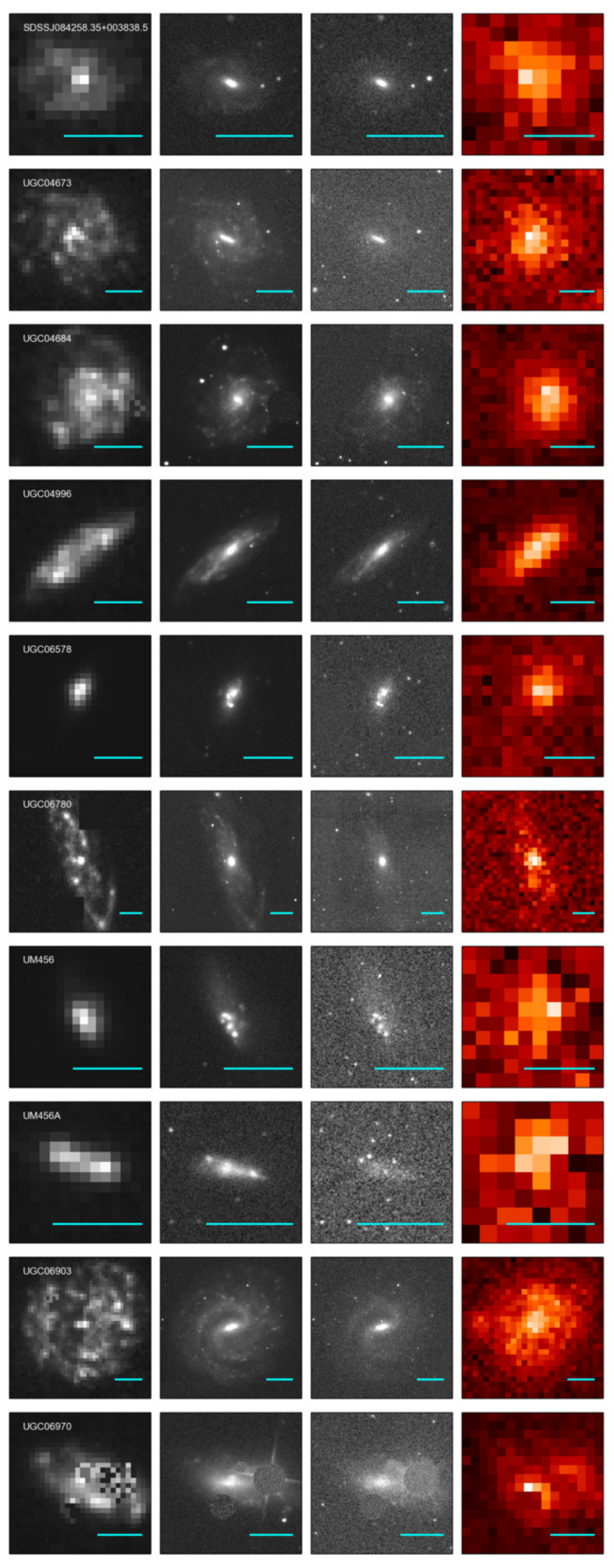}
  \includegraphics[width=0.49\textwidth]{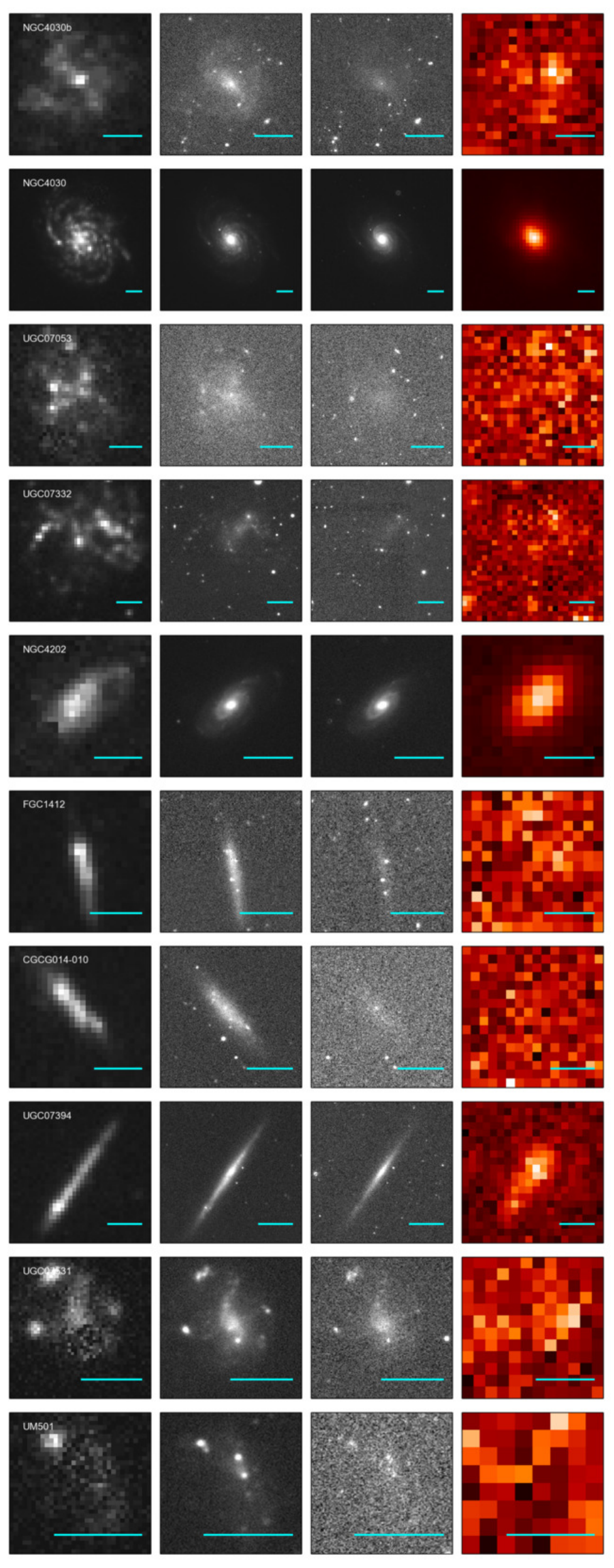}
  \caption{Multiwavelength images for all the H{\sc i}-selected sources. The brightest foreground stars and background galaxies have been subtracted and replaced by adjacent pixels. The bands displayed, from left-to-right, are: {\em GALEX} FUV, SDSS \textit{r}-band, VIKING \textit{K$_S$}-band, and \textit{Herschel} 250 $\mu$m. The size of each cutout is 1.5 times the semi-major axis of the aperture and a scale bar with a length of 30" is shown on each image in cyan.}
  \label{images}
\end{figure*}
\addtocounter{figure}{-1}

\begin{figure*}
  \includegraphics[width=0.49\textwidth]{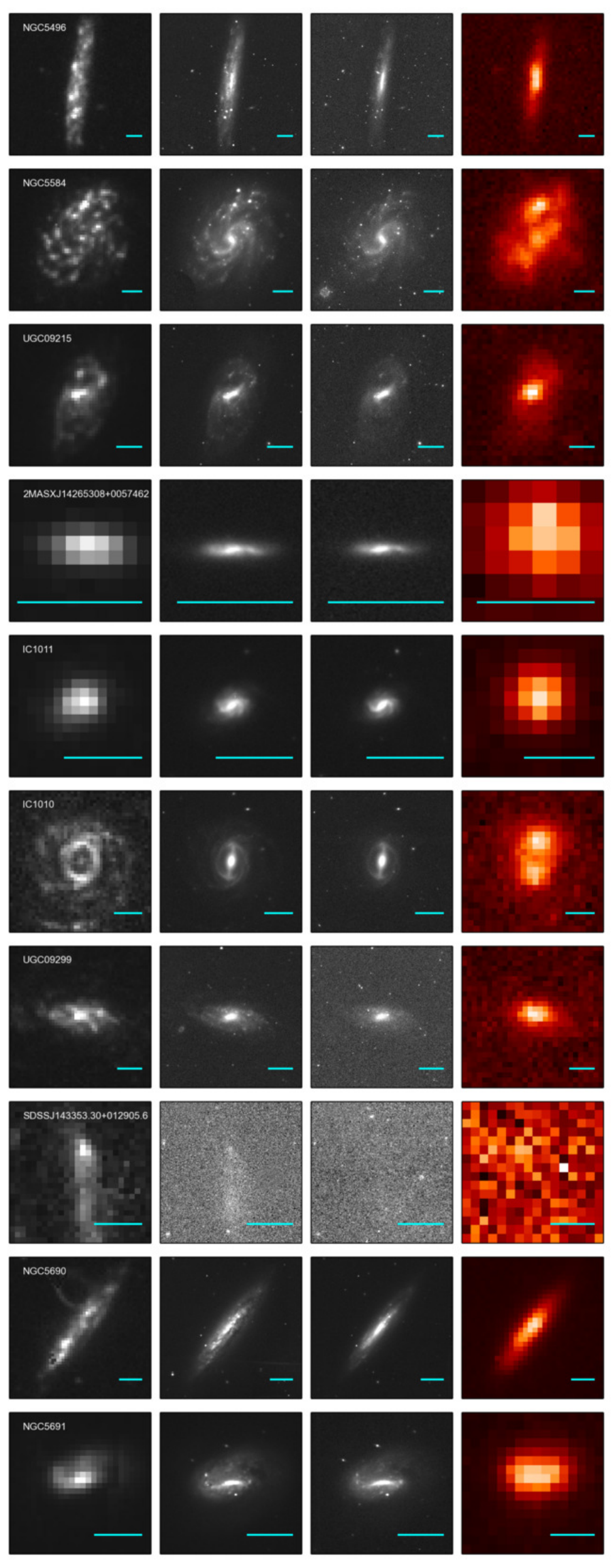}
  \includegraphics[width=0.49\textwidth]{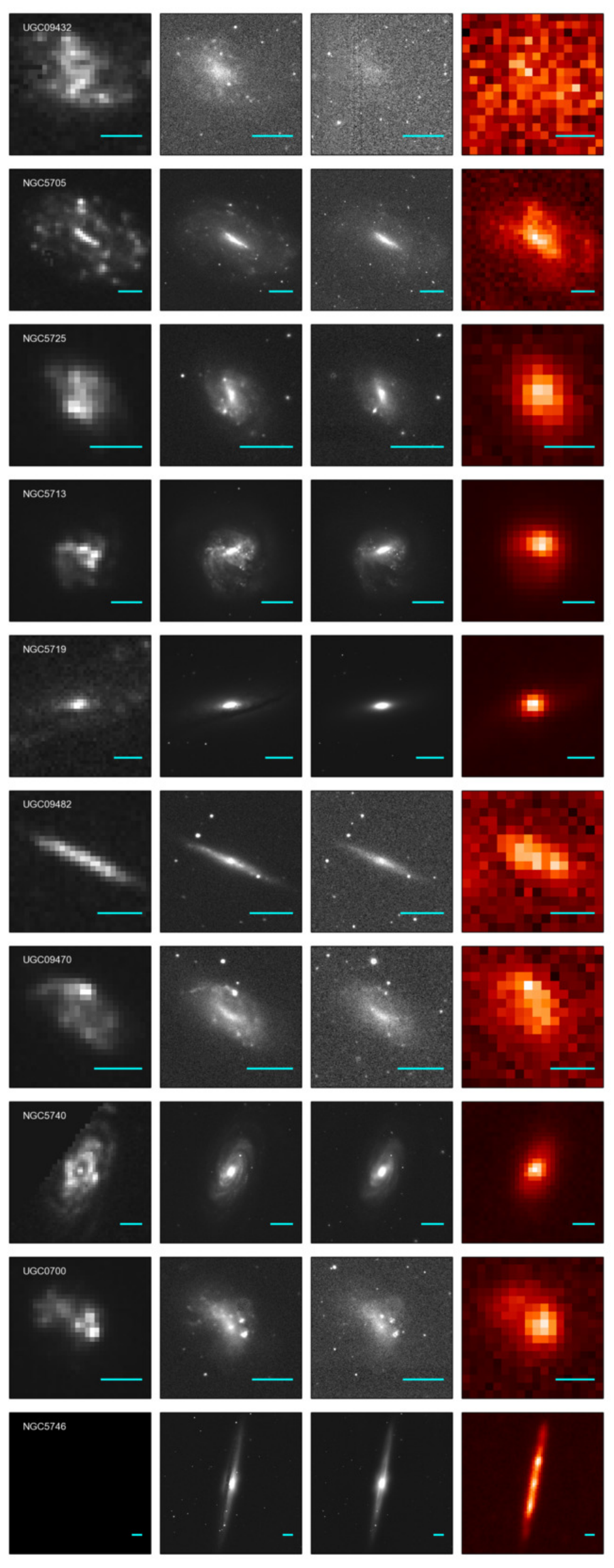}
    \caption{ - \textit{continued}}
\end{figure*}

\begin{figure*}
  \includegraphics[width=0.49\textwidth]{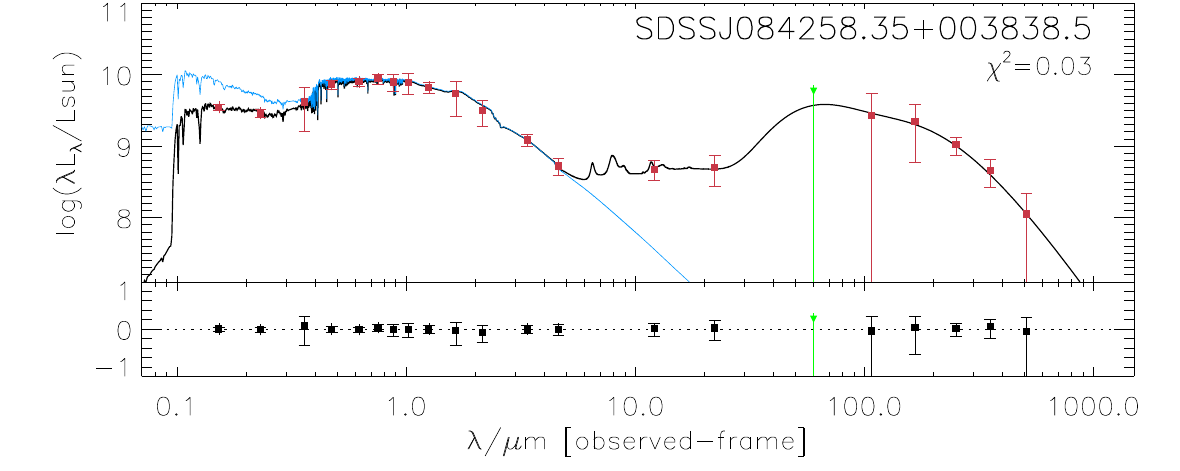}
  \includegraphics[width=0.49\textwidth]{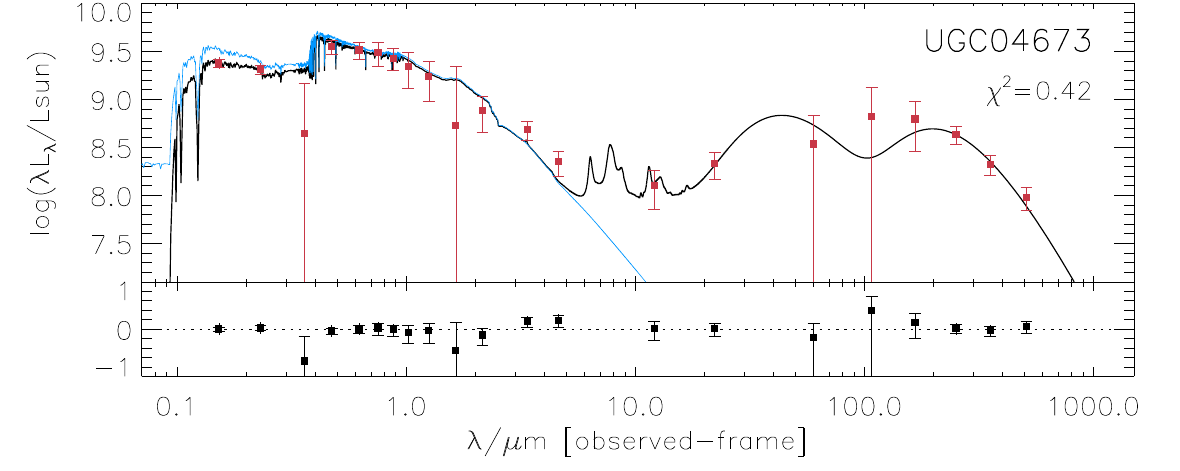}
  \includegraphics[width=0.49\textwidth]{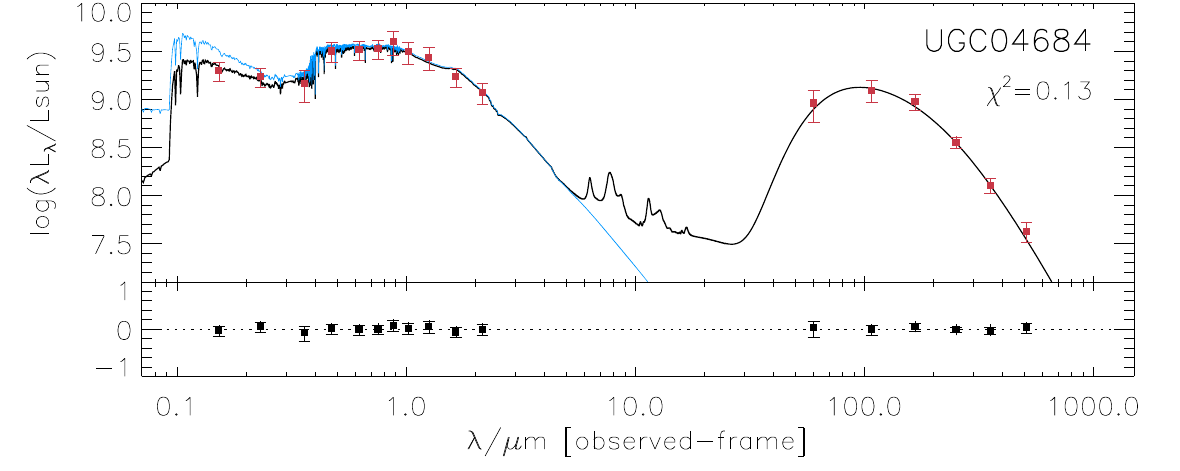}
  \includegraphics[width=0.49\textwidth]{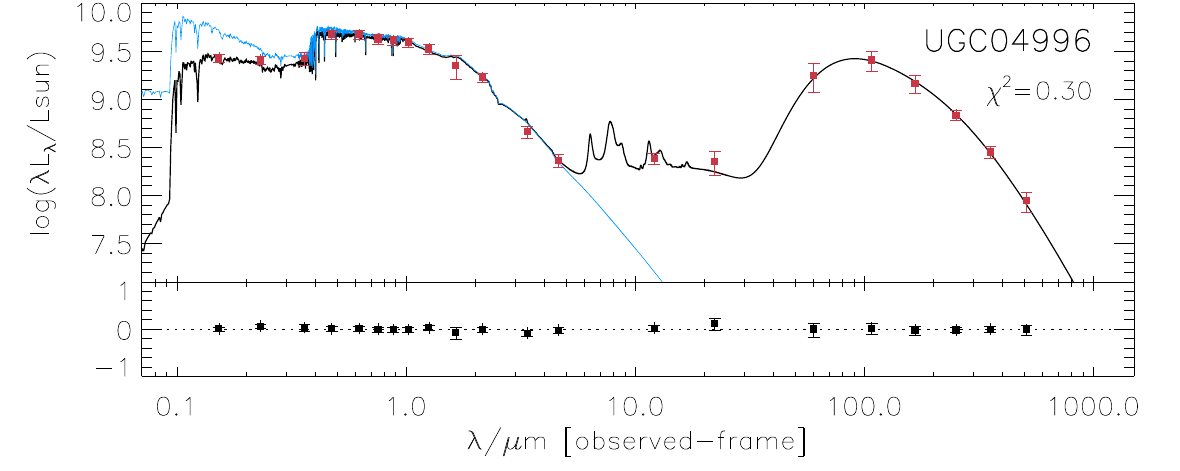}
  \includegraphics[width=0.49\textwidth]{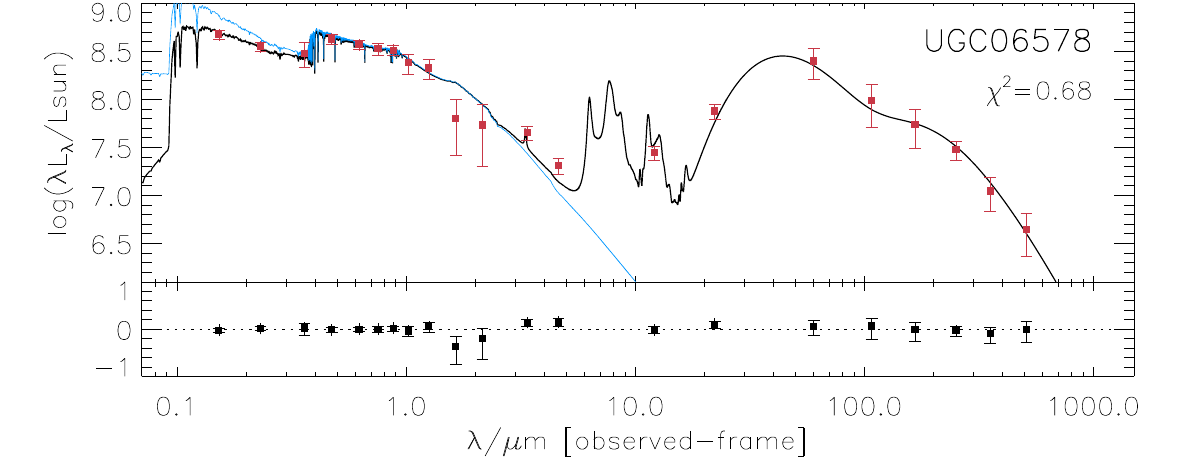}
  \includegraphics[width=0.49\textwidth]{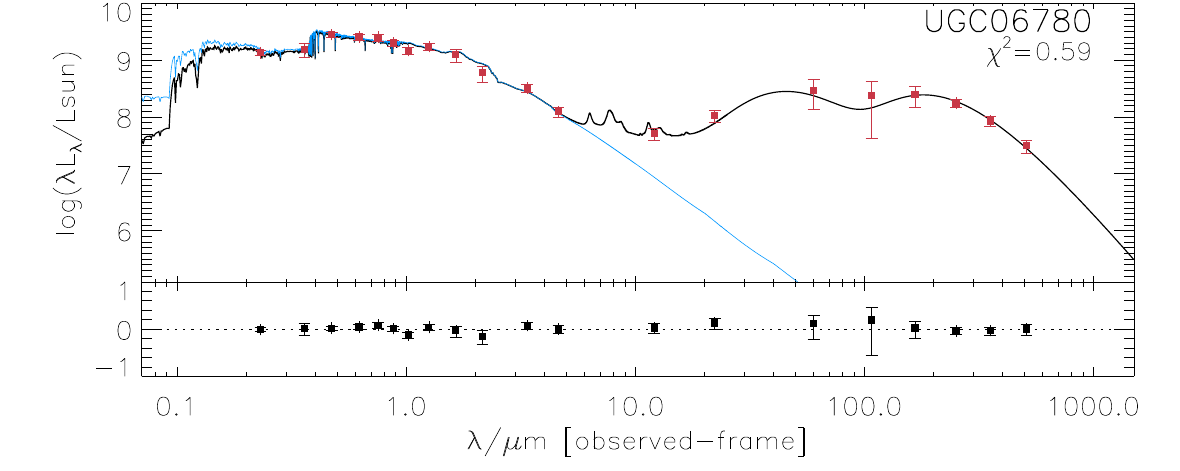}
  \includegraphics[width=0.49\textwidth]{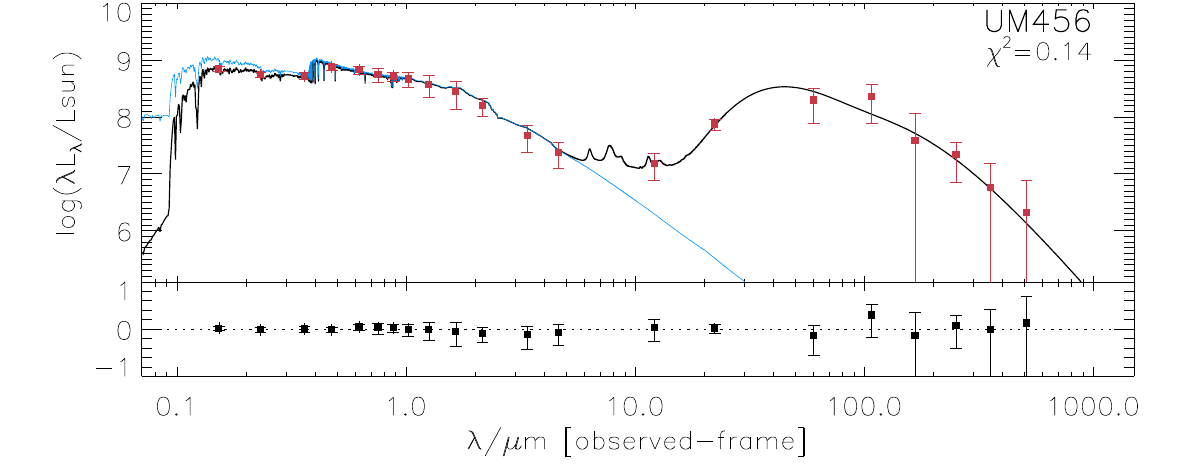}
  \includegraphics[width=0.49\textwidth]{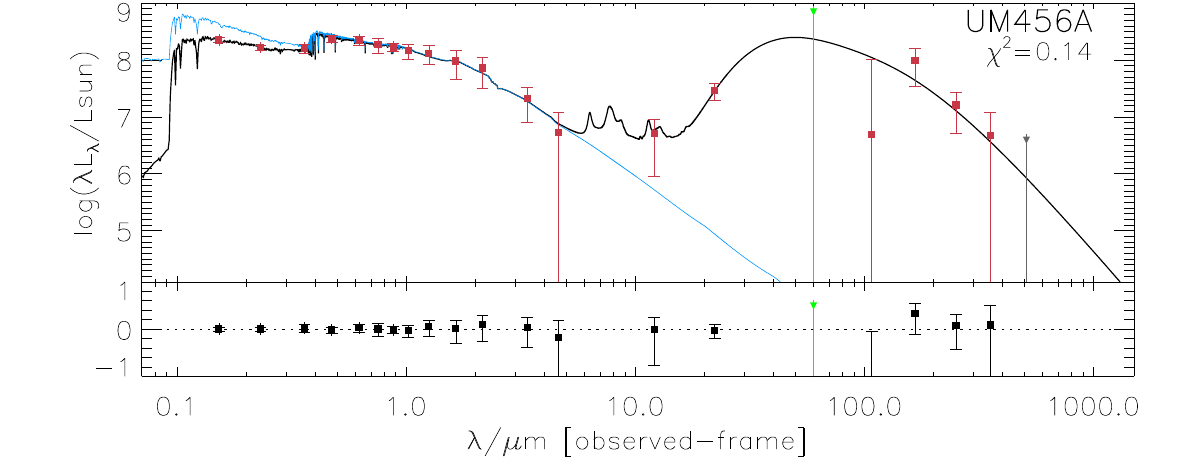}
  \includegraphics[width=0.49\textwidth]{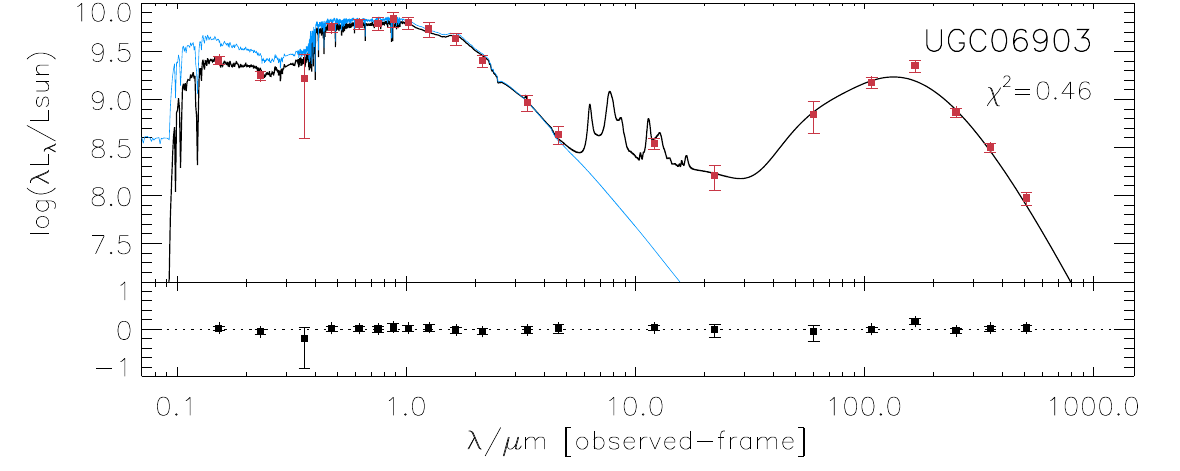}
  \includegraphics[width=0.49\textwidth]{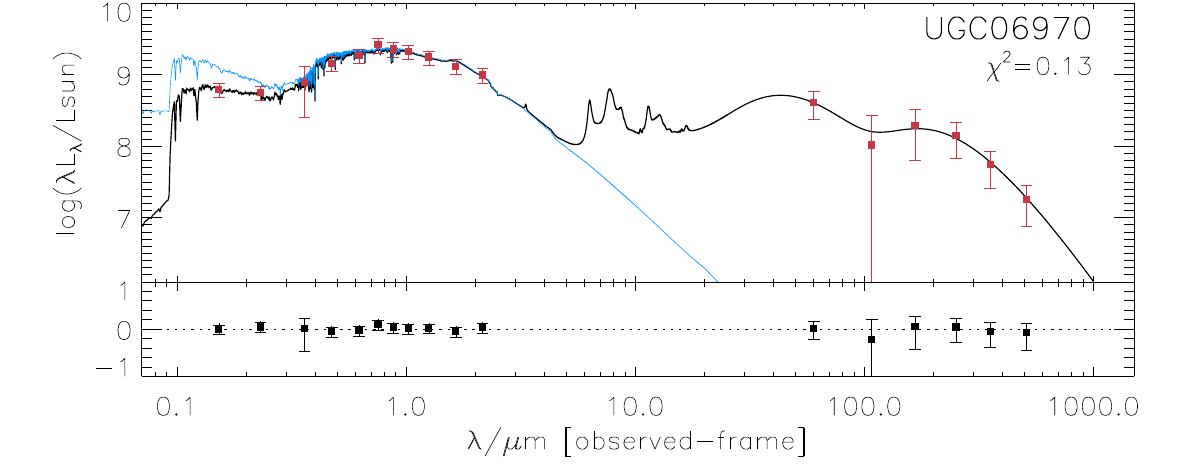}
  \includegraphics[width=0.49\textwidth]{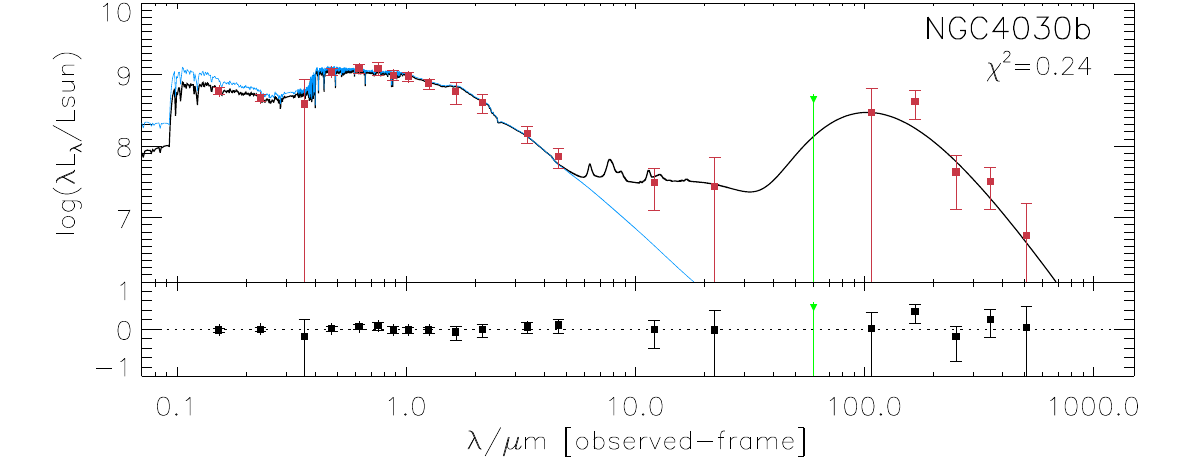}
  \includegraphics[width=0.49\textwidth]{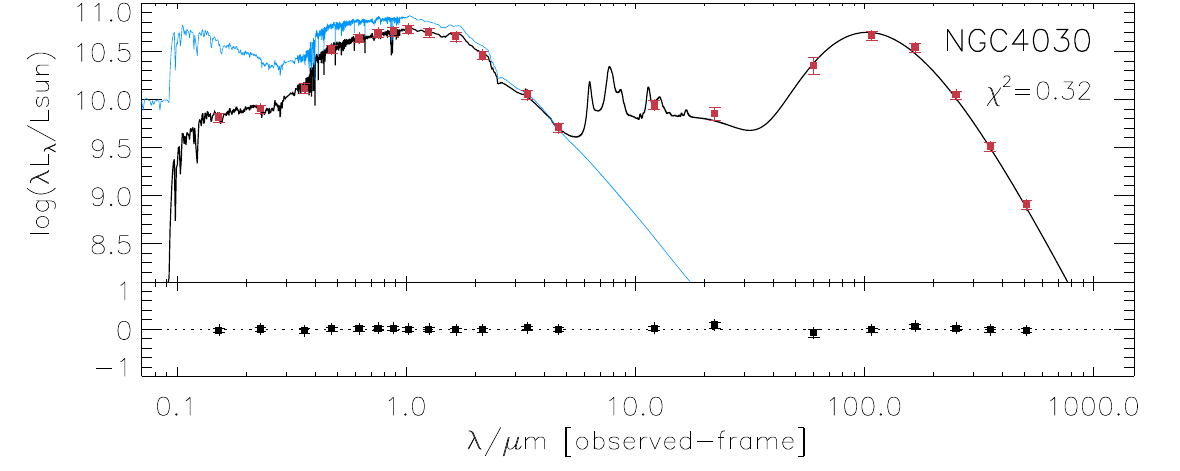}
  \caption{Multiwavelength SEDs of the 40 H{\sc i}-selected sources in H{\sc i}GH, with observed photometry (red points) from FUV to the submillimetre. The photometry process (including determination of errors) is described in Section \ref{photometrysection}. IRAS60 $3\sigma$-upper limits are shown as green triangles. Since negative fluxes cannot be plotted on a logarithmic scale, we have plotted the $1\sigma$ upper limits as orange triangles. The solid black line is the best-fit model SED and the solid blue line is the unattenuated optical model. The residuals of the fit are shown in the panel below each SED. The shown $\chi^2$ are the total $\chi^2$ divided by the number of bands, as given by the standard version of {\sc magphys}.}
  \label{SED}
\end{figure*}

\addtocounter{figure}{-1}

\begin{figure*}
  \includegraphics[width=0.47\textwidth]{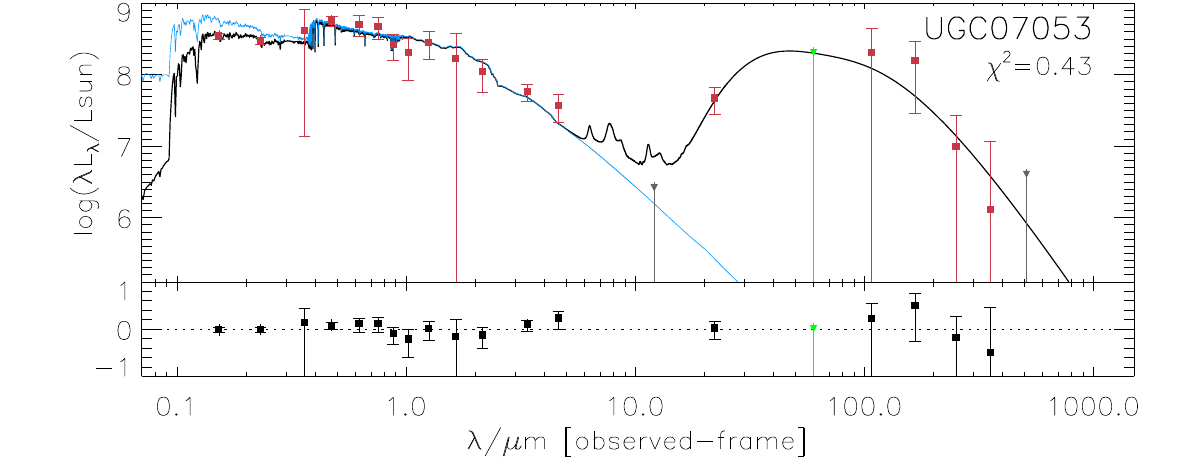}
  \includegraphics[width=0.47\textwidth]{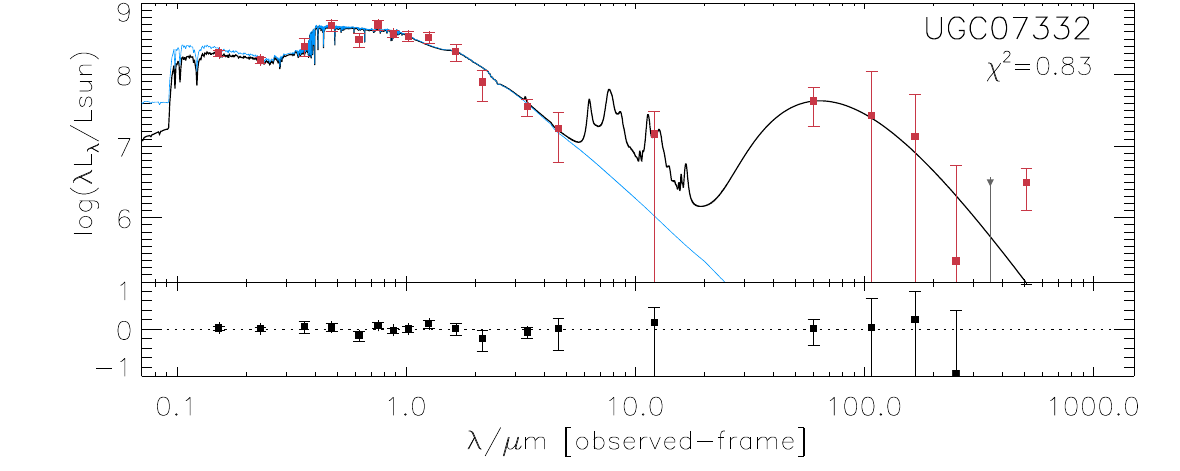}
  \includegraphics[width=0.47\textwidth]{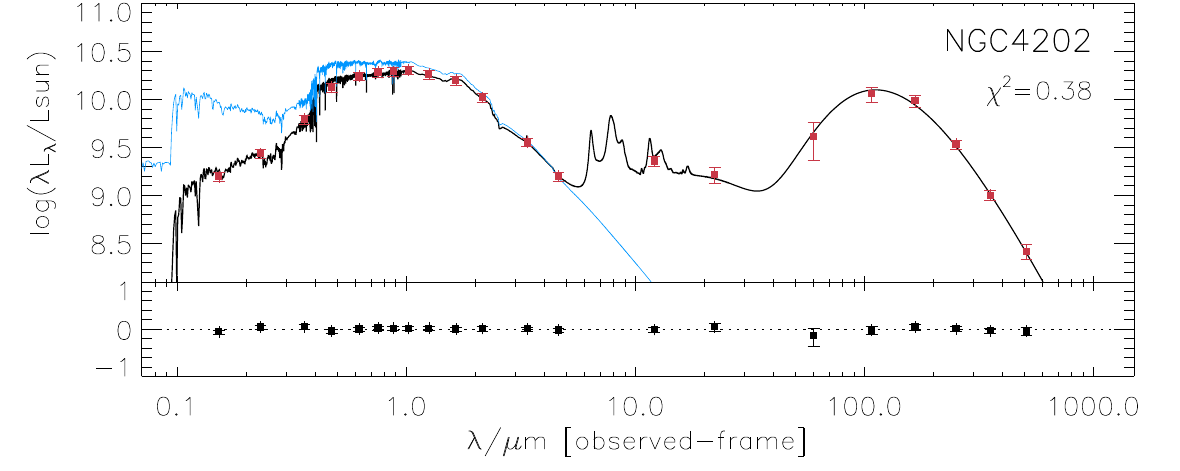}
  \includegraphics[width=0.47\textwidth]{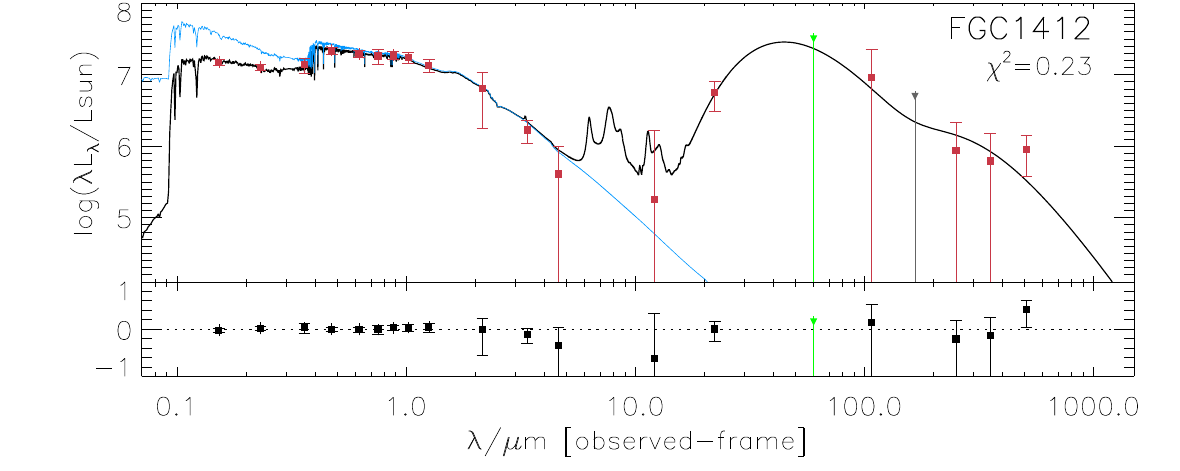}
  \includegraphics[width=0.47\textwidth]{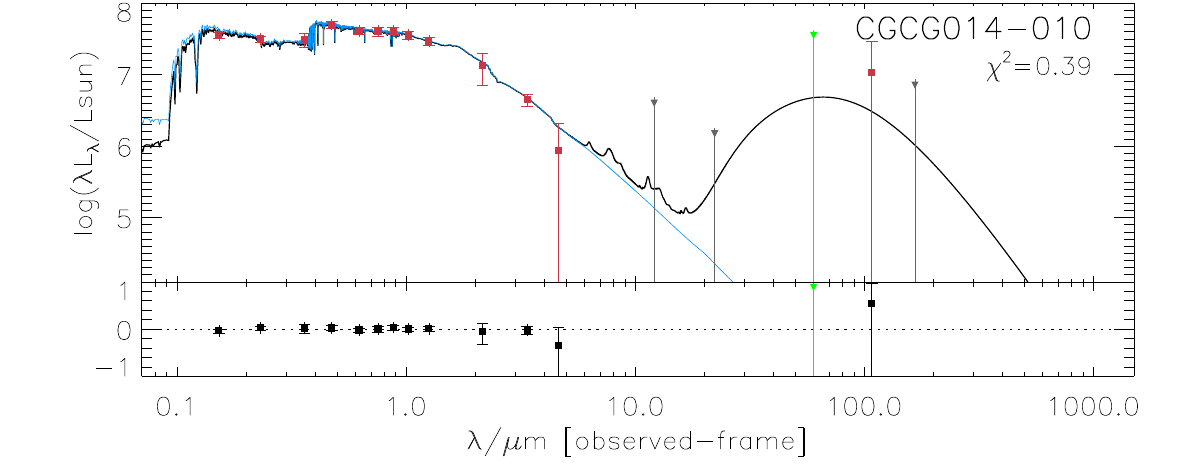}
  \includegraphics[width=0.47\textwidth]{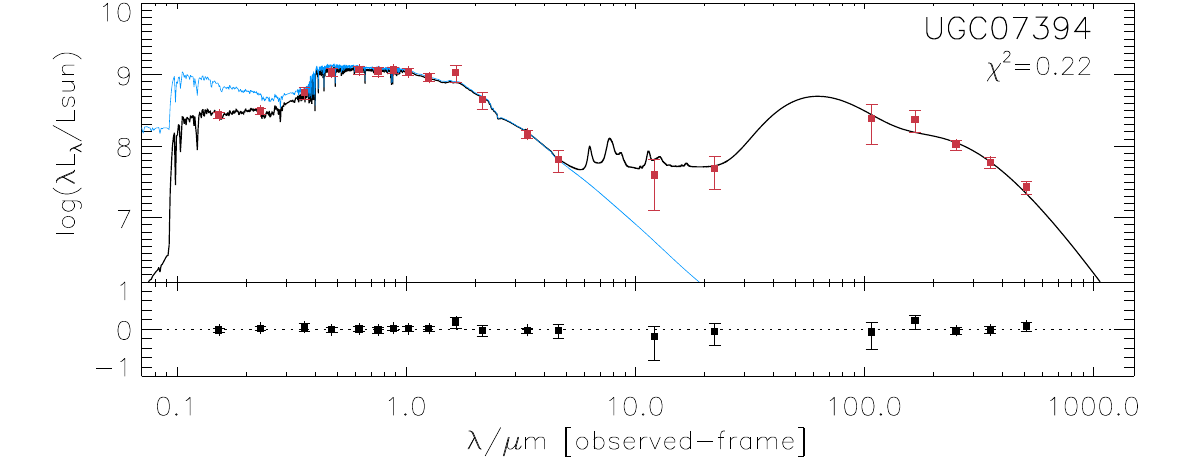}
  \includegraphics[width=0.47\textwidth]{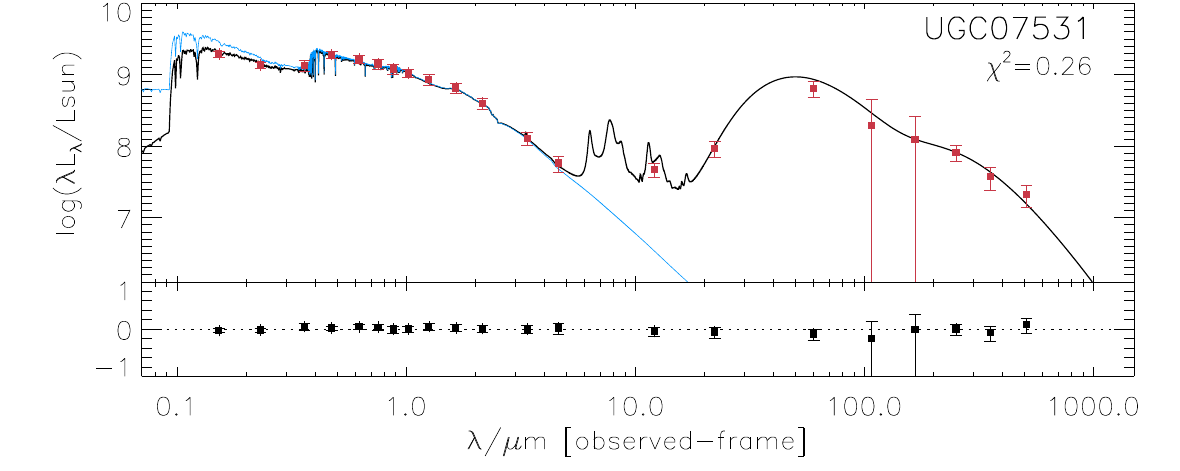}
  \includegraphics[width=0.47\textwidth]{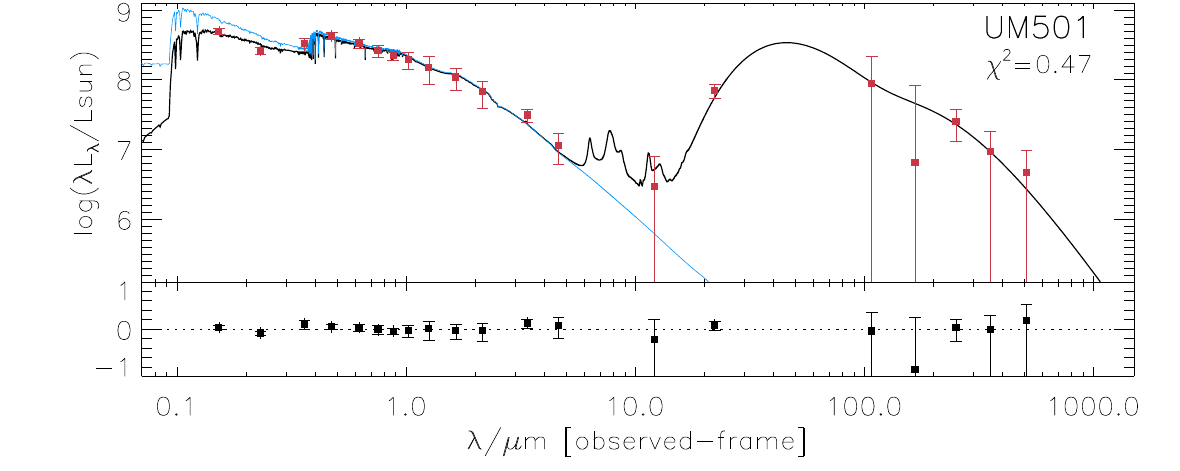}
  \includegraphics[width=0.47\textwidth]{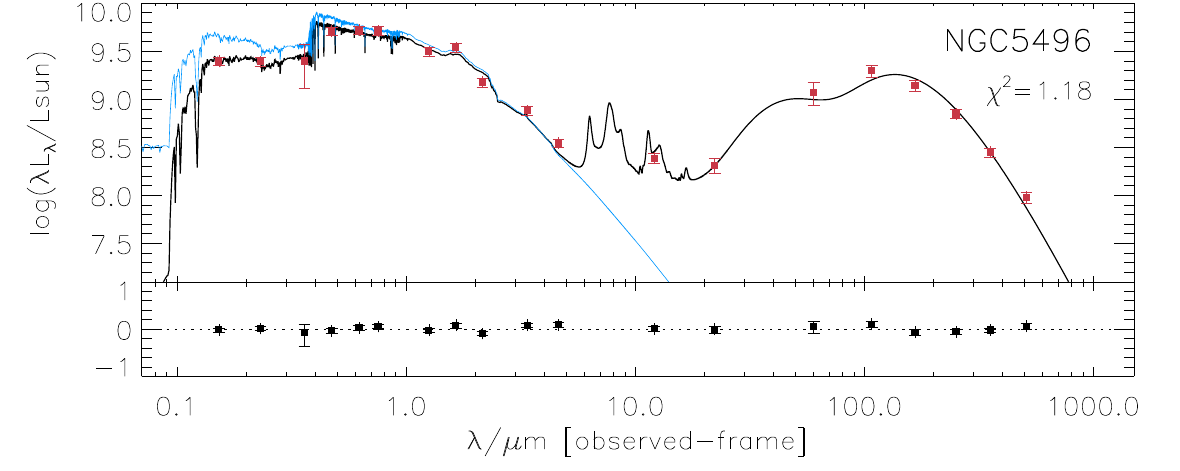}
  \includegraphics[width=0.47\textwidth]{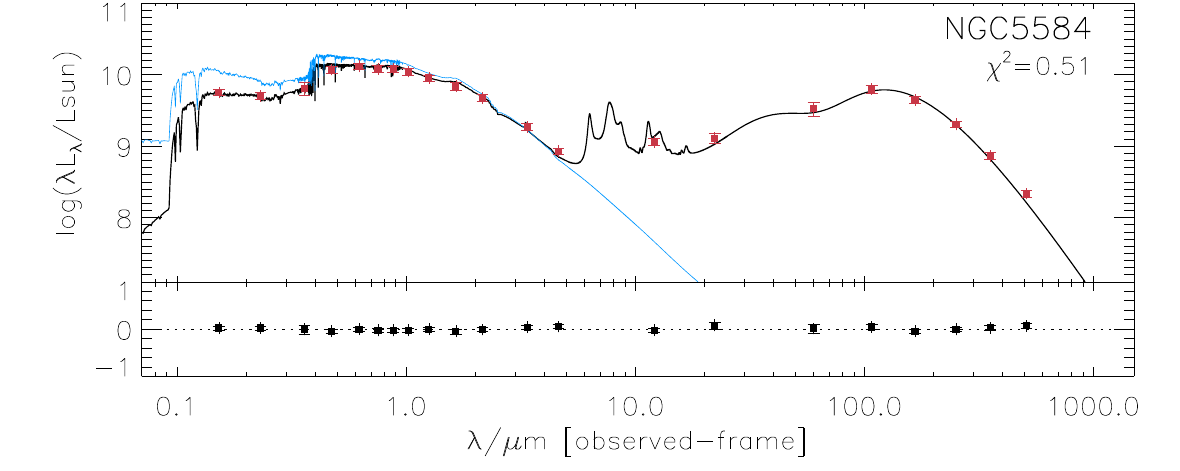}
  \includegraphics[width=0.47\textwidth]{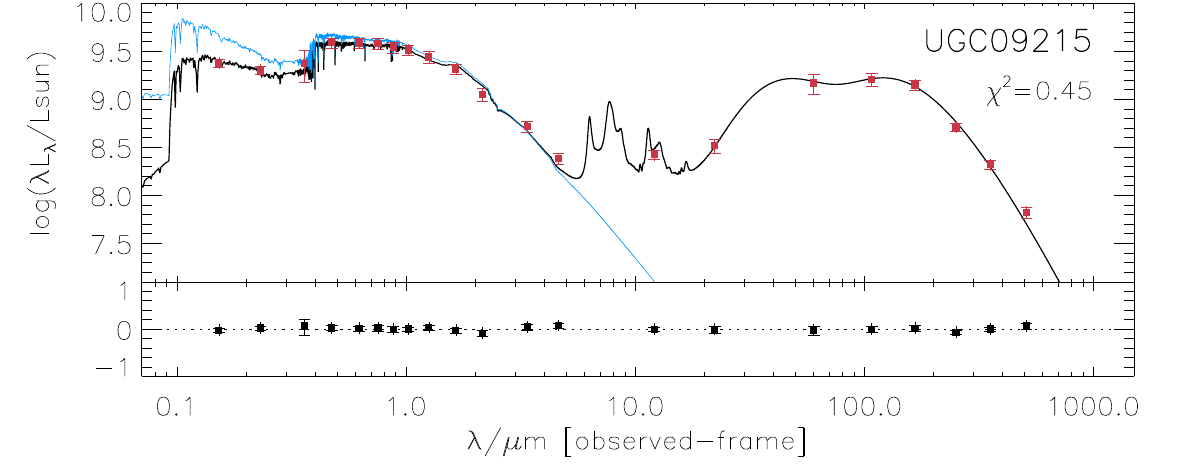}
  \includegraphics[width=0.47\textwidth]{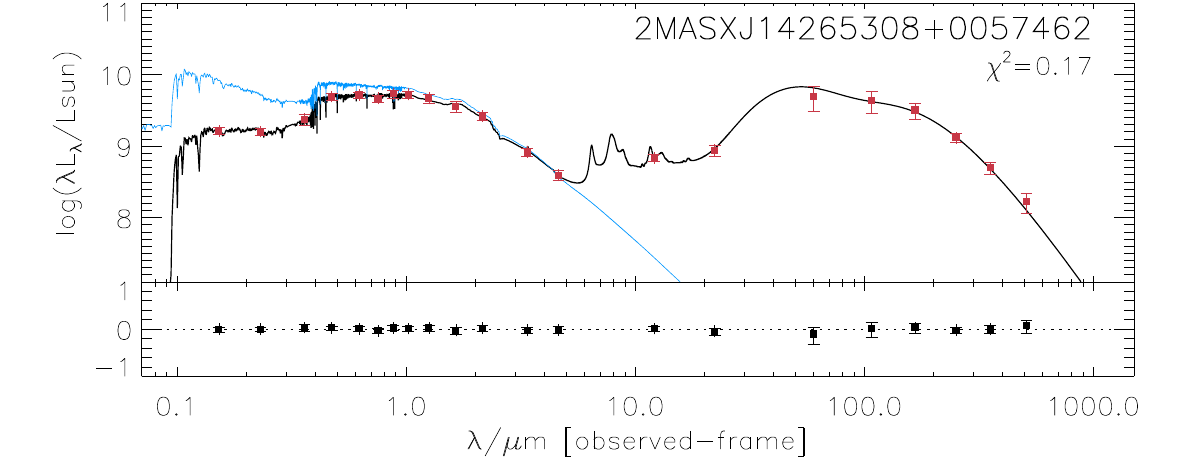}
  \includegraphics[width=0.47\textwidth]{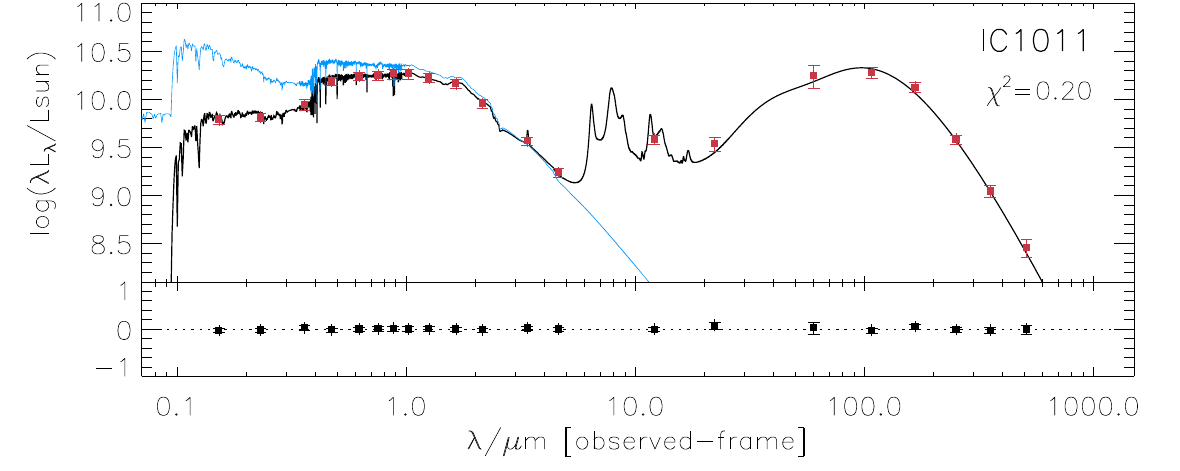}
  \includegraphics[width=0.47\textwidth]{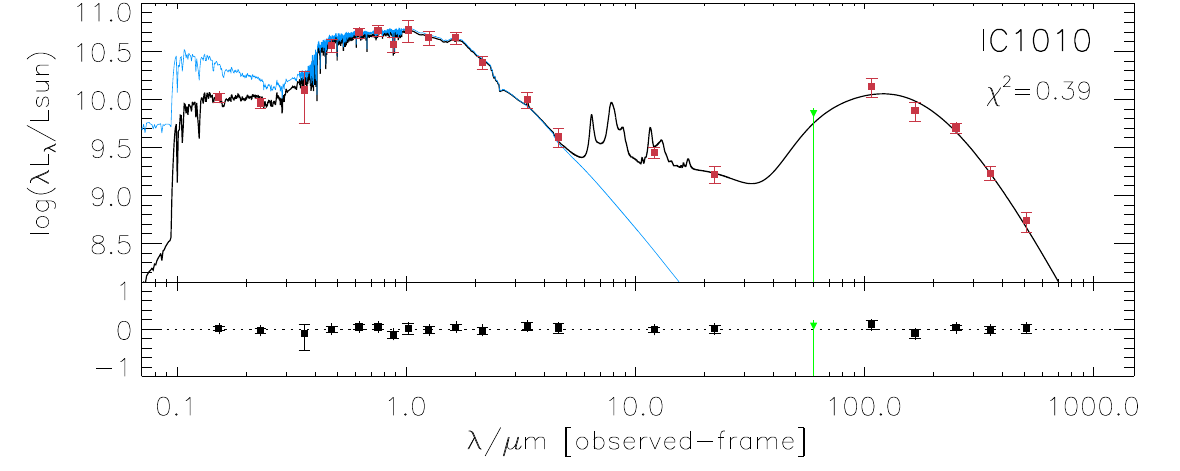}
  \caption{ - \textit{continued}}
\end{figure*}

\addtocounter{figure}{-1}

\begin{figure*}
  \includegraphics[width=0.47\textwidth]{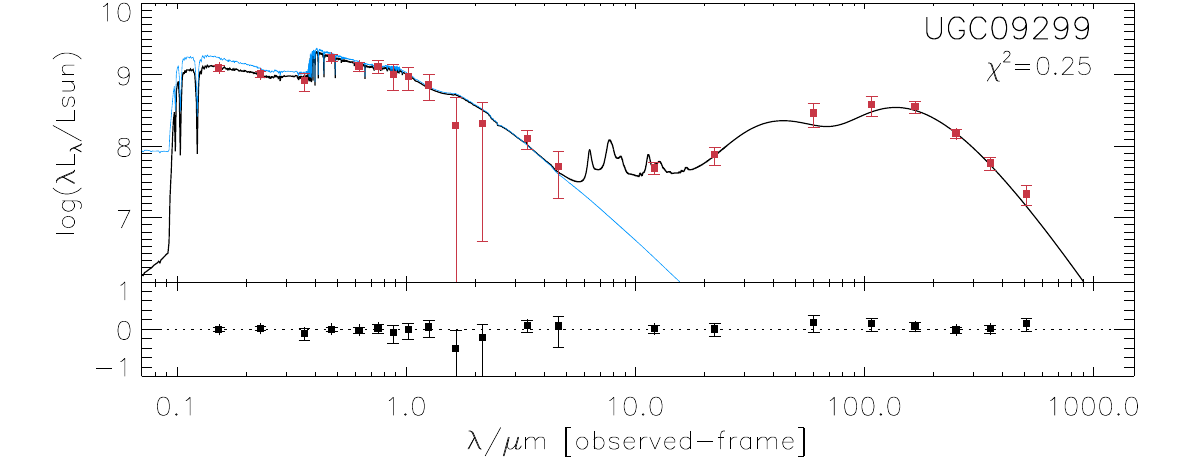}
  \includegraphics[width=0.47\textwidth]{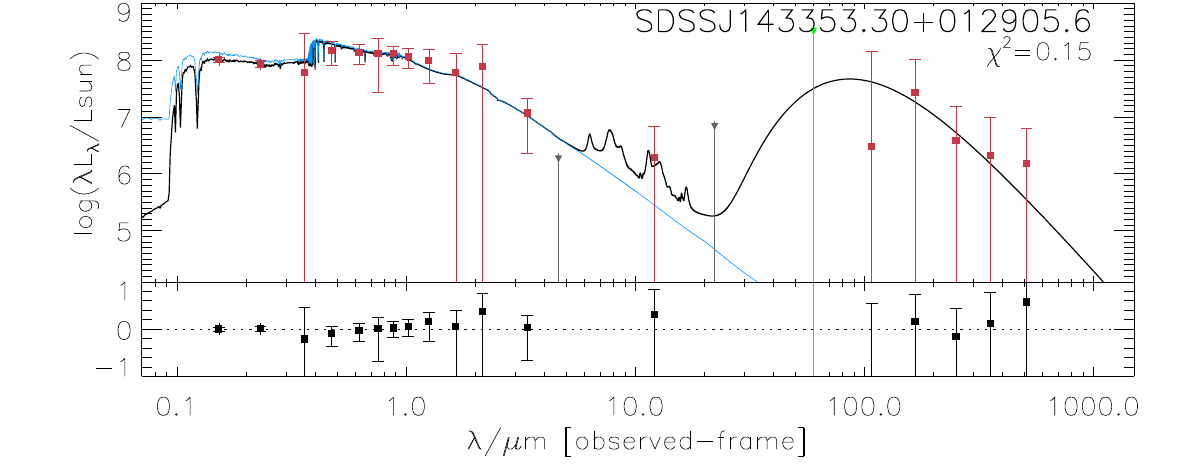}
  \includegraphics[width=0.47\textwidth]{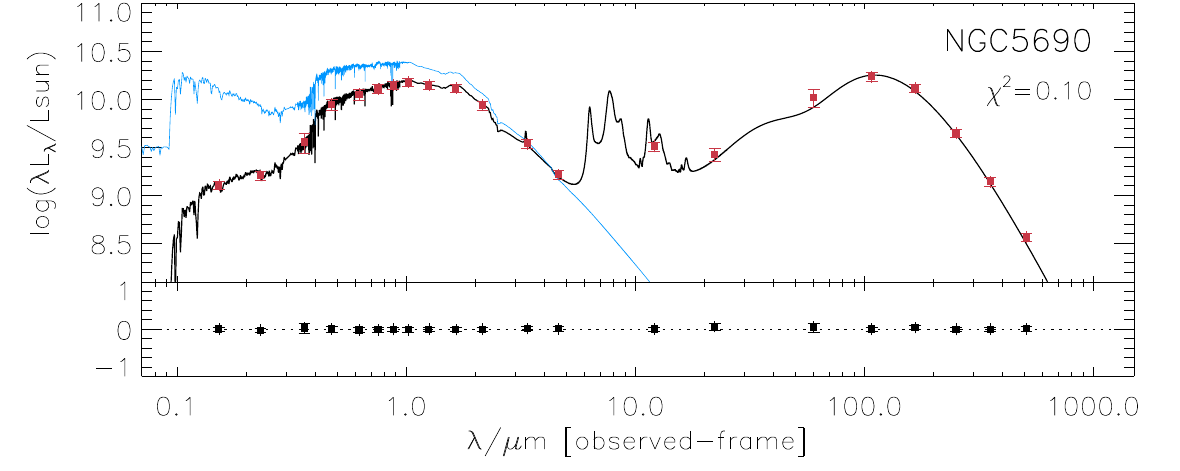}
  \includegraphics[width=0.47\textwidth]{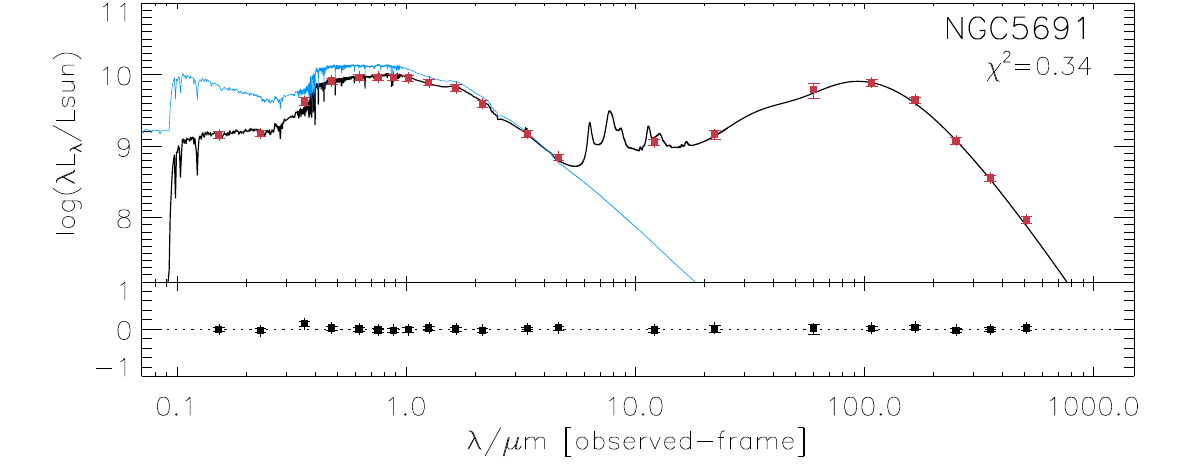}
  \includegraphics[width=0.47\textwidth]{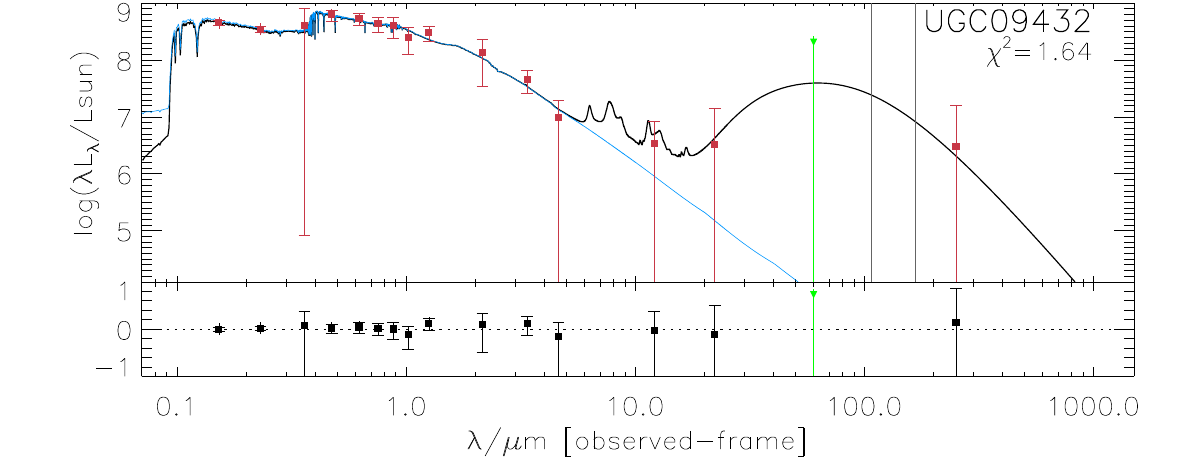}
  \includegraphics[width=0.47\textwidth]{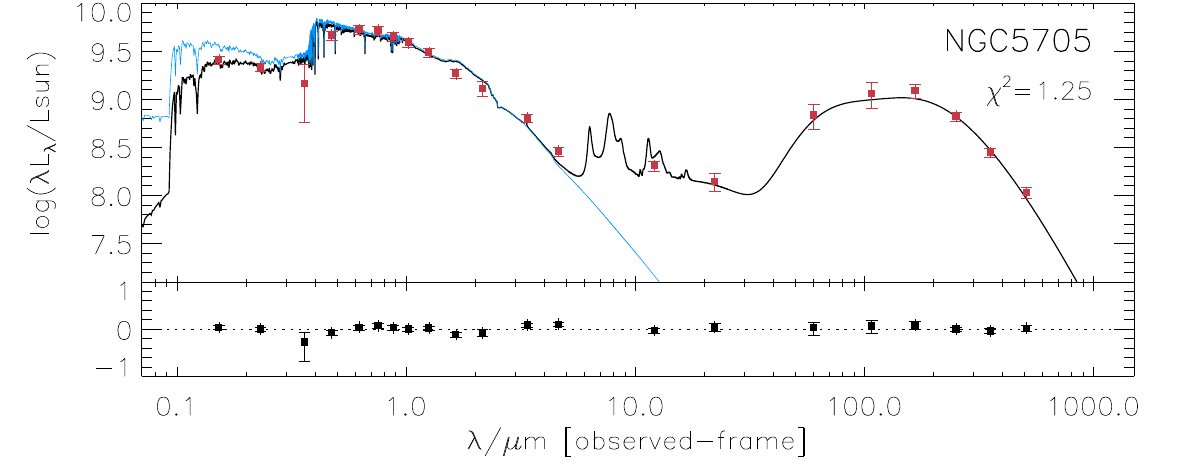}
  \includegraphics[width=0.47\textwidth]{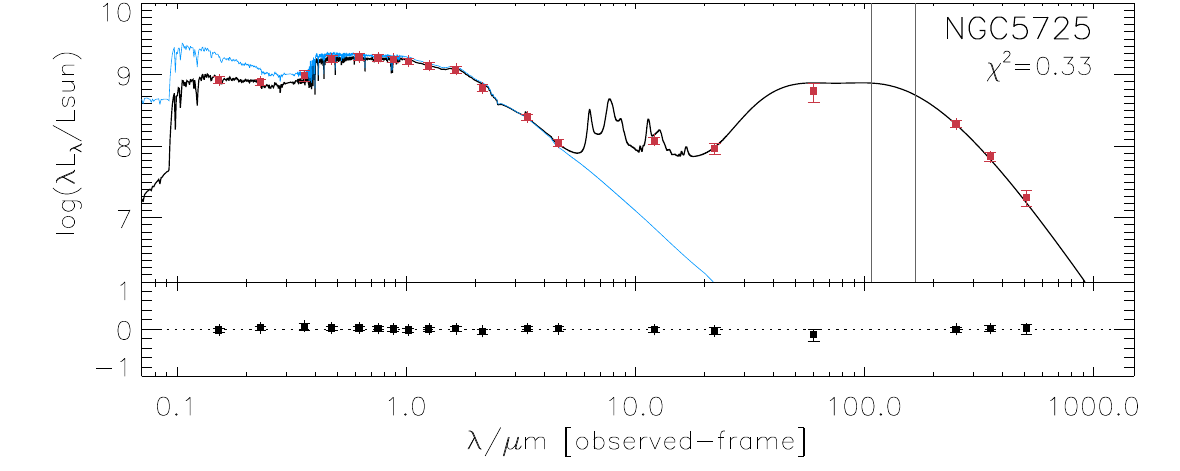}
  \includegraphics[width=0.47\textwidth]{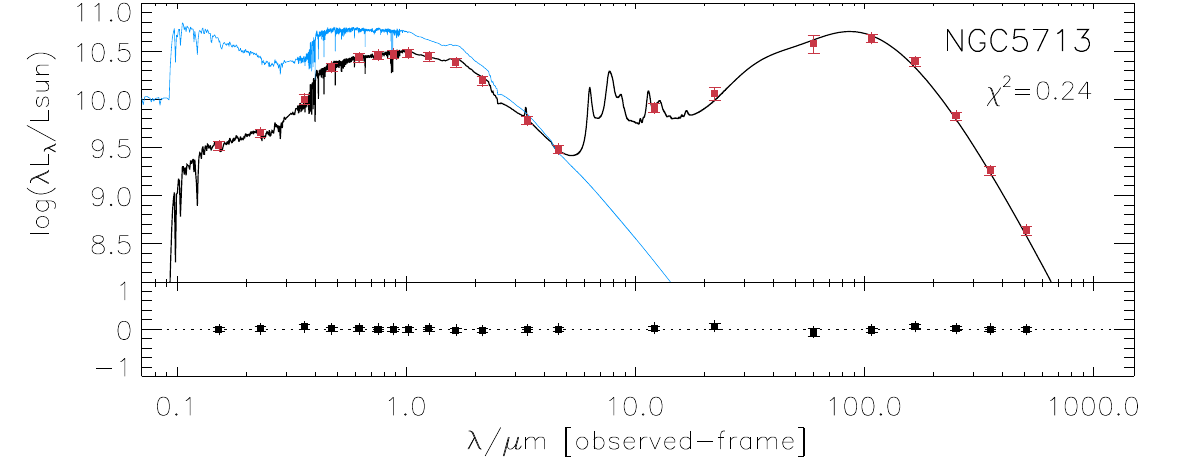}
  \includegraphics[width=0.47\textwidth]{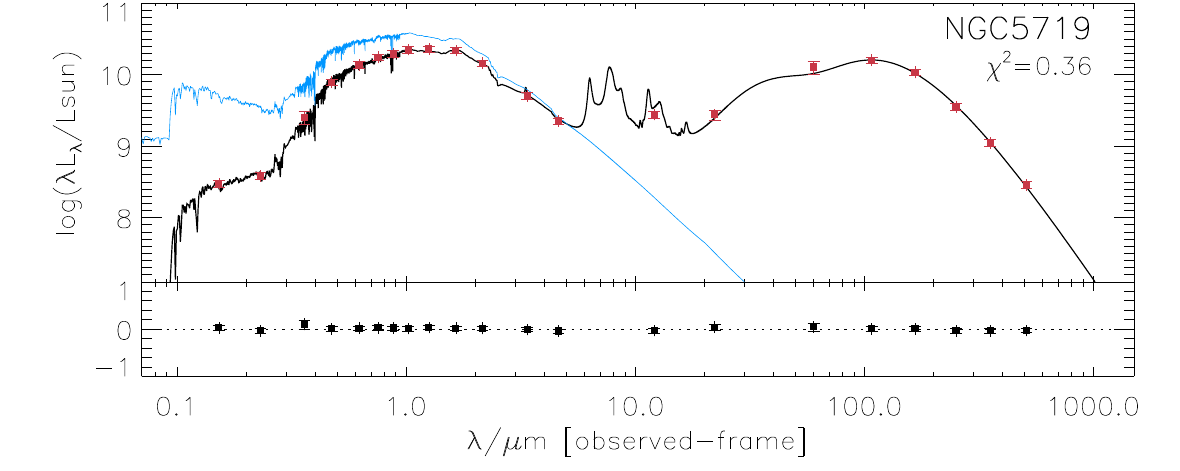}
  \includegraphics[width=0.47\textwidth]{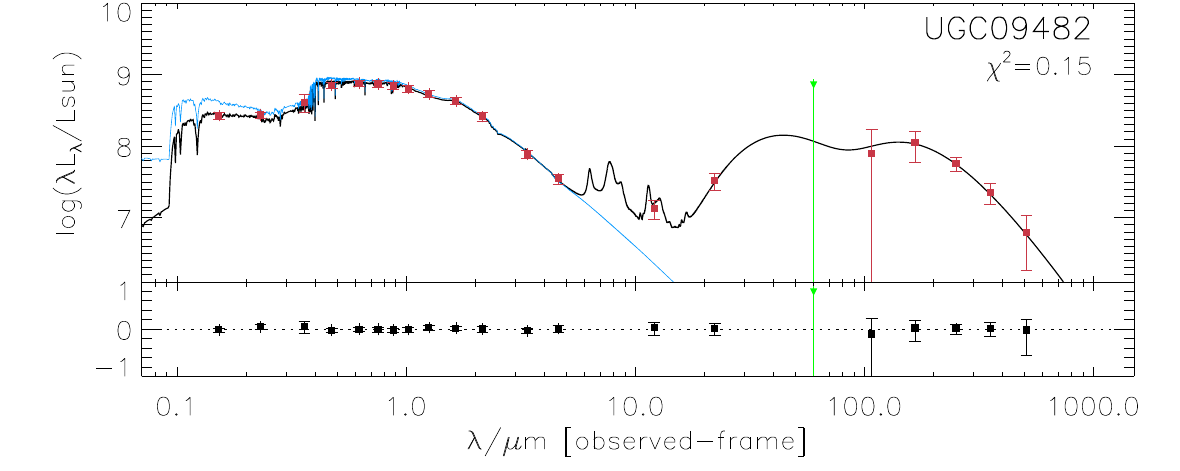}
  \includegraphics[width=0.47\textwidth]{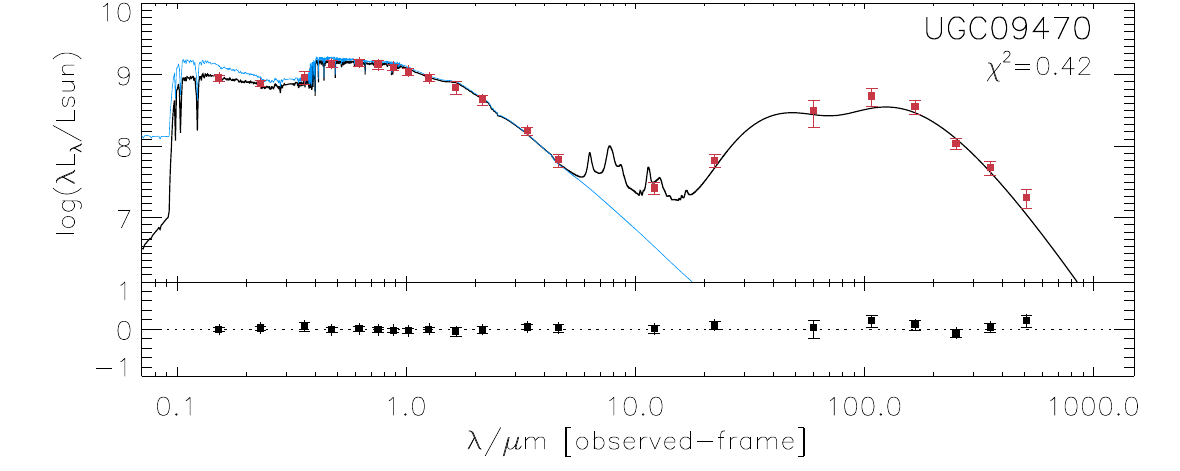}
  \includegraphics[width=0.47\textwidth]{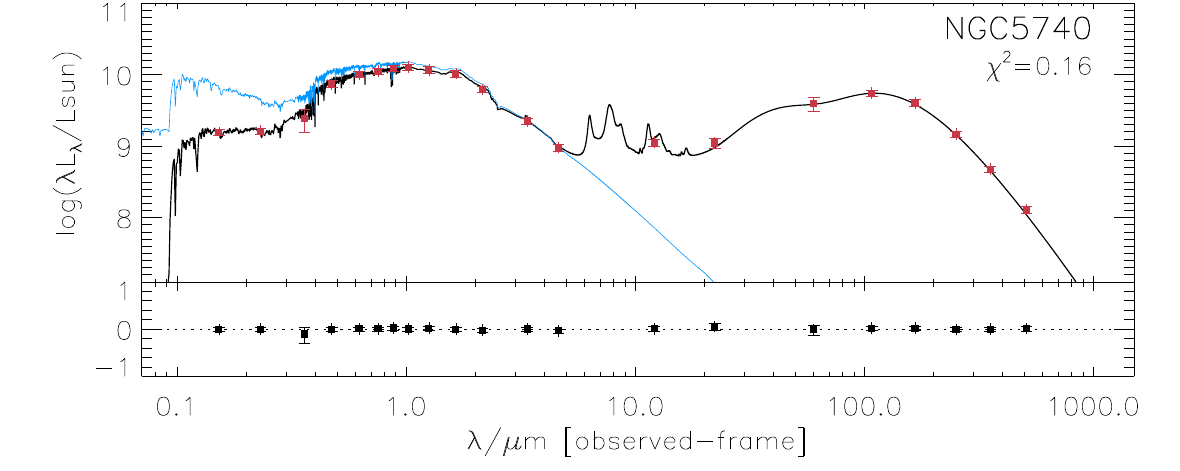}
  \includegraphics[width=0.47\textwidth]{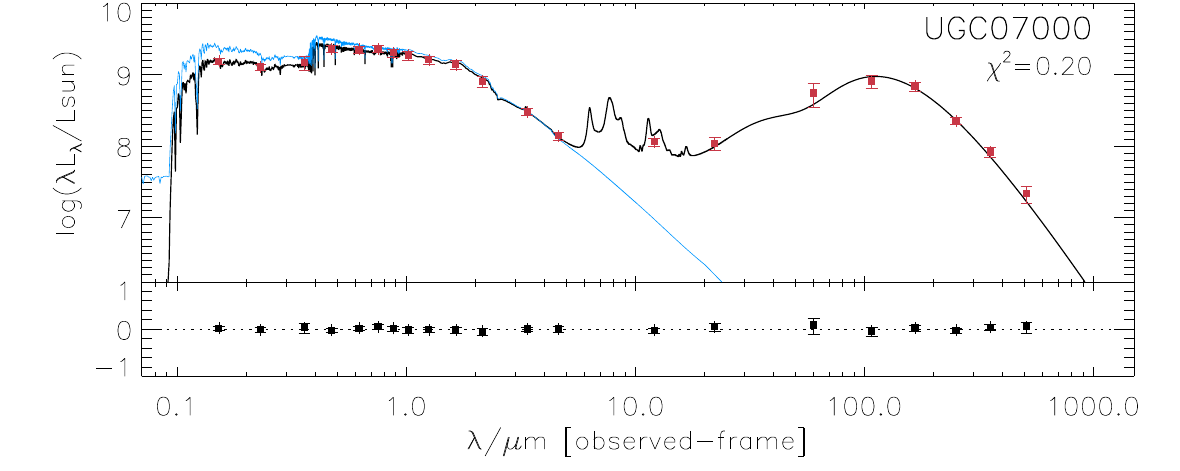}
  \includegraphics[width=0.47\textwidth]{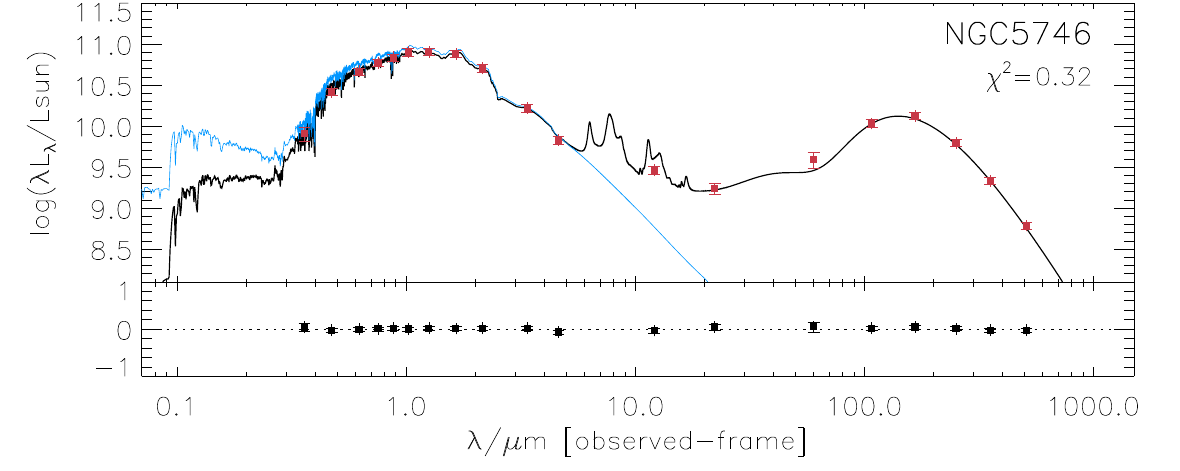}
    \caption{ - \textit{continued}}
\end{figure*}

\end{document}